\documentclass[aps,prl,preprint,superscriptaddress,nobibnotes]{revtex4-2}
\usepackage{graphicx,color}
\usepackage{amsmath}
\usepackage{bm}
\usepackage{dcolumn}
\usepackage{ifthen}
\usepackage{multirow}
\usepackage{threeparttable}
\usepackage{siunitx}
\usepackage{hyperref}
\hypersetup{backref=true, pdfnewwindow=true, linkcolor=blue, colorlinks=true, anchorcolor=blue, citecolor=blue, filecolor=blue,  menucolor=blue, urlcolor=blue}
\usepackage{setspace}
\begin{document}
\title{Which chromium-sulfur compounds exist as 2D material?}

\author{Affan Safeer}
\affiliation{II. Physikalisches Institut, Universität zu Köln, Zülpicher Str. 77, D-50937 Köln, Germany}
\author{Mahdi Ghorbani-Asl}
\affiliation{Helmholtz-Zentrum Dresden-Rossendorf, Institute of Ion Beam Physics and Materials Research, D-01328 Dresden, Germany}
\author{Wouter Jolie}
\affiliation{II. Physikalisches Institut, Universität zu Köln, Zülpicher Str. 77, D-50937 Köln, Germany}
\author{Arkady V. Krasheninnikov}
\affiliation{Helmholtz-Zentrum Dresden-Rossendorf, Institute of Ion Beam Physics and Materials Research, D-01328 Dresden, Germany}
\author{Thomas Michely}
\affiliation{II. Physikalisches Institut, Universität zu Köln, Zülpicher Str. 77, D-50937 Köln, Germany}
\author{Jeison Fischer}
\email{jfischer@ph2.uni-koeln.de}
\affiliation{II. Physikalisches Institut, Universität zu Köln, Zülpicher Str. 77, D-50937 Köln, Germany}

\begin{abstract}

Two-dimensional (2D) chromium-sulfides are synthesized by molecular beam epitaxy using graphene as a substrate. Structure characterization by employing scanning tunneling microscopy and low energy electron diffraction indicates that there are two 2D phases, Cr$_2$S$_3$-2D and Cr$_{2\frac{2}{3}}$S$_4$-2D, which have not been reported before. Cr$_{2\frac{2}{3}}$S$_4$-2D is related to bulk Cr$_5$S$_6$, but thinner than a bulk unit cell. For Cr$_2$S$_3$-2D, an even thinner material, no bulk counterpart exists. Both 2D materials are found to be structurally stable under ambient conditions and exhibit interesting electronic properties. Extensive first-principles calculations provide further insight into the electronic structure of these systems and indicate that they should be magnetic. Although single layers of CrS$_2$ were predicted to be stable by density functional theory calculations and reported in previous experimental studies, we were unable to synthesize CrS$_2$ under our range of experimental conditions.
\end{abstract}

\maketitle
\newpage

\section{Introduction}

Research on chromium-chalcogene compound 2D materials has recently flourished, with significant advancements both theoretically in the prediction of phases and properties as well as experimentally in the preparation and investigation of new phases \cite{Liu2024a,Fan2023,Rajan2024}. One element of excitement, shared with chromium halides \cite{Huang2017,Chen2019,Bedoya-Pinto2021}, is the interesting magnetic properties of these new 2D materials \cite{Chua2021,Lasek2022,Saha2022,Xian2022,Zhang2023,Khatun24,Cui2024,Lu2024,Kushwaha2024}.
Focusing on the case of chromium-sulfur 2D materials \cite{Chu2019,Cui2019,Zhou2019,Xie2021,Moinuddin2021,Cui2022,Liu2022,Yao2023,Song2024,Habib2019,Nair2022,Xiao2022a,Yao2024,Khan2024,Shivayogimath2019,Chen2021,Habibb2019,Lei2024,Liu2024b,Sun2020,Wang2018,Zhuang2014,Gao2014,Li2023,Zhang2021,Zhang2019} it seems not yet to be settled which phases exist as 2D materials. Neither were all compounds predicted from DFT calculations experimentally fabricated nor is the experimental research comprehensive enough to make additional discoveries unlikely. This situation is one of the motivations for the present study.  

While the structure and composition of 2D materials created by exfoliation are known from their bulk crystal, the situation is far more intricate for 2D materials grown by molecular beam epitaxy (MBE) or chemical vapor deposition (CVD). These 2D materials may possess no bulk counterpart; thus, their structure and composition are not known a priori. 
The thinness and lack of signal make standard techniques like X-ray diffraction or Raman spectroscopy difficult to apply. The unknown arrangement of planes of different species, depth-dependent attenuation, and vacancy-containing planes impair even composition analysis methods like X-ray photoelectron spectroscopy (XPS). Frequently, density functional theory (DFT) calculations are used as a guideline for interpreting experimental compounds, although stability in DFT calculations guarantees neither a pathway for creation nor the existence of a specific phase. All of this is relevant to 2D materials based on chromium and sulfur.

The bulk chromium sulfur phase diagram is complex
\cite{Massalski2007,Jellinek1957,Popma1969,Rau1977}, but no Cr-rich phases have been documented. Unquestioned S-rich phases are limited in stoichiometry by Cr$_2$S$_3$, except of the more S-rich Cr${_5}$S$_8$ phase which can be prepared under high pressure. Noteworthy, no bulk phase with van der Waals gaps has been reported. 

\begin{figure}
\includegraphics[width=\linewidth]{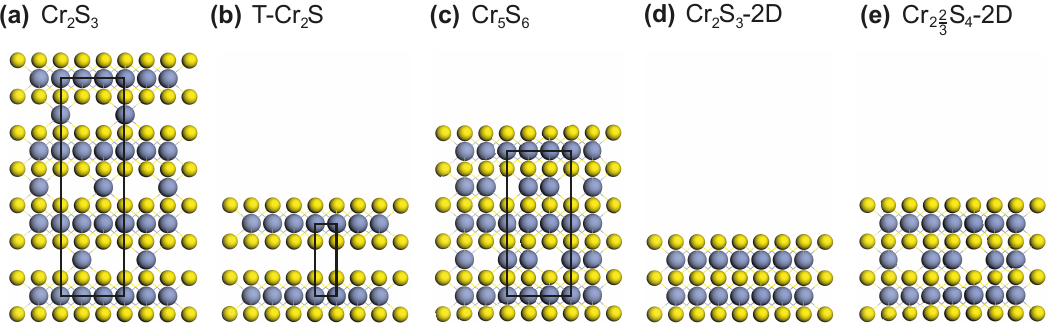}
  \caption{Schematic side view ball models of Cr$_x$S$_y$ materials. (a) Rhombohedral Cr$_2$S$_3$. (b) T-phase CrS$_2$. (c) Cr$_5$S$_6$. (d) Cr$_2$S$_3$-2D. (e) Cr$_{2\frac{2}{3}}$S$_4$-2D. In the ball models of bulk materials (a)-(c), a unit cell is indicated, in (d) and (e) single unit cell thick 2D materials are shown. Additional schematic top and side view ball model views visualizing the full structure are given in Figure~S1 (Supporting Information). 
  }
  \label{fgr:fig_ball_models}
\end{figure}

Bulk Cr$_2$S$_3$ can be considered to be composed of hypothetical CrS$_2$ layers with octahedral (T) Cr coordination, where the interlayer gaps are filled by 1/3 of the possible sites establishing chemical bonds between the CrS$_2$ layers. Depending on the precise conditions of preparation and composition, the distribution of interlayer Cr atoms may be disordered or give rise to trigonal or rhombohedral stacking [see  Figure~\ref{fgr:fig_ball_models}(a)] \cite{Jellinek1957,Yuzuri1964,Mikami1972}. For the latter two, the arrangement of the interlayer Cr repeats every two or three CrS$_2$ layers, respectively. Complex magnetically ordered structures are reported, which critically depend on the arrangement of Cr interlayer atoms and the exact composition \cite{Yuzuri1964,vanLaar1967,Popma1971,Mikami1972,Maignan1912}. The magnetic moments are in-plane, normal to the c-axis \cite{vanLaar1967,Mikami1972}. Rhombohedral Cr$_2$S$_3$ is a semiconductor, paramagnetic above and ferrimagnetic below 120\,K with a maximum magnetization around 90\,K \cite{Popma1971,Yuzuri1964,Mikami1972,Maignan1912}. Negative magnetostriction and photoconductivity were also reported for rhombohedral Cr$_2$S$_3$ \cite{Maignan1912,Anedda1982}.

Rhombohedral \cite{Chu2019,Cui2019,Zhou2019,Xie2021,Moinuddin2021,Cui2022,Liu2022,Yao2023,Song2024} and trigonal \cite{Fan2024} Cr$_2$S$_3$ were synthesized as thin films,either using chemical vapor deposition or sulfurization of a metallic film. 2D material of few unit cell thickness for trigonal \cite{Fan2024} and single unit cell thickness for rhombohedral Cr$_2$S$_3$  \cite{Chu2019,Cui2019,Xie2021,Liu2022,Song2024} was reported. For the latter, the unit cell consists of four CrS$_2$ layers with three planes of 1/3 Cr sites occupied in the interlayer gaps, just as shown in Figure~\ref{fgr:fig_ball_models}(a)]. However, in a transmission electron microscopy study the arrangement of Cr interlayer atoms in as-grown "rhombohedral" CVD films was found to be random \cite{Liu2022}. Similar to the bulk counterpart, semiconducting behavior, magnetization with a T$_\mathrm{C} = 120$\,K, negative magnetostriction, and good photosensitivity were found \cite{Cui2019,Moinuddin2021,Xie2021}. 

No bulk crystals of composition CrS$_2$ have yet been reported. The closest to the existence of bulk CrS$_2$ are misfit layer compounds. These crystals consist of layers of monoclinic CrS$_2$ alternating  with layers of LaS, BiS or Y \cite{Kato1990, Lafond1992, Lafond1994}. Although bulk CrS$_2$ is absent, DFT calculations proposed the existence of CrS$_2$ as a typical transition metal dichalcogenide with the S atomic planes sandwiching the Cr atomic plane [see Figure~\ref{fgr:fig_ball_models}(b)] \cite{Zhuang2014,Wang2018}. H- (trigonal prismatic Cr coordination), T-, and T'- phases were identified as possible variants, with ambiguity on the most stable phase being either the H- \cite{Zhuang2014,Chen2021} or the T-phase \cite{Wang2018}. Recently, also two different orthorhombic CrS$_2$ phases were identified in calculations \cite{Lei2024,Xiao2022b}. DFT based research for CrS$_2$ materials is flourishing due to exciting predicted properties ranging from ferromagnetism \cite {Wang2018,Lei2024,Chen2021} with Curie temperatures up to 1000\,K, over a direct band gap with valley polarization \cite{Zhuang2014} to excellent performance for the hydrogen evolution reaction \cite{Sun2020}, to name a few.

Despite the absence of the bulk compound, the synthesis of CrS$_2$ as 2D material has been communicated \cite{Habib2019,Shivayogimath2019,Xiao2022a,Nair2022,Yao2024}, in the T-phase \cite{Habib2019,Xiao2022a,Nair2022,Yao2024,Zhang2021,Li2023}, but also in the H- and T'-phases \cite{Habib2019}. CrS$_2$ islands tend to grow thicker than single unit cell, but in two cases minimal heights in the range of $\approx 0.75$\,nm were measured \cite{Habib2019,Yao2024}. Transport in CrS$_2$ field effect devices was characterized as p-type \cite{Habib2019}, n-type \cite{Nair2022}, or metallic \cite{Xiao2022a,Yao2024}. Weak ferromagnetic behavior up to room temperature \cite{Xiao2022a} and use as a floating gate in photoelectric non-volatile memory \cite{Yao2024} were described. 

In bulk Cr$_5$S$_6$, 2/3 of the possible Cr sites in the interlayer gaps are filled, i.e., it is a material in the NiAs structure with 1/3 of Cr vacancies in every second Cr layer [see Figure~\ref{fgr:fig_ball_models}(c)] \cite{Jellinek1957}. Cr$_5$S$_6$ is antiferromagnetic below 168\,K and ferrimagnetic up to 303\,K \cite{vanLaar1967}. Until now, to the authors' best knowledge, preparation of Cr$_5$S$_6$ as a 2D material has not yet been reported. 

Lastly, there is a yet unconfirmed DFT prediction for a new Cr$_x$S$_y$ 2D material, namely Cr$_2$S$_3$-2D [see Figure~\ref{fgr:fig_ball_models}(d)] \cite{Zhang2021,Li2023}. To avoid confusion with bulk materials with the same composition but different structure, "-2D" is attached to the stoichiometric formulas in the 2D case from hereon.  Cr$_2$S$_3$-2D is unrelated to bulk Cr$_2$S$_3$ and consists of five planes of atoms in the sequence S-Cr-S-Cr-S in the NiAs structure, which can be related to bulk CrS, although it lacks the monoclinic distortion observed there \cite{Jellinek1957}. Based on DFT calculations, Cr$_2$S$_3$-2D is proposed to be a magnetic semiconductor with a transition temperature above room temperature (405\,K) \cite{Lu2023}. The key difference of Cr$_2$S$_3$-2D compared to other 2D magnetic semiconductors, such as CrI$_3$ (ordering temperature $\approx 45$~K~\cite{Huang2017}) and CrSBr (ordering temperature $\approx 146$~K~\cite{Lee2021}), is the presence of both intralayer and interlayer exchange coupling \cite{Zhang2021}. The additional presence of interlayer exchange coupling might be a decisive factor enhancing the magnetic ordering temperature of Cr$_2$S$_3$-2D. Increasing covalency of bonds is an additional factor enhancing superexchange interactions \cite{Wang2022}. Therefore one may expect an increasing ordering temperature from CrI$_3$ over CrSBr to Cr$_2$S$_3$-2D.

Given the diversity of materials and predictions, it remains worthwhile to carefully characterize and investigate Cr$_x$S$_y$-2D phases. In this paper, we use molecular beam epitaxy (MBE) to prepare Cr$_x$S$_y$-2D on graphene and subsequently characterize the materials in situ by low energy electron diffraction (LEED), scanning tunneling microscopy (STM) and spectroscopy (STS). Two phases can be prepared phase pure and are identified as Cr$_2$S$_3$-2D [Figure~\ref{fgr:fig_ball_models}(d)] and  Cr$_{2\frac{2}{3}}$S$_4$-2D [see Figure~\ref{fgr:fig_ball_models}(e)]. While Cr$_2$S$_3$-2D has already been predicted to exist as mentioned above, Cr$_{2\frac{2}{3}}$S$_4$-2D lacks such a DFT based forecast. It can be considered the minimum thickness variant of bulk Cr$_5$S$_6$, i.e., it consists of two CrS$_2$ layers with additional Cr filling 2/3 of the sites in the interlayer gap. Moreover, based on our experiments, we tentatively assert that the existence of CrS$_2$ as 2D-material is questionable, consistent with the non-existence of this material in bulk.

\section{Results and discussion}

\subsection{Structure from experiment}
\begin{figure}
\includegraphics[width=\linewidth]{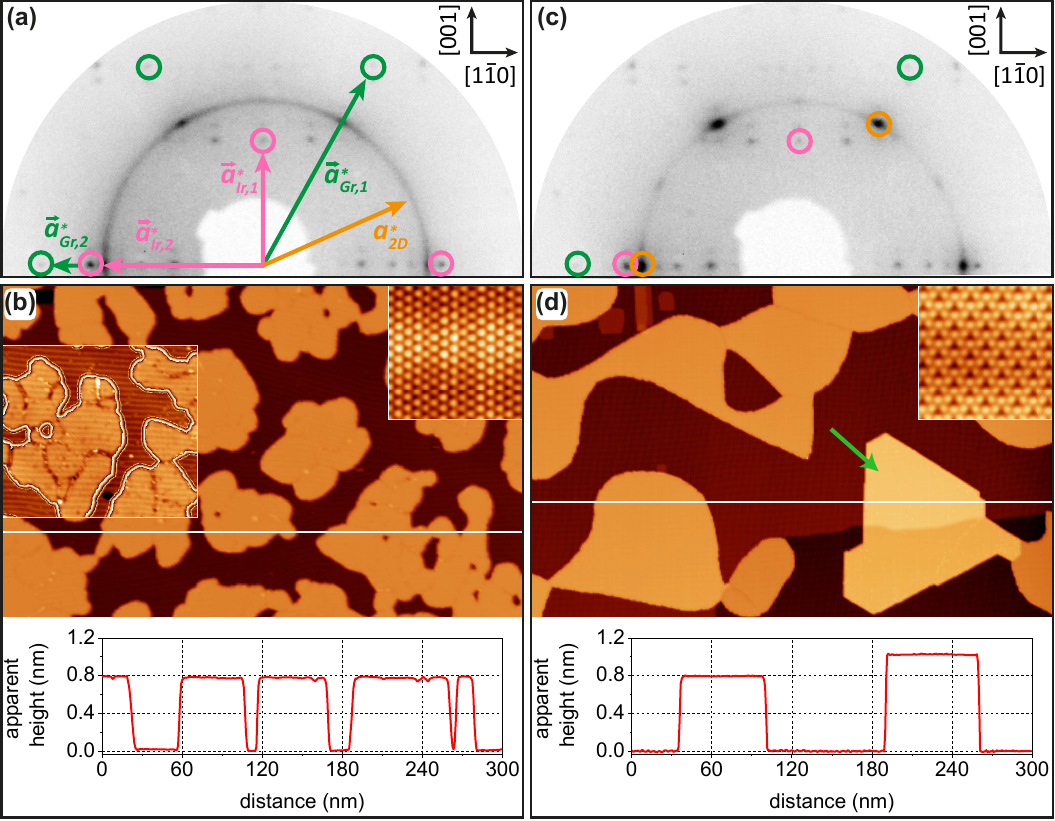}
  \caption{Characterization of Cr$_x$S$_y$-2D by LEED and STM. (a) Contrast-inverted 120\,eV LEED pattern of Cr$_2$S$_3$-2D on Gr/Ir(110). Annealing temperature 750\,K. First-order Ir and Gr reflections are encircled magenta and green, respectively. Reciprocal Ir and Gr primitive translations are indicated. Modulated diffraction ring of radius $a_\mathrm{2D}^*$ is due to Cr$_2$S$_3$-2D. (b) STM topograph corresponding to sample in (a). Inset displays atomically resolved lattice of Cr$_2$S$_3$-2D. Contrast-enhanced box makes grain boundaries (dark lines) and $\approx 3.3$\,nm Gr/Ir(110) moiré periodicity and its imprint on Cr$_2$S$_3$-2D visible. The lower panel shows the height profile along the white line in the topography. (c) Contrast-inverted 120\,eV LEED pattern corresponding to sample in (a) and (b), but after additional annealing to 950\,K. The diffraction ring has fragmented into elongated spots. (d) STM topograph corresponding to the sample of (c). Green arrow points to Cr$_{2\frac{2}{3}}$S$_4$-2D island, which displays $(\sqrt{3} \times \sqrt{3})$-R30$^{\circ}$ superstructure as visible in atomically resolved inset. Note the presence of a single Ir(110) substrate step covered by Gr in the lower right corner. Lower panel shows height profile along the white line in the topograph. Compared to Cr$_{2}$S$_3$-2D, the apparent height of Cr$_{2\frac{2}{3}}$S$_4$-2D is larger. STM topographs are acquired with $V_\mathrm{b} = 1.0$\,V and $I_\mathrm{t} = 50$\,pA at 1.7\,K. Insets atomically resolved STM insets obtained with (b) -750\,mV, $I_\mathrm{t} = 100$\,pA and (d) 50\,mV and  $I_\mathrm{t} = 200$\,pA. STM topographs are 300\,nm $\times$ 200\,nm and insets are 4\,nm $\times$ 4\,nm.}
  \label{fgr:Fig_Growth}
\end{figure}

Cr$_x$S$_y$-2D compounds are grown on graphene (Gr) prepared on low index Ir single crystal surfaces. Gr is a suitable inert van der Waals substrate that enables the facile growth of transition-metal sulfide compounds, leaving their intrinsic properties largely unaffected \cite{Hall2018}. On Ir(111) and Ir(110) Gr can be grown prior to Cr$_x$S$_y$-2D under ultrahigh vacuum (UHV) conditions in single crystal quality \cite{Busse2011,Kraus2022}. Thus, all measurements shown here result from samples prepared under UHV conditions that have never left the UHV environment, and are consequently ultimately clean. The presentation here will focus on the results using the Gr/Ir(110) substrate. The results for Gr/Ir(111) show only minute differences, as will be mentioned in more detail on the sideline below.

For the LEED pattern in Figure~\ref{fgr:Fig_Growth}(a) after deposition of Cr in S vapor at room temperature, the sample was post-annealed in S vapor at 750\,K (see Methods for details). In addition to the Gr/Ir(110) substrate reflections \cite{Kraus2022}, a diffraction ring is also visible. It has a sixfold symmetric intensity modulation with the highest intensity aligned with the first-order Gr reflections. We assign this diffraction ring with radius $a_\mathrm{2D}^*$ to the newly formed Cr$_x$S$_y$-2D compound. Based on the sixfold symmetry of the diffraction ring, a threefold or sixfold in-plane symmetry of the new compound can be assumed, resulting in an in-plane lattice constant of $a_\mathrm{2D} = 0.341 \pm 0.001$~nm.  

The typical topography of the sample is presented through the STM image of Figure~\ref{fgr:Fig_Growth}(b). Gr/Ir(110) is covered with an area fraction of about $0.6$ by coalesced Cr$_x$S$_y$-2D islands with wavy edges and typical linear dimensions of 50\,nm. Faint dark lines are visible within the islands, which we interpret as grain boundaries formed during annealing because of the coalescence of almost randomly oriented smaller islands. They are best visible in the contrast-enhanced box in Figure~\ref{fgr:Fig_Growth}(b). There, also the larger $\approx 3.3$\,nm Gr/Ir(110) moiré periodicity can be recognized, which is imprinted from the substrate onto the island. The atomically resolved STM image of Cr$_x$S$_y$-2D in the inset displays a surface lattice of periodicity of $0.340\pm 0.002$\,nm consistent with $a_\mathrm{2D}$ determined from LEED. As shown by the height profile along the white line in Figure~\ref{fgr:Fig_Growth}(b), the islands display a uniform apparent height of $0.79$\,nm at a sample bias $V_\mathrm{b} = 1.0$\,~V. Regardless of the precise preparation conditions, we never observed islands of lower height. As justified below, we interpret this Cr$_x$S$_y$-2D compound as Cr$_2$S$_3$-2D of which the side view ball model is shown in Figure~\ref{fgr:fig_ball_models}(d).

Annealing to successively higher temperatures using the same sulfur pressure improves the epitaxial orientation of Cr$_2$S$_3$-2D and gradually transforms it into thicker islands. The onset of this process is visualized in Figure~\ref{fgr:Fig_Growth}(c) and (d), obtained after annealing to 950\,K. The LEED pattern of Figure~\ref{fgr:Fig_Growth}(c) displays now a pronounced modulation of the ring intensity. Spot-like ring maxima are aligned to the first-order Gr reflections while the ring intensity in between has faded away. The STM topograph in Figure~\ref{fgr:Fig_Growth}(d) displays now extended Cr$_2$S$_3$-2D islands with less wavy steps. One island, marked by a green arrow, has a somewhat different appearance and makes straight edges with angles of 60$^\circ$ or 120$^\circ$ at its corners. As shown by the profile along the white line in Figure~\ref{fgr:Fig_Growth}(d), this island displays a uniform apparent height of $1.02$\,nm at $V_\mathrm{b} = 1.0$\,V, i.e., its apparent height is about 0.23\,nm larger than Cr$_2$S$_3$-2D islands. The atomically resolved inset of the higher island makes plain that the structure has also changed. Compared to Cr$_2$S$_3$-2D a trimerization of the surface atoms took place, leading to triangular depressions with the $(\sqrt{3} \times \sqrt{3})$-R30$^{\circ}$ periodicity. The underlying lattice parameter did not change within the limits of error, i.e., $a_\mathrm{2D-\sqrt{3}} = \sqrt{3}\, a_\mathrm{2D}$. As argued below, we interpret the Cr$_x$S$_y$-2D compound represented by the 1.02\,nm high island with the $(\sqrt{3} \times \sqrt{3})$ superstructure in Figure~\ref{fgr:fig_ball_models}(d) as Cr$_{2\frac{2}{3}}$S$_4$-2D of which the side view is shown in Figure~\ref{fgr:fig_ball_models}(e).

While phase pure Cr$_2$S$_3$-2D is formed up to 800\,K, upon increasing the annealing temperature, the transformation of Cr$_2$S$_3$-2D is not limited to Cr$_{2\frac{2}{3}}$S$_4$-2D, but thicker and thicker layers are formed [see Figure~S2, Supporting Information]. In the absence of an S vapor during annealing or for larger initial deposited amounts, the transformation of Cr$_2$S$_3$-2D to thicker layers sets in already at lower temperatures (see Figure~S3, Supporting Information). While the Cr$_x$S$_y$-2D materials thicker than Cr$_{2\frac{2}{3}}$S$_4$-2D are beyond the scope of the present manuscript, we point out that the apparent heights for $V_\mathrm{b} = 1.0$\,V increase in the sequence 0.79\,nm, 1.02\,nm, 1.29\,nm, 1.56\,nm, 1.82\,nm and so forth (see Figure~S4, Supporting Information), i.e., the levels above Cr$_2$S$_3$-2D increase in steps of 0.23\,nm to 0.27\,nm. 

\begin{figure}
\includegraphics[width=\linewidth]{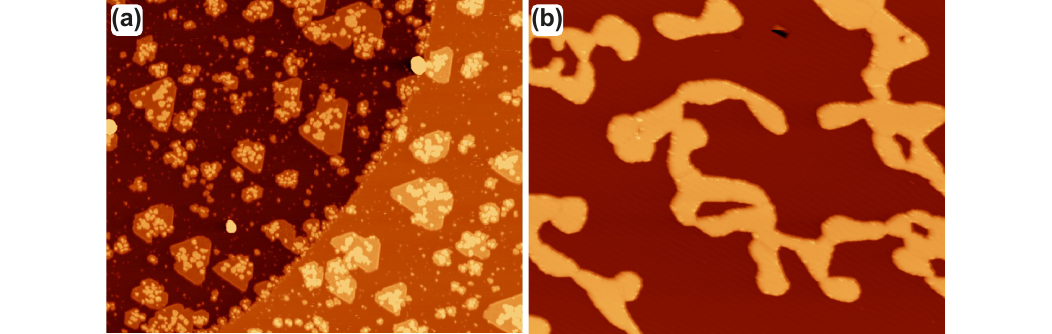}
  \caption{(a) STM topograph of pseudomorphic Cr islands on Ir(111) after 210\,s of Cr deposition. An atomic step of Ir(111) is crossing the image. (b) Cr$_2$S$_3$-2D islands on Gr/Ir(110) grown by 210\,s of Cr deposition in S background pressure and additional annealing to 750\,K. The same Cr evaporator with identical settings and identical evaporator-sample geometry was used. STM images obtained at 1.7\,K with (a) $V_\mathrm{b} = 1.0$\,V and $I_\mathrm{t} = 20$\,pA and (b) $V_\mathrm{b} = 1.0$\,V and $I_\mathrm{t} = 50$\,pA. Image sizes are 200\,nm $\times$ 150\,nm   
  }
  \label{fgr:Fig_Cal}
\end{figure}

A key piece of information still missing in order to pinpoint the structure of Cr$_2$S$_3$-2D is its composition. To this end, we conducted a set of Cr calibration experiments, of which one is visualized by Figure~\ref{fgr:Fig_Cal}. After 210\,s of Cr deposition the total coverage of pseudomorphic single and double layer Cr islands on Ir(111) was 0.39 monolayers as shown in Figure~\ref{fgr:Fig_Cal}(a). That is 0.39 of the surface atomic density of Ir(111) being $1.57\times10^{19}$\,atoms/m$^{-2}$. Next, Cr$_2$S$_3$-2D was formed by Cr deposition with the same evaporator for exactly the same time and with exactly the same settings in a S background pressure on Gr/Ir(110) and subsequent annealing to 750\,K. As shown in Figure~\ref{fgr:Fig_Cal}(b) the resulting coverage of Cr$_2$S$_3$-2D is 0.31 of the surface area. Considering the different lattice parameters of Ir(111) [0.2715\,nm] and  Cr$_2$S$_3$-2D [0.341\,nm], we arrive at a Cr content of $1.98\pm{0.05}$ Cr atoms per Cr$_2$S$_3$-2D unit cell. For a single layer of CrS$_2$ one would have expected 1 Cr atom per unit cell, not 2.

\begin{figure}
  \includegraphics[width=\linewidth]{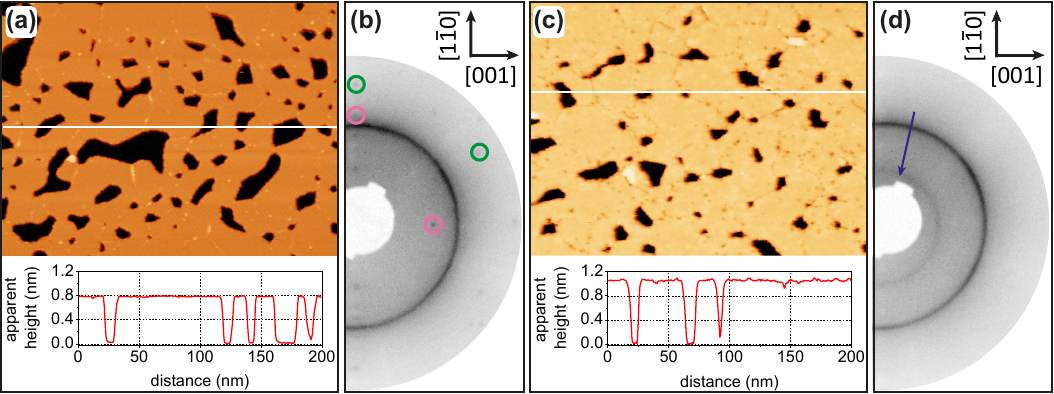}
  \caption{Phase-pure Cr$_2$S$_3$-2D and Cr$_{2\frac{2}{3}}$S$_4$-2D. (a) STM topograph of a coalesced layer of  Cr$_2$S$_3$-2D with a coverage of 0.85 and deposition time 360\,s. Lower panel displays the height profile along white line in STM topograph. (b) Constrast-inverted 130\,eV LEED patterns of  Cr$_2$S$_3$-2D sample. Few first-order Ir and Gr reflections are encircled magenta and green, respectively. Diffraction ring is due to randomly oriented Cr$_2$S$_3$-2D islands. (c) STM topograph of a coalesced layer of Cr$_{2\frac{2}{3}}$S$_4$-2D with a coverage of 0.92 after additional growth step on sample in (a) involving 150\,s Cr deposition. Lower panel displays height profile along the white line in the STM topograph. (d) Constrast-inverted 130\,eV LEED pattern of Cr$_{2\frac{2}{3}}$S$_4$-2D sample in (c). The blue arrow indicates diffraction ring resulting from the ($\sqrt3\times\sqrt3$)-R30$^{\circ}$ superstructure. See text. STM images are acquired at room temperature with $V_\mathrm{b} = 1.0$\,V, $I_\mathrm{t} = 100$\,pA. Image sizes are 200\,nm $\times$ 150\,nm.
  } 
  \label{fgr:Fig_Phase_pure}
\end{figure}

Both phases, Cr$_2$S$_3$-2D and Cr$_{2\frac{2}{3}}$S$_4$-2D, can be grown phase pure with nearly complete coverage. In a growth experiment similar to the one represented by Figure~\ref{fgr:Fig_Growth}(a) and (b), but using a larger Cr amount, a coverage of 0.85 Cr$_2$S$_3$-2D is achieved [see  Figure~\ref{fgr:Fig_Phase_pure}(a)]. The layer is fully coalesced, and only some holes are left. The height profile in the lower panel of Figure~\ref{fgr:Fig_Phase_pure}(a) and the LEED pattern of Figure~\ref{fgr:Fig_Phase_pure}(b) are consistent with our assignment. The Cr deposition time was 360\,s. 

After preparation of the Cr$_2$S$_3$-2D represented by Figure~\ref{fgr:Fig_Phase_pure}(a) and (b) we conducted a second growth step involving additional Cr deposition for 150\,s with identical evaporation parameters. The resulting sample is represented by Figure~\ref{fgr:Fig_Phase_pure}(c) and (d). The coverage increased to 0.92 of the surface area, and the height profile in the lower panel of Figure~\ref{fgr:Fig_Phase_pure}(c) displays uniformly the characteristic height of Cr$_{2\frac{2}{3}}$S$_4$-2D around 1.02\,nm. The presence of Cr$_{2\frac{2}{3}}$S$_4$-2D is corroborated by a second LEED-ring with a radius corresponding to a $\sqrt{3} \times\sqrt{3})$-R30$^{\circ}$ superstructure and highlighted in Figure~\ref{fgr:Fig_Phase_pure}(d) by a blue arrow. 

Considering that Cr$_2$S$_3$-2D contains 2 Cr atoms per unit cell as established above, by taking into account the deposition times and coverages measured, we obtain a Cr content of $2.62\pm{0.08}$ for Cr$_{2\frac{2}{3}}$S$_4$-2D. The experimentally found Cr content is thus consistent with Cr$_{2\frac{2}{3}}$S$_4$-2D containing $2\frac{2}{3}$ or 2.67 Cr atoms per original $ 1 \times 1$ unit cell or 8 Cr atoms for the $\sqrt 3 \times \sqrt3$ superstructure unit cell.  

Based on the experimental information and considering that Cr$_x$S$_y$-2D phases are alternating sequences of S and partial or complete Cr planes \cite{Jellinek1957}, we are ready to make an informed hypothesis about the structure of the two Cr$_x$S$_y$-2D phases. For Cr$_2$S$_3$-2D we assume a structure as depicted in Figure~\ref{fgr:fig_ball_models}(d). It contains 2 Cr atoms per unit cell, as found experimentally. In principle, one S plane sandwiched between two Cr planes would be consistent with our experiments. However, the reactivity of the metal and growth under sulfur-rich conditions excludes the presence of bare metal at the Gr interface (see ref. \citenum{Knispel2025}) or the vacuum. Therefore, we assume the presence of two additional sulfur planes, making the compound inert and resulting in the stoichiometry Cr$_2$S$_3$-2D. Although we cannot guess the precise stacking, it is plausible to be similar to the bulk compounds Cr$_2$S$_3$ or Cr$_5$S$_6$ [Figure~\ref{fgr:fig_ball_models}(a) and (c)], when disregarding the presence of Cr vacancies in these compounds. The observed hexagonal surface symmetry with a lattice parameter of 0.341\,nm matches reasonably with the lattice parameter of the Cr deficient bulk compounds Cr$_2$S$_3$ and Cr$_5$S$_6$, which is in the range from 0.343-0.345\,nm. Previous DFT calculations for Cr$_2$S$_3$-2D yield lattice parameters of 0.343-0.348\,nm \cite{Li2023,Zhang2021} in acceptable agreement with our experiment. We emphasize that the absence of any superstructure together with 2 Cr atoms per unit cell is strong evidence for complete Cr planes in Cr$_2$S$_3$-2D.

An additional partial or complete Cr plane plus a saturating S plane has for bulk Cr$_x$S$_y$ materials an average height of 0.28\,nm \cite{Jellinek1957}, in close agreement with the step height of 0.26-0.27\,nm measured for thicker Cr$_x$S$_y$-2D by us. The slightly smaller apparent step height between Cr$_2$S$_3$-2D and Cr$_{2\frac{2}{3}}$S$_4$-2D of 0.23\,nm fits reasonably well, and the slight deviation may be of electronic origin. It is, therefore, most plausible to assume now the presence of an additional 2/3 Cr plane (based on the measured additional Cr amount of 0.62 of a full plane) and an additional S plane giving rise to the proposed stoichiometry Cr$_{2\frac{2}{3}}$S$_4$-2D. The presence of a 2/3 Cr plane within the stack provides a natural explanation for the observed $(\sqrt{3} \times \sqrt{3})$-R30$^{\circ}$ superstructure, although the position of the 2/3 Cr plane is uncertain. Based on the surface sensitivity of LEED signals, it is unlikely that the 2/3 Cr plane is the lowermost in the stack. Whether the partial Cr plane is in the center as indicated in Figure~\ref{fgr:fig_ball_models}(e) or at the top cannot be decided based on the experimental data provided up to now. 

\begin{figure}
  \includegraphics[width=0.5\linewidth]{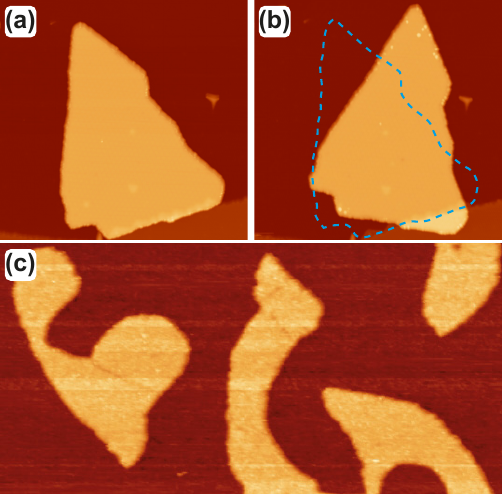}
  \caption{Moving of Cr$_2$S$_3$-2D islands and their air stability. (a)  Cr$_2$S$_3$-2D island on Gr/Ir(111). (b) After interaction with STM tip (see debris at upper tip of island). Dashed blue line indicates initial position of island, as in (a). (c) Cr$_2$S$_3$-2D on Gr/Ir(110) after exposure for 3 hours to ambient conditions and subsequent pump-down without annealing. STM images (a-c) are taken at 1.7\,K with $V_\mathrm{b}$ = 1 V and $I_\mathrm{t}$ = 50 pA. Image sizes of (a) and (b) are 100\,nm $\times$ 100\,nm, and (c) is 200\,nm $\times$ 100\,nm.} 
  \label{fgr:Fig_binding_stability}
\end{figure}

Lastly, the stability and inertness of the covalently bound Cr$_2$S$_3$-2D is highlighted through Figure~\ref{fgr:Fig_binding_stability}. As documented by the two subsequent STM topographs of Figure~\ref{fgr:Fig_binding_stability}(a) and (b), a large Cr$_2$S$_3$-2D island can be laterally shifted by the STM-tip interacting with the island edge without significant defect formation. This is a direct proof of a weak van der Waals interaction between Cr$_2$S$_3$-2D and Gr/Ir(111). Figure~\ref{fgr:Fig_binding_stability}(c) displays an STM topograph of Cr$_2$S$_3$-2D islands taken in vacuum after exposure for three hours to ambient conditions [see Figure~S5, Supporting Information, for a before and after exposure comparison and a large-scale topography of the same sample]. Not unexpectedly, adsorption on Gr and Cr$_2$S$_3$-2D is apparent, but the structural integrity of the islands is obvious.

\subsection{Absence of CrS$_2$}

Despite substantial efforts, the growth of CrS$_2$ using our molecular beam epitaxy approach remained unachievable, both as mono- as well as multilayer. In all experiments, including those with variations in growth and annealing procedure, we always observed a plain hexagonal surface lattice with 0.341\,nm as the lattice parameter. This rules out the formation of the predicted orthorhombic CrS$_2$ phases \cite{Lei2024,Xiao2022b}, the formation of the T'-phase \cite{Habib2019,Chen2021}, and distorted T-phases \cite{Wang2018,Chen2021}. H- and undistorted T-CrS$_2$ display a hexagonal lattice, but with a significantly smaller lattice parameter of $\approx 0.30-0.31$\,nm \cite{Chen2021,Habib2019,Liu2024b,Wang2018}.

Moreover, for the H-,T-, and T'-phases of CrS$_2$ one would expect a significantly smaller single layer height than the 0.79\,nm measured for Cr$_2$S$_3$ at our reference voltage. This is because CrS$_2$ is a three-plane S-Cr-S system rather than a five-plane S-Cr-S-Cr-S system as for Cr$_2$S$_3$-2D. Based on the height difference between Cr$_2$S$_3$-2D and  Cr$_{2\frac{2}{3}}$S$_4$-2D of 0.23\,nm, one would expect a single layer of CrS$_2$ to display an apparent height well below 0.60\,nm. However, this was never the case and at the reference voltage we invariably measured an apparent height of 0.79\,nm. 

Growth variations involved lowering [see Figure~S6(a), Supporting Information] and increasing the sulfur pressure during growth and annealing, as well as changing the growth temperature in the range from 160\,K [see Figure~S6(b), Supporting Information] to 600\,K [see Figure~S6(c), Supporting Information]. 
Particularly, we also changed from Gr/Ir(110) to Gr/Ir(111) as growth substrate [see Figure~S7, Supporting Information], without being able to detect an indication for CrS$_2$ formation. Noteworthy, when grown on Gr/Ir(111) the apparent heights of the islands at $V_\mathrm{b} = 1.0$\,V are 0.86\,nm and 1.09\,nm for Cr$_2$S$_3$-2D and Cr$_{2\frac{2}{3}}$S$_4$-2D, respectively. This increase by about 0.07\,nm compared to Gr/Ir(110) is primarily attributed to a larger van der Waals gap between Cr$_x$S$_y$-2D and Gr. In fact, Gr is physisorbed on Ir(111) with weak chemical modulation \cite{Busse2011}, whereas it is chemisorbed on Ir(110) \cite{Kraus2022}. This leads to stronger interactions with Cr$_x$S$_y$-2D in the latter case and consequently to a reduced height above Gr. 

The fact that the same phases of Cr$_x$S$_y$-2D result on these two different substrates suggests that the outcome of growth does not depend on the substrate, be it metallic, insulating, or semiconducting, as long as it is sufficient inert to S and Cr and thus does not suppress the reaction of S and Cr with each other. Support for this view is from the fact single layer NbSe$_2$ has the same structure and a nearly identical electronic structure, irrespective of whether grown on bi-layer Gr/SiC(0001), hexagonal boron nitride (h-BN) on Ir(111) or on bulk WSe$_2$ \cite{Dreher2021}. Likewise, also MoS$_2$ forms in the same way and under comparable conditions on Gr/Ir(111) and h-BN/Ir(111) \cite{Hall2018}.  

The MBE growth method used here has so far enabled us to reproduce whatever transition metal disulfide described in the literature that it was applied to \cite{Hall2018,VanEfferen2024,Knispel2025}. Therefore, the inability to grow CrS$_2$-2D using the same method indicates that it may be worthwhile to reevaluate the evidence for the existence of CrS$_2$. 

TEM images of CrS$_2$ are provided in four publications \cite{Habib2019,Nair2022,Xiao2022a,Yao2024} and display a perfectly hexagonal lattice ruling out the observation of orthorhombic, distorted T- or T'-phases. Based on the specified net plane distances or scale bars provided, the lattice parameter is in the range of 0.33-0.36\,nm, in descent agreement with the lattice parameter of 0.341\,nm measured for Cr$_2$S$_3$-2D and Cr$_{2\frac{2}{3}}$S$_3$-2D, but at variance with the calculated lattice parameters of $\approx 0.30-0.31$\,nm \cite{Chen2021,Habib2019,Liu2024b,Wang2018,Zhuang2014} for the undistorted T- or H-phase. The measured CrS$_2$ lattice parameters also agree with the reported lattice parameters of  Cr$_2$S$_3$ thin films [see Figure~1(c)] ranging from 3.4\,\AA~to 3.51\,\AA \cite{Zhou2019,Liu2022,Chu2019,Cui2019}. 

The minimum heights of CrS$_2$ flakes measured by atomic force microscopy are 2\,nm \cite{Nair2022}, 1.5\,nm \cite{Xiao2022a}, 0.75\,nm \cite{Habib2019}, and 
0.682\,nm \cite{Yao2024}. Despite the three significant digits provided in the latter height, the noise level on each terrace is 0.2\,nm. Given the experimental uncertainties, the measured heights are also compatible with  Cr$_2$S$_3$-2D, or Cr$_{2\frac{2}{3}}$S$_3$-2D and Cr$_2$S$_3$ thin films for the thicker layers.

The characteristic Raman peaks of CrS$_2$ scatter with a few wave numbers around 175\,cm$^{-1}$, 252\,cm$^{-1}$, 283\,cm$^{-1}$, and 359\,cm$^{-1}$ \cite{Nair2022,Xiao2022a,Habib2019,Yao2024}. It also has to be noted that Raman peaks are measured for thicker CrS$_2$ films for signal reasons \cite{Xiao2022a}. Peak number and positions agree well with the ones reported for Cr$_2$S$_3$ thin films, specifically when considering their thickness-dependent variability \cite{Zhou2019,Xie2021,Cui2019}.

XPS characterization through the core level doublets S 2p$_{3/2}$/S 2p$_{1/2}$ and Cr 2p$_{3/2}$/Cr 2p$_{1/2}$ does neither provide a clear distinction between between CrS$_2$ \cite{Habib2019,Nair2022,Yao2024}and thin films of Cr$_2$S$_3$ \cite{Zhou2019,Chu2019,Moinuddin2021,Liu2022,Cui2019,Cui2022}. For instance, the data provided by Nair et al. \cite{Nair2022} for CrS$_2$ agree perfectly with the one provided by Cui et al. \cite{Cui2022} for Cr$_2$S$_3$ thin films. Although XPS can be used for composition analysis, precise composition analysis is not feasible without proper calibration, taking depth-dependent attenuation of the signal into account and considering the depth distribution of the species in a layered material.
It also has to be noted that the top three layers of S-Cr-S in thin Cr$_2$S$_3$ films and CrS$_2$ are identical. Lastly, the prominent intensity peak at $2\theta \approx 16^\circ$ in X-ray diffraction of CrS$_2$ \cite{Habib2019,Yao2024,Xiao2022a} is also present for thin Cr$_2$S$_3$ films \cite{Chu2019,Moinuddin2021}.   

Given the absence of CrS$_2$ as a bulk material and the ambiguity surrounding the experimental evidence for its existence as a 2D material, it may be valuable to consider how to obtain more conclusive experimental proof for it.

One might be tempted to use additional arguments for the existence of CrS$_2$ as a 2D material from the analysis of the convex hull of the DFT-calculated formation energies in the Cr--S phase diagram. In fact the Computational 2D Materials Database C2DB \cite{Haastrup2018,Gjerding2021} finds the H-phase of single layer CrS$_2$ on the convex hull, as do our own calculations discussed below (compare Figure~\ref{fgr:Fig_convex_hull}). However, the relevance of this prediction is questioned by the fact that also the bulk compound of CrS$_2$ is on the convex hull (see Materials Project database \cite{Jain13} and Figure~\ref{fgr:Fig_convex_hull}), although its experimental synthesis has not been reported.

\subsection{Electronic structure from experiment}
\begin{figure}
  \includegraphics[width=\linewidth]{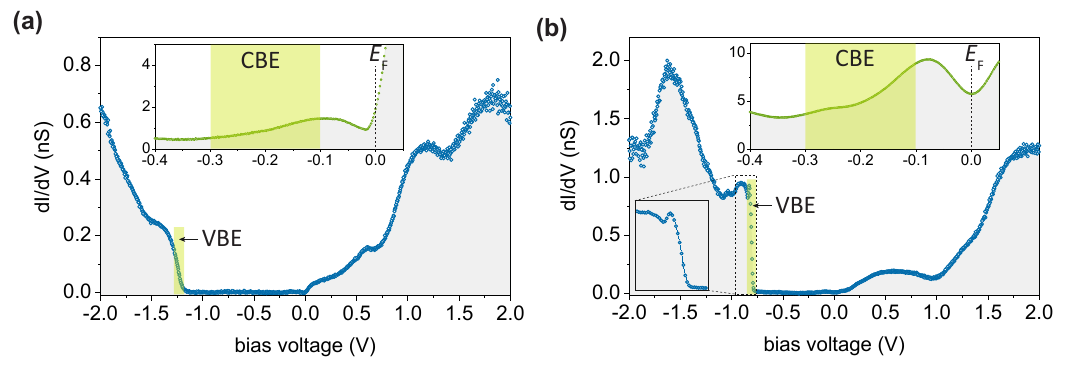}
  \caption{Electronic structure of single layer Cr$_2$S$_3$-2D and Cr$_{2\frac{2}{3}}$S$_4$-2D on Gr/Ir(110). (a) and (b) are $\mathrm{d}I/\mathrm{d}V$ spectra of Cr$_2$S$_3$-2D and Cr$_{2\frac{2}{3}}$S$_4$-2D, respectively. The upper insets in (a) and (b) are smaller range high-resolution $\mathrm{d}I/\mathrm{d}V$ spectra. Yellow boxes indicate estimated positions and uncertainties of valence band edges (VBEs) and conduction band edges (CBEs). Inset in the lower left of (b) is a magnified view of the VBE. The spectra are obtained at 1.7\,K with (a) $V_{st}$ = 2 V, $I_{st}$ = 1\,nA, $f_{mod}$ = 667\,Hz, and $V_{mod}$ = 20\,mV, and for the insets $V_{st}$ = 100\,mV, $I_{st}$ = 1\,nA, $f_{mod}$ = 667\,Hz, $V_{mod}$ = 20\,mV.}
  \label{fgr:Fig_STS}
\end{figure}

The electronic structure of Cr$_2$S$_3$-2D and Cr$_{2\frac{2}{3}}$S$_4$-2D was investigated by STS. Figure~\ref{fgr:Fig_STS}(a) and~\ref{fgr:Fig_STS}(b) present differential conductance ($\mathrm{d}I/\mathrm{d}V$) spectra of Cr$_2$S$_3$-2D and Cr$_{2\frac{2}{3}}$S$_4$-2D in the energy range $ -2\,\rm{V} \leq \textit{V}_\mathrm{b} \leq +2\,\rm{V}$. The spectra were measured far away ($>10$\,nm) from island edges and are highly reproducible. The most remarkable feature in both spectra is the band gap, which seems to range from the Fermi level at $\textit{V}_\mathrm{b}$ = 0\,V to the valence band edges (VBEs) at $V_\mathrm{b} = -1.20 \pm 0.05$\,V and $\textit{V}_\mathrm{b} = -0.80 \pm 0.05\,V$ in Cr$_2$S$_3$-2D and Cr$_{2\frac{2}{3}}$S$_4$-2D, respectively. While high-resolution spectra confirm the sharp rises at negative voltages and thus the assignment of the VBEs, the apparent conduction band edges (CBEs) at the Fermi energy $E_\mathrm{F}$  are suspicious. High-resolution spectra near $E_\mathrm{F}$ shown as insets in Figure~\ref{fgr:Fig_STS}(a) and~\ref{fgr:Fig_STS}(b) make plain that there is a finite density of states below $E_\mathrm{F}$. 

Such a smeared-out behavior of the conduction band edge (CBE) has been observed in metallic MoS$_2$ (see Figure~S5 of the Supplementary Data of ref. \citenum{VanEfferen2022}). Detection of a sharp CBE for MoS$_2$ in $\mathrm{d}I/\mathrm{d}V$ spectra was hampered by the fact that the conduction band minimum is located at the K-point of the first Brillouin zone (BZ). For such electrons with a large parallel momentum STS is relatively insensitive. As a result, precise energy assignment to the CBE is challenging. Here we assume a similar origin of the smeared-out CBE, which would imply a large parallel momentum of the electrons at the conduction band minimum. As an estimate, we suggest the CBE at $V_\mathrm{b} = - 0.2$\,V for both materials, yielding approximately band gaps of $1.0$\,eV and $0.6$\,eV for Cr$_2$S$_3$-2D and Cr$_{2\frac{2}{3}}$S$_4$-2D, respectively. Furthermore, both spectra exhibit a pronounced dip at $E_\mathrm{F}$. Features at $E_\mathrm{F}$, including dips and peaks, are commonly observed in $\mathrm{d}I/\mathrm{d}V$ spectra of 2D materials \cite{Knispel2024} and are likely related to electronic correlations. However, the exact origin remains uncertain in this instance. 

Whether a 2D material with a full band gap is a semiconductor or a metal depends on the position of $E_\mathrm{F}$ which may be affected by the substrate. In MoS$_2$, a semiconductor-metal transition was driven by doping-induced work function lowering of the Gr substrate \cite{VanEfferen2022}. Vacuum level alignment of Gr and MoS$_2$ then implied a downshift of the MoS$_2$ band structure until the CBE moved into the Fermi level and an interface dipole was formed by charge transfer from Gr to MoS$_2$. We consider here for Cr$_2$S$_3$-2D and Cr$_{2\frac{2}{3}}$S$_4$-2D a similar scenario. That is while the freestanding materials might be semiconducting, we speculate that a lower Gr work function implies, through vacuum level alignment for the Cr$_x$S$_y$-2D compounds, could lead to a downward shift of their bands and, eventually, the formation of an interface dipole due to charge transfer into their conduction bands. Consequently, they become metallic. Such a scenario would imply that the work function of the Cr$_x$S$_y$-2D compounds to be larger than the one of Gr. To substantiate this speculation, we estimated the work functions of the Cr$_x$S$_y$-2D compounds using the method of field emission resonances \cite{Binnig1985,Becker1985} and found them in fact to be larger by more than one electronvolt compared the one of Gr/Ir(110) (see Figure~S8, Supporting Information).

A notable feature in the $\mathrm{d}I/\mathrm{d}V$ spectrum of and Cr$_{2\frac{2}{3}}$S$_4$-2D is a sharp rise peak at the VBE as highlighted by the inset in the lower left corner of Figure~\ref{fgr:Fig_STS}(b). Such a sharp feature might be related to a van Hove singularity of a surface state \cite{Li1998}. 

Work function differences and the density of states also affect the measured apparent height in STM. 
The apparent heights of Cr$_2$S$_3$-2D and Cr$_{2\frac{2}{3}}$S$_4$-2D were up to now only specified for $V_\mathrm{b} = 1.0$\,V. This choice is meaningful because at positive bias the apparent heights are stable and vary only marginally with $V_\mathrm{b}$, as expected for a metal, for which the apparent height usually agrees rather well with the geometric height. Not unexpectedly, at negative bias in the range of the band gap, apparent heights up to 0.15\,nm lower are measured (see Figure~S9, Supporting Information). 

Finally, it is found that the electronic structure remains invariant with respect to the substrate, whether Gr/Ir(110) or Gr/Ir(111), as evidenced by comparison in Figure Figure~S10 (Supporting Information). The invariance of the measured electronic structure is consistent with a weak interaction and the absence of significant hybridization with the two different substrates.

\subsection{Density functional theory calculations}
Based on the experiments, a structure S-Cr-S-Cr-S consisting of five atomic planes is inferred for Cr$_2$S$_3$-2D, but the stacking sequence could not be established.  In our DFT calculations, we tested the typical trigonal prismatic (1H) and octahedral (1T) coordination of Cr with respect to S, with stackings displaying aligned and shifted Cr positions between the two Cr planes (see Table\,S1, Supporting Information). 
To identify the most stable structure for Cr$_2$S$_3$-2D among the resulting eight possible crystal configurations, their total energies were calculated by DFT. The results, presented in Table\,S1 (Supporting Information), give the lowest energy for the crystal structure with aligned 1T/1T Cr coordination, as illustrated in Figure~\ref{fgr:Fig_DFT_Cr2S3}(a) to (c). This crystal structure corresponds to the classical NiAs-structure and is the same as for bulk Cr$_2$S$_3$ and Cr$_5$S$_6$, when disregarding the presence of Cr vacancies in the bulk structures. The NiAs-structure of Cr$_2$S$_3$-2D is preferred by 0.17\,eV per formula unit compared to a structure with the same octahedral coordination of the Cr atoms, but shifted Cr positions on the atomic planes, and by more than 0.6\,eV compared to all other structures. The calculated lattice parameter of 0.344\,nm agrees reasonably well with the experimental value of 0.341\,nm. Furthermore, no superstructure is observed, consistent with the experimental results. 

\begin{figure}[hbt!]
  \includegraphics[width=0.5\linewidth]{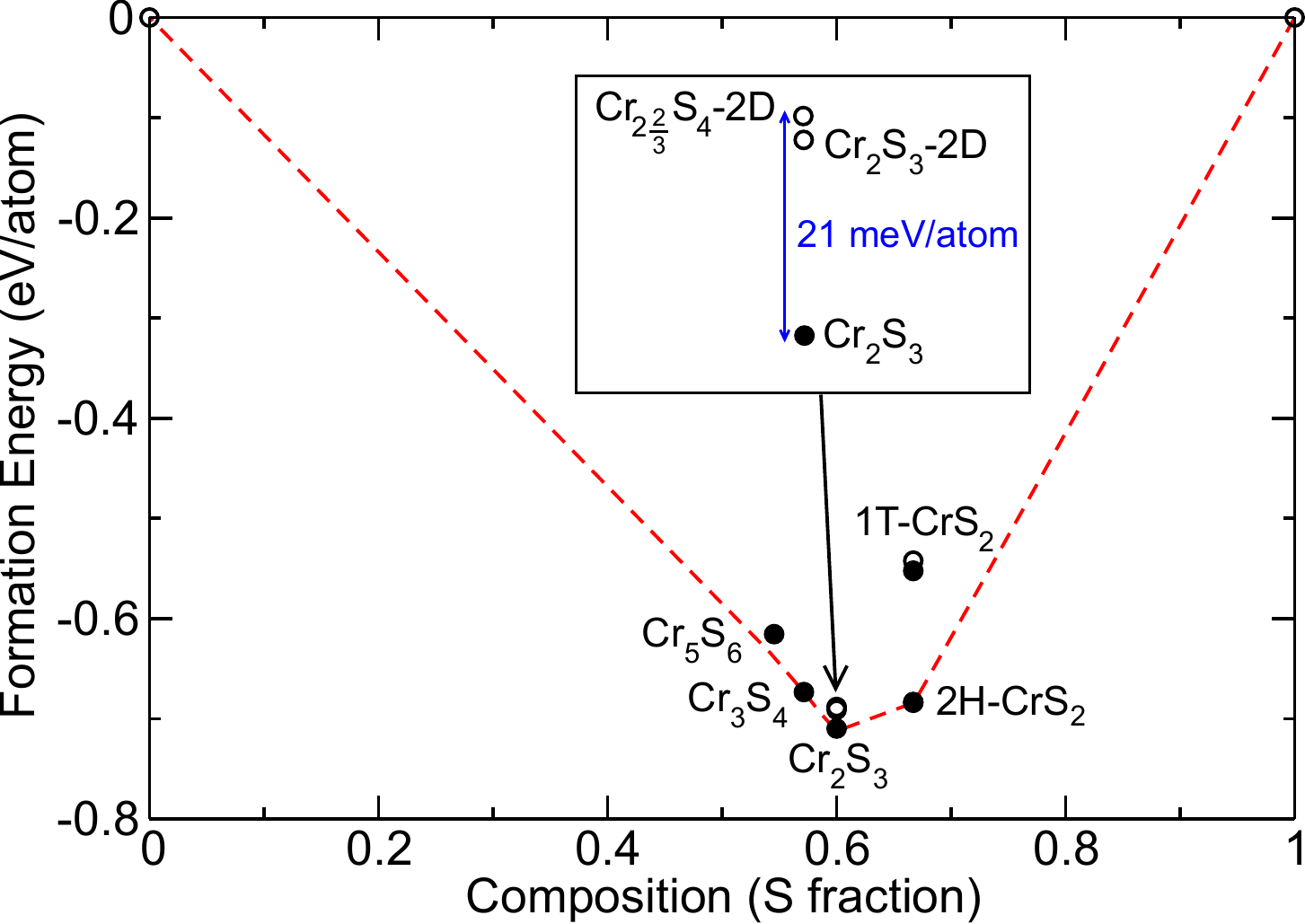}
  \caption{Formation energies of Cr-S compounds. The convex hull is determined by bulk Cr$_2$S$_3$ and bulk 2H-CrS$_2$. Bulk phases are indicated by full circles, 2D phases by open circles. 2D H-CrS$_2$ is in energy indistinguishable in the plot from bulk 2H-CrS$_2$.} 
  \label{fgr:Fig_convex_hull}
\end{figure}

In the convex hull plot of Figure~\ref{fgr:Fig_convex_hull}, the formation energy of the Cr$_2$S$_3$-2D is seen to be just disfavored by about 20\,meV/atom to the bulk phase with the same stoichiometry (but different structure). The DFT-calculated van der Waals binding energy of Cr$_2$S$_3$-2D to Gr of 36\,meV/atom (17.5\,meV/\AA$^2$) is larger than the lack in formation energy and thus favors the formation of the 2D phase over the bulk phase with the same composition. However, we note that these energy differences are not significantly outside the error margin of DFT calculations. Additionally, the thermal stability of the new 2D-phases was also confirmed through molecular dynamics simulations conducted at 700\,K for 10~ps (see video files as Supporting Information). The absence of imaginary frequencies in the DFT calculated phonon dispersion, shown in Figure 2 of ref. \citenum{Zhang2021}, is also in line with the experimentally observed stability of Cr$_2$S$_3$-2D.

\begin{figure}
  \includegraphics[width=\linewidth]{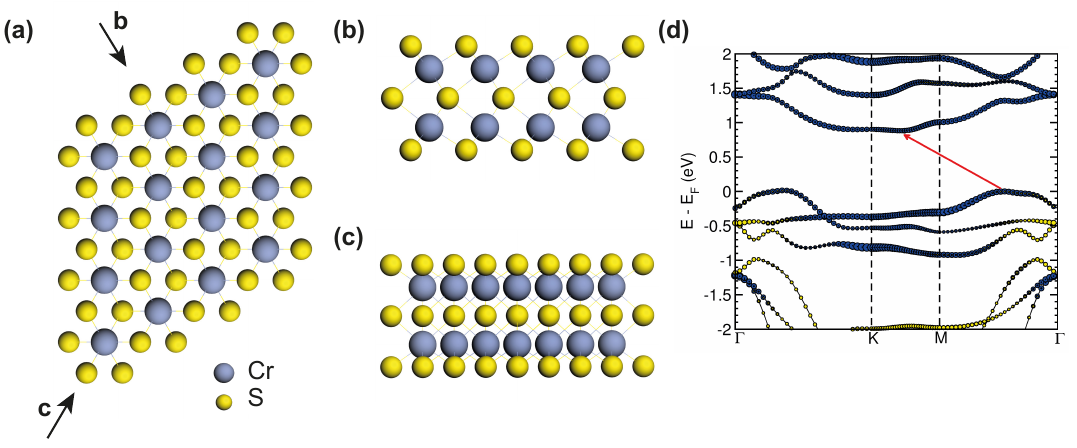}
  \caption{Atomic and electronic structure of Cr$_2$S$_3$-2D obtained from DFT calculations.  (a) top and (b),(c) side views of the lowest energy crystal structure for Cr$_2$S$_3$-2D. Side views (b) and (c) are along the arrows labeled \textit{b} and \textit{c} in (a), respectively. (d) Band structure of free-standing Cr$_2$S$_3$-2D. The total band structure (black) is projected onto the Cr (blue) and S (yellow) states. The size of the circles also represents the weight of the projected states.
  }
  \label{fgr:Fig_DFT_Cr2S3}
\end{figure}

A noteworthy feature of the DFT-optimized Cr$_2$S$_3$-2D structure is the asymmetry of the Cr atomic planes with respect to the S ones. As well visible in the side view ball models of the final DFT geometry in Figures~\ref{fgr:Fig_DFT_Cr2S3}(b) and (c), the Cr planes are relaxed outwards by about $\Delta x = 0.018$\,nm away from the center position between the S planes.  Consequently the bond length of the surface S atoms of 0.233\,nm is considerably shorter than the bond length of the center S atoms of length 0.253\,nm. The bond shortening $\Delta b = 0.020$\,nm of the surface S atoms is interpreted as a bond strengthening in consequence of their lower coordination. In fact, an energy gain of 0.26\,eV per formula unit is associated with the relaxation. The energy gain due to bond strengthening of the surface S atoms may explain in part, why the formation energies of the 2D compounds are only higher by about 20\,meV/atom as compared to a bulk phase with the same composition.

The calculated DFT (for Hubbard $U = 0$) band structure is shown in Figure~\ref{fgr:Fig_DFT_Cr2S3}(d). It exhibits an indirect band gap of 0.9\,eV in reasonable agreement with our experimental estimate of 1.0\,eV. The conduction band minimum is located between the K and M points of the first BZ. The large parallel momentum of the electrons at the conduction band minimum is consistent with the insensitivity of STS to its position and its smeared-out appearance in the $\mathrm{d}I/\mathrm{d}V$ spectra [see Figure~\ref{fgr:Fig_STS}(a)]. The valence band maximum lies close to the center ($\Gamma$) of the first BZ. Correspondingly, the smaller parallel momentum makes plain why the VBE appears rather sharp in the Cr$_2$S$_3$-2D $\mathrm{d}I/\mathrm{d}V$ spectrum of Figure~\ref{fgr:Fig_STS}(a). 

Recalculating the band structure with $U > 0$ does not change the overall picture, but the magnitude of the band gap is affected. It increases with increasing $U$, reaches a maximum around 1.32\,eV for $U=2.0$\,eV and then drops to 0.84\,eV for $U = 3.5$\,eV. Taking into account spin orbit coupling (SOC) decreases the bandgap size only slightly by 20~meV, for the complete list of bandgap sizes as a function of $U$ see Table~S2 (Supporting Information). 

Given the semiconducting nature of free-standing Cr$_2$S$_3$-2D, one can estimate the associated exciton binding energy. Using Green’s function–Bethe–Salpeter equation, we calculated this value to be of the order of 0.38~eV, see Figure~S11 (Supporting Information). The exciton binding energy for Cr$_2$S$_3$-2D lies between the reported values for CrSBr bulk (0.25~eV) \cite{Telford2020} and monolayer (0.5~eV) \cite{Wilson2021}.

A metallic state of Cr$_2$S$_3$-2D is obtained when the calculations are carried out considering the Gr substrate. Indeed, the CBE shifts below $E_\mathrm{F}$, indicating a semiconductor-metal transition of Cr$_2$S$_3$-2D (see projected density of states (PDOS) in Figure~S12, Supporting Information).

Next, we assess the magnetic properties of the Cr$_2$S$_3$-2D structure via DFT calculations. We find that the magnetic moment of the Cr$^{+3}$ atoms (i.e., 3\,$\mu_\mathrm{B}$) is affected by SOC and reduced to 1.73\,$\mu_\mathrm{B}$ per Cr atom, indicating the presence of magnetism in the structure.
To infer the magnetic ground state of Cr$_2$S$_3$-2D, intralayer Cr-Cr coupling was considered ferromagnetic (FM). This assumption is consistent with expectation based on the Goodenough-Kanamori-Anderson rule, which favors FM coupling when the bond angle Cr-S-Cr is close to 90\textdegree~(in Cr$_2$S$_3$-2D it is 85.77\textdegree). For the interlayer coupling, we considered FM or A-type (or layerwise) antiferromagnetic (AFM) states, for a schematic representation, see Figure~S13 (Supporting Information). 
The energy difference $E_\mathrm{AFM} - E_\mathrm{FM}$ between these two states is $-51.78$\,meV, that is the A-type AFM configuration is energetically preferred.

The interlayer magnetic exchange splitting calculated by GGA is notably weak, emphasizing the importance of accurately accounting for electronic correlation effects when estimating magnetism. To address this, we analyzed the energetics of FM and AFM states by varying the $U$ parameters in GGA+U calculations. As the on-site Coulomb term $U$ increased, a transition from the AFM to the FM state was observed, as shown in Table~S2 (Supporting Information). This shift occurs because the super-exchange interlayer coupling is inversely related to the hopping parameter; thus, raising $U$ reduces the stability of AFM coupling relative to FM coupling.

\begin{figure}
  \includegraphics[width=\linewidth]{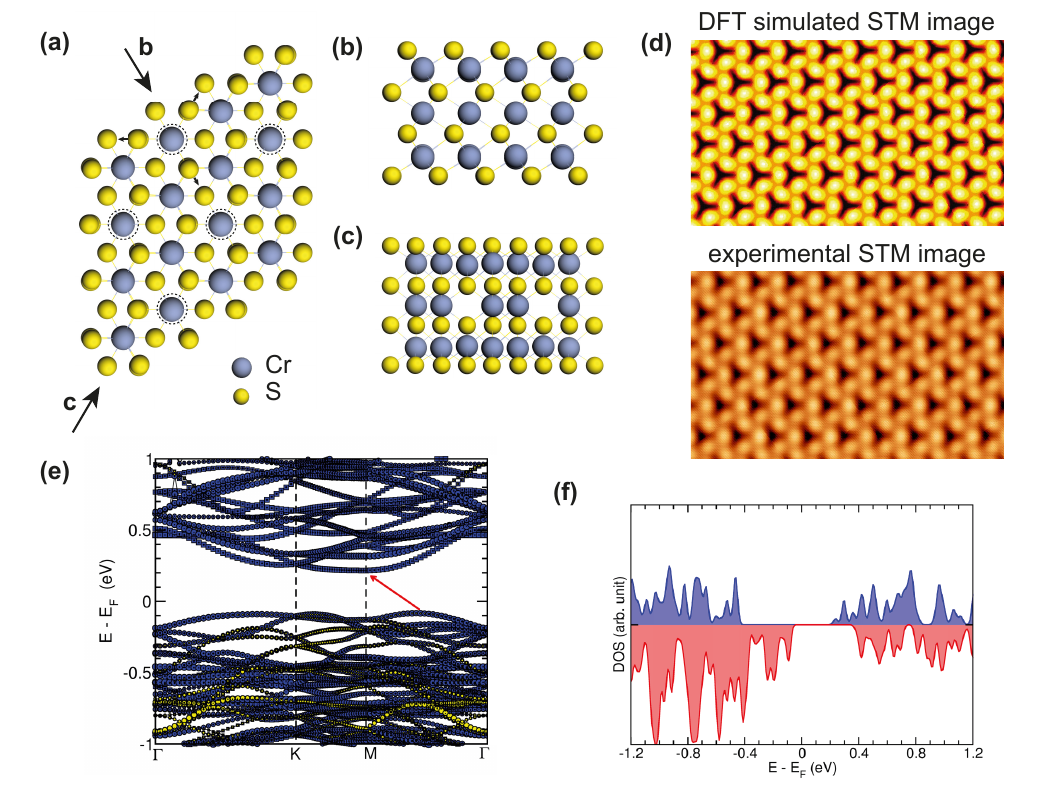}
  \caption{Atomic and electronic structure of Cr$_{2\frac{2}{3}}$S$_4$-2D obtained from DFT calculations. (a) top and (b),(c) side views of the lowest energy crystal structure of Cr$_{2\frac{2}{3}}$S$_4$-2D.  Dotted circles around Cr atoms in (a) indicate where Cr atoms are missing in the middle layer. Side views (b) and (c) are along arrows labeled \textit{b} and \textit{c} in (a), respectively. (d) Comparison between DFT simulated and STM topograph. The DFT simulated topograph takes states between $E_\mathrm{F}$ and $E_\mathrm{F}$ + 0.35\,eV into account. STM topograph is taken at 1.7\,K with V$_b$ = 50\,mV and I$_t$ = 200\,pA. Image sizes are 5\,nm $\times$ 3\,nm. Gr/Ir(110) moiré is removed from the STM topograph by using an FFT filter. (e), (f) Band structure and PDOS of free-standing Cr$_{2\frac{2}{3}}$S$_4$-2D, respectively. The total band structure (black) is projected onto the Cr (blue) and S (yellow) states. The size of the circles also represents the weight of the projected states.
  } 
  \label{fgr:Fig_DFT_Cr2.66S4}
\end{figure}

For Cr$_{2\frac{2}{3}}$S$_4$-2D not only the stacking sequence, but also the position of the fractional Cr plane was uncertain from experimental data. Experimentally, LEED ruled out the fractional Cr plane to be the lowermost. Since for Cr$_2$S$_3$-2D octahedral (T) coordination of Cr atoms was preferred, we computed for Cr$_{2\frac{2}{3}}$S$_4$-2D the four different structures consistent with these two constraints as outlined in Table~S3 (Supporting Information). We find the minimum energy configuration to be a NiAs-type structure with the fractional Cr plane in the center, as represented in Figures~\ref{fgr:Fig_DFT_Cr2.66S4}(a) to (c). This structure is preferred by 0.6\,eV per formula unit. The DFT optimized configuration reflects the Cr$_5$S$_6$ motif of intermediate Cr layers filled only by $\frac{2}{3}$ [see Figures~1(c) and 1(e)]. The relaxed structure has a lattice constant of 0.347\,nm, which agrees with an acceptable error to the experimental 0.341\,nm. The thickness difference between the relaxed crystal structures of Cr$_2$S$_3$-2D and Cr$_{2\frac{2}{3}}$S$_4$-2D is 0.26\,nm, a value that aligns closely with the apparent height difference of 0.23\,nm observed in STM. 

Similar to  Cr$_2$S$_3$-2D, for Cr$_{2\frac{2}{3}}$S$_4$-2D the gain in van der Waals binding of 33\,meV/atom (19.8\,meV/\AA$^2$) outweighs the lack in binding energy of about 21\,meV/atom compared to a bulk crystal with the same composition. As for Cr$_2$S$_3$-2D, the thermal stability of  Cr$_{2\frac{2}{3}}$S$_4$-2D was confirmed through molecular dynamics simulations (see video file as Supporting Information). Lastly, also for  Cr$_{2\frac{2}{3}}$S$_4$-2D clear surface relaxation effects are visible. Again the outer Cr planes are relaxed outwards towards the surface by $\Delta x = 0.018$\,nm, making the S surface bonds shorter by $\Delta b = 0.017$\,nm. 

The ($\sqrt3\times\sqrt3$)-R30° superstructure visible in LEED and STM is according to our DFT calculations due to distortions caused by the missing Cr atoms in the middle Cr plane. The missing Cr atoms affect the position of S in the topmost plane. For three S atoms surrounding one Cr vacancy, their displacements are highlighted by arrows in Figure~\ref{fgr:Fig_DFT_Cr2.66S4}(a). The DFT simulated constant-current STM topograph of Cr$_{2\frac{2}{3}}$S$_4$-2D shown in the upper panel of Figure~\ref{fgr:Fig_DFT_Cr2.66S4}(d) closely resembles the experimental STM constant-current topograph shown in the lower panel.

Band structure and PDOS of Cr$_{2\frac{2}{3}}$S$_4$-2D are presented in Figure~\ref{fgr:Fig_DFT_Cr2.66S4}(e,f). The unfolded band structure has the conduction band minimum at the boundary of the original first BZ at the M-point, consistent with the smeared out CBE observed in the $\mathrm{d}I/\mathrm{d}V$ spectra in Figure~\ref{fgr:Fig_STS}(b) and as discussed above. The valence band maximum is located roughly in the middle between the M- and the $\Gamma$-point. Thus, the electrons at the VBE have smaller parallel momentum, consistent with the sharp VBE seen in STS. The calculated band gap of Cr$_{2\frac{2}{3}}$S$_4$-2D is with 0.33\,eV clearly smaller than the one of Cr$_2$S$_3$-2D, consistent with experiment. However, a quantitative agreement of the $U = 0$ band gap with the experiment is not expected. When $U$ is increased in the calculations, also the band gap increases and for a reasonable value of $U = 2$\,eV it amounts to 0.62\,eV, close to the experimental result (see Table~S4, Supporting Information).

For Cr$_{2\frac{2}{3}}$S$_4$-2D, the oxidation state of Cr atoms is also $+3$ thanks to the identical stoichiometry of the 2D compound. The magnetic moment is retained and takes identical values to  Cr$_2$S$_3$-2D, i.e., 3\,$\mu_\mathrm{B}$ without and 1.73\,$\mu_\mathrm{B}$ per Cr atom with SOC. To explore the magnetic ground state of Cr$_{2\frac{2}{3}}$S$_4$-2D, we assumed similarly to Cr$_2$S$_3$-2D FM intralayer coupling. For the interlayer magnetic order three configurations were considered: one FM and two ferrimagnetic (FiM) ones. Two adjacent FM coupled Cr planes (one outer and the inner) couple AFM to the third (FiM-I), or the two outer Cr planes are both coupled AFM to the inner partial Cr plane (FiM-II), (see Figure~S14, Supporting Information). The relative energies between FM and FiM-I/FiM-II configurations are listed in Table~S4 (Supporting Information). The results indicate that the ground state of Cr$_{2\frac{2}{3}}$S$_4$-2D is FiM-I. Cr$_{2\frac{2}{3}}$S$_4$-2D has an odd number of Cr planes and is a highly uncompensated antiferromagnet, as the magnetic moment of both outer Cr planes point in the same direction, while the fractional Cr plane contains magnetic moments pointing in the opposite direction. It results in ferrimagnet behavior displaying clearly spin-polarized bands, best visualized in Figure~\ref{fgr:Fig_DFT_Cr2.66S4}(e). Interestingly, CBE is dominated by spin-up states and VBE is composed of spin-down electrons. Like Cr$_2$S$_3$-2D, the interlayer magnetic coupling of Cr$_{2\frac{2}{3}}$S$_4$-2D changes from AFM to FM as the $U$ is considered. The relative energies between FM and FiM-I/FiM-II configurations as a function of $U$ are given in Table\,S4 (Supporting Information). 

\section{Conclusion}
{
We demonstrate the growth of two new chromium-sulfur 2D materials, Cr$_2$S$_3$-2D and Cr$_{2\frac{2}{3}}$S$_4$-2D, in single unit cell thickness. Cr$_2$S$_3$-2D exhibits threefold symmetry with an in-plane lattice constant of $0.341 \pm 0.001$\,nm, an apparent STM height of 0.79\,nm, and 2 Cr atoms per unit cell. Upon annealing to 850\,K or growing over the complete layer, Cr$_2$S$_3$-2D transforms into Cr$_{2\frac{2}{3}}$S$_4$-2D with the same lattice parameter, but a thickness of 1.02\,nm. This transition results in an additional $(\sqrt{3} \times \sqrt{3})$-R30$^{\circ}$ superstructure in respect to the original lattice. Both materials can also be grown phase-pure to almost a full layer. Cr$_2$S$_3$-2D and Cr$_{2\frac{2}{3}}$S$_4$-2D display a NiAs-type structure terminated by S planes. However, the middle Cr layer in  Cr$_{2\frac{2}{3}}$S$_4$-2D displays Cr vacancies with 1/3 of the Cr positions missing, resulting in a ($\sqrt3\times\sqrt3$)-R30° superstructure. 

A key factor to develop a proper structure model was the precise STM-based calibration of the amount of Cr per unit cell. This was accomplished by growing and measuring the area of Cr islands on Ir(111). While our method could not be applied to determine the S content in the new materials, we envision that the same type of calibration may be possible for the less volatile Se and Te. Thereby the full stoichiometry for 2D materials of unknown structure could be provided with precision, making our calibration method generally relevant in 2D material structure determination.

Despite considerable variations in the MBE growth procedure, none of the predicted CrS$_2$ phases could be synthesized. Based on a critical assessment of the experimental evidence for the existence of CrS$_2$ and considering the absence of CrS$_2$ as bulk material, it is concluded that additional evidence may be needed to make the existence of CrS$_2$ unambiguous.

Combining the STM and LEED data with DFT calculations enabled us to establish the structure of the two new 2D materials in detail. Remarkably, surface induced relaxation of the Cr layers is apparent and may be a factor stabilizing the new 2D materials. Our DFT calculations predict that both 2D materials are indirect bandgap semiconductors in their free-standing form. Interestingly, we find that when supported on Gr, they exhibit metallic behavior due to their relatively large work function compared to Gr. DFT calculations find sizable magnetic moments for both phases, possessing magnetic order with coupling depending on the degree of Coulomb repulsion, a characteristic that requires experimental validation. The phase-pure, large-area growth of Cr$_2$S$_3$-2D and Cr$_{2\frac{2}{3}}$S$_4$-2D enables detailed investigation of their magnetic properties using averaging techniques and establishes these compounds as promising 2D magnetic materials for potential applications.

\section{Experimental Section}
\subsection{Experimental methods}

In situ growth and analysis of Cr$_x$S$_y$-2D were performed in a variable temperature and a low temperature bath cryostat ultra-high vacuum STM system, both with base pressure of  $1\times10^{-10}$\,mbar. Each system is equipped with standard sample preparation tools, MBE growth, LEED or microchannel plate LEED, and STM operating at 300\,K and at 1.7\,K. 

\subsection{Substrate preparation}
Ir(110) is cleaned and prepared in its unreconstructed state by cycles of 4.5\,keV Xe$^+$ ion sputtering and flashing annealing to 1510\,K followed by cooling to 400\,K in $1\times10^{-7}$\,mbar oxygen pressure. Ir(111) is cleaned by cycles of 1\,keV Ar$^+$ ion sputtering and flash annealing to 1510\,K.

To grow single crystal Gr on Ir(110), clean unreconstructed Ir(110) is heated to 1510\,K and exposed at this temperature for 240\,s to $3\times10^{-7}$\,mbar ethylene \cite{Kraus2022}. 
To grow single crystal Gr on Ir(111), clean Ir(111) is exposed to $1\times10^{-7}$\,mbar ethylene at room temperature for 120\,sec. After flash annealing to 1470\,K without ethylene, subsequently the sample is exposed to $3\times10^{-7}$\,mbar ethylene at 1370\,K for 600\,sec for completion of Gr growth \cite{Coraux2009}. The quality of the as-grown Gr on Ir(110) and Ir(111) is checked by LEED and STM.

\subsection{Sample preparation}
To grow Cr$_x$S$_y$-2D, Gr/Ir(110) is exposed to a Cr flux in the range of $2-5 \times 10^{16}$\,atoms m$^{-2}$ s$^{-1}$ from an e-beam evaporator simultaneously with a sulfur flux obtained by decomposing pyrite (FeS$_2$) in a Knudsen cell in a distance of 8\,cm from the sample. The S flux is characterized by a distant ion gauge measuring a S pressure of $5\times10^{-8}$\,mbar. This growth step was typically performed at 300\,K, if not specified differently. To complete Cr$_x$S$_y$-2D formation, the sample is subsequently annealed for half the initial deposition time in the same sulfur flux at temperatures ranging from 500\,K to 1050\,K and specified in the respective captions.

\subsection{LEED measurements}
LEED patterns were acquired at 300\,K. Distortions in the reciprocal space due to the flat microchannel plate are corrected for the microchannel plate LEED using a Phython script.    

\subsection{STM measurements}
STM data is acquired at 300\,K in the variable temperature STM system and at 1.7\,K in the bath cyrostat STM system. Constant current topographies are presented with sample bias $V_\mathrm{b}$ and tunneling current $I_\mathrm{t}$ specified in the captions. Image processing was conducted using WSxM \cite{Horcas_WSxM_2007}. 

\subsection{STS measurements}
For STS, W tips were further prepared on Cu(111) until the signature of the Cu(111) surface state was well reproduced. The $\mathrm{d}I/\mathrm{d}V$ spectra are recorded at 1.7\,K with stabilization bias $V_{st}$ and stabilization current $I_{st}$ using a standard lock-in technique with a modulation frequency $f_{mod}$ and modulation voltage $V_{mod}$ with respective values specified in the captions. 

\subsection{Theoretical calculations}
The energetics and magnetic properties of all considered systems were investigated using spin-polarized density functional theory (DFT) as implemented in the VASP code \cite{vasp1,vasp2}.
All the calculations were carried out using the Perdew-Burke-Ernzerhof exchange-correlation functional \cite{perdew_1996}. To account for the high correlation effects among d-orbital electrons of chromium atoms, an effective Hubbard value ($U$) was considered for Cr atoms within the DFT + U approach \cite{Dudarev1998}. A plane-wave cut-off energy of 500 eV and force tolerance of 0.01 eV/Å was set for geometry optimization. Van der Waals interactions were considered using the DFT-D2 method proposed by Grimme \cite{DFT-D2}. A vacuum space of approximately 30 Å was introduced to model the slabs between the periodic images in the confinement direction.  

The magnetic configuration does not influence the structure outcome, as the magnetic order energy dependence is an order of magnitude smaller than the stacking dependence. The magnetic order was calculated only for the NiAs-type structure.

Note that, due to the minimal contribution of the SOC effect to the total energy, structural changes arising from the SOC effect were not included in the calculations. Furthermore, the energy differences among magnetic states typically change by less than 11 meV with the SOC effect. Therefore, SOC effects are unlikely to significantly alter the results presented in this work.
\\
\begin{acknowledgments}
	\noindent
    Funding from Deutsche Forschungsgemeinschaft (DFG) through CRC 1238 (project number 277146847, subprojects A01 and B06) is acknowledged. J.F. and W.J. acknowledge financial support from the DFG through project FI 2624/1-1 (project No. 462692705) within the SPP 2137 and project JO 1972/2-1 (project No. 535290457) within SPP 2244, respectively. A.V.K. acknowledges funding from DFG projects KR 4866/11-1, and the collaborative research center “Chemistry of Synthetic 2D Materials” CRC-1415-417590517. Generous CPU time grants from the Technical University of Dresden computing cluster (TAURUS) and Gauss Centre for Supercomputing e.V. (www.gauss-centre.eu), Supercomputer HAWK at Höchstleistungsrechenzentrum Stuttgart (www.hlrs.de), are greatly appreciated.
\end{acknowledgments}

\section*{Conflict of Interest}
\noindent
The authors declare no conflict of interest.

\section*{Data Availability Statement}
\noindent
The data that support the findings of this study are available from the
corresponding author upon reasonable request.

\bibliographystyle{apsrev4-2}
\bibliography{Ref}

\begin{thebibliography}{82}%
\makeatletter
\providecommand \@ifxundefined [1]{%
 \@ifx{#1\undefined}
}%
\providecommand \@ifnum [1]{%
 \ifnum #1\expandafter \@firstoftwo
 \else \expandafter \@secondoftwo
 \fi
}%
\providecommand \@ifx [1]{%
 \ifx #1\expandafter \@firstoftwo
 \else \expandafter \@secondoftwo
 \fi
}%
\providecommand \natexlab [1]{#1}%
\providecommand \enquote  [1]{``#1''}%
\providecommand \bibnamefont  [1]{#1}%
\providecommand \bibfnamefont [1]{#1}%
\providecommand \citenamefont [1]{#1}%
\providecommand \href@noop [0]{\@secondoftwo}%
\providecommand \href [0]{\begingroup \@sanitize@url \@href}%
\providecommand \@href[1]{\@@startlink{#1}\@@href}%
\providecommand \@@href[1]{\endgroup#1\@@endlink}%
\providecommand \@sanitize@url [0]{\catcode `\\12\catcode `\$12\catcode `\&12\catcode `\#12\catcode `\^12\catcode `\_12\catcode `\%12\relax}%
\providecommand \@@startlink[1]{}%
\providecommand \@@endlink[0]{}%
\providecommand \url  [0]{\begingroup\@sanitize@url \@url }%
\providecommand \@url [1]{\endgroup\@href {#1}{\urlprefix }}%
\providecommand \urlprefix  [0]{URL }%
\providecommand \Eprint [0]{\href }%
\providecommand \doibase [0]{https://doi.org/}%
\providecommand \selectlanguage [0]{\@gobble}%
\providecommand \bibinfo  [0]{\@secondoftwo}%
\providecommand \bibfield  [0]{\@secondoftwo}%
\providecommand \translation [1]{[#1]}%
\providecommand \BibitemOpen [0]{}%
\providecommand \bibitemStop [0]{}%
\providecommand \bibitemNoStop [0]{.\EOS\space}%
\providecommand \EOS [0]{\spacefactor3000\relax}%
\providecommand \BibitemShut  [1]{\csname bibitem#1\endcsname}%
\let\auto@bib@innerbib\@empty
\bibitem [{\citenamefont {Liu}\ \emph {et~al.}(2024{\natexlab{a}})\citenamefont {Liu}, \citenamefont {Gebredingle}, \citenamefont {Guo}, \citenamefont {Zhang},\ and\ \citenamefont {Kim}}]{Liu2024a}%
  \BibitemOpen
  \bibfield  {author} {\bibinfo {author} {\bibfnamefont {X.}~\bibnamefont {Liu}}, \bibinfo {author} {\bibfnamefont {Y.}~\bibnamefont {Gebredingle}}, \bibinfo {author} {\bibfnamefont {X.}~\bibnamefont {Guo}}, \bibinfo {author} {\bibfnamefont {F.}~\bibnamefont {Zhang}},\ and\ \bibinfo {author} {\bibfnamefont {N.}~\bibnamefont {Kim}},\ }\href {https://doi.org/https://doi.org/10.1002/adfm.202316834} {\bibfield  {journal} {\bibinfo  {journal} {Adv. Funct. Mater.}\ }\textbf {\bibinfo {volume} {34}},\ \bibinfo {pages} {2316834} (\bibinfo {year} {2024}{\natexlab{a}})}\BibitemShut {NoStop}%
\bibitem [{\citenamefont {Fan}\ \emph {et~al.}(2023)\citenamefont {Fan}, \citenamefont {Xin}, \citenamefont {Li}, \citenamefont {Zhang}, \citenamefont {Li}, \citenamefont {Zhou}, \citenamefont {Chen}, \citenamefont {Zhang}, \citenamefont {OuYang},\ and\ \citenamefont {Zhou}}]{Fan2023}%
  \BibitemOpen
  \bibfield  {author} {\bibinfo {author} {\bibfnamefont {X.}~\bibnamefont {Fan}}, \bibinfo {author} {\bibfnamefont {R.}~\bibnamefont {Xin}}, \bibinfo {author} {\bibfnamefont {L.}~\bibnamefont {Li}}, \bibinfo {author} {\bibfnamefont {B.}~\bibnamefont {Zhang}}, \bibinfo {author} {\bibfnamefont {C.}~\bibnamefont {Li}}, \bibinfo {author} {\bibfnamefont {X.}~\bibnamefont {Zhou}}, \bibinfo {author} {\bibfnamefont {H.}~\bibnamefont {Chen}}, \bibinfo {author} {\bibfnamefont {H.}~\bibnamefont {Zhang}}, \bibinfo {author} {\bibfnamefont {F.}~\bibnamefont {OuYang}},\ and\ \bibinfo {author} {\bibfnamefont {Y.}~\bibnamefont {Zhou}},\ }\href {https://doi.org/10.1007/s11467-023-1342-y} {\bibfield  {journal} {\bibinfo  {journal} {Front. Phys.}\ }\textbf {\bibinfo {volume} {19}},\ \bibinfo {pages} {23401} (\bibinfo {year} {2023})}\BibitemShut {NoStop}%
\bibitem [{\citenamefont {Rajan}\ \emph {et~al.}(2024)\citenamefont {Rajan}, \citenamefont {Buchberger}, \citenamefont {Edwards}, \citenamefont {Zivanovic}, \citenamefont {Kushwaha}, \citenamefont {Bigi}, \citenamefont {Nanao}, \citenamefont {Saika}, \citenamefont {Armitage}, \citenamefont {Wahl}, \citenamefont {Couture},\ and\ \citenamefont {King}}]{Rajan2024}%
  \BibitemOpen
  \bibfield  {author} {\bibinfo {author} {\bibfnamefont {A.}~\bibnamefont {Rajan}}, \bibinfo {author} {\bibfnamefont {S.}~\bibnamefont {Buchberger}}, \bibinfo {author} {\bibfnamefont {B.}~\bibnamefont {Edwards}}, \bibinfo {author} {\bibfnamefont {A.}~\bibnamefont {Zivanovic}}, \bibinfo {author} {\bibfnamefont {N.}~\bibnamefont {Kushwaha}}, \bibinfo {author} {\bibfnamefont {C.}~\bibnamefont {Bigi}}, \bibinfo {author} {\bibfnamefont {Y.}~\bibnamefont {Nanao}}, \bibinfo {author} {\bibfnamefont {B.~K.}\ \bibnamefont {Saika}}, \bibinfo {author} {\bibfnamefont {O.~R.}\ \bibnamefont {Armitage}}, \bibinfo {author} {\bibfnamefont {P.}~\bibnamefont {Wahl}}, \bibinfo {author} {\bibfnamefont {P.}~\bibnamefont {Couture}},\ and\ \bibinfo {author} {\bibfnamefont {P.~D.~C.}\ \bibnamefont {King}},\ }\href {https://doi.org/https://doi.org/10.1002/adma.202402254} {\bibfield  {journal} {\bibinfo  {journal} {Adv. Mater.}\ }\textbf {\bibinfo {volume} {36}},\ \bibinfo {pages} {2402254} (\bibinfo {year} {2024})}\BibitemShut
  {NoStop}%
\bibitem [{\citenamefont {Huang}\ \emph {et~al.}(2017)\citenamefont {Huang}, \citenamefont {Clark}, \citenamefont {Navarro-Moratalla}, \citenamefont {Klein}, \citenamefont {Cheng}, \citenamefont {Seyler}, \citenamefont {Zhong}, \citenamefont {Schmidgall}, \citenamefont {McGuire}, \citenamefont {Cobden}, \citenamefont {Yao}, \citenamefont {Xiao}, \citenamefont {Jarillo-Herrero},\ and\ \citenamefont {Xu}}]{Huang2017}%
  \BibitemOpen
  \bibfield  {author} {\bibinfo {author} {\bibfnamefont {B.}~\bibnamefont {Huang}}, \bibinfo {author} {\bibfnamefont {G.}~\bibnamefont {Clark}}, \bibinfo {author} {\bibfnamefont {E.}~\bibnamefont {Navarro-Moratalla}}, \bibinfo {author} {\bibfnamefont {D.~R.}\ \bibnamefont {Klein}}, \bibinfo {author} {\bibfnamefont {R.}~\bibnamefont {Cheng}}, \bibinfo {author} {\bibfnamefont {K.~L.}\ \bibnamefont {Seyler}}, \bibinfo {author} {\bibfnamefont {D.}~\bibnamefont {Zhong}}, \bibinfo {author} {\bibfnamefont {E.}~\bibnamefont {Schmidgall}}, \bibinfo {author} {\bibfnamefont {M.~A.}\ \bibnamefont {McGuire}}, \bibinfo {author} {\bibfnamefont {D.~H.}\ \bibnamefont {Cobden}}, \bibinfo {author} {\bibfnamefont {W.}~\bibnamefont {Yao}}, \bibinfo {author} {\bibfnamefont {D.}~\bibnamefont {Xiao}}, \bibinfo {author} {\bibfnamefont {P.}~\bibnamefont {Jarillo-Herrero}},\ and\ \bibinfo {author} {\bibfnamefont {X.}~\bibnamefont {Xu}},\ }\href {https://doi.org/https://doi.org/10.1038/nature22391} {\bibfield  {journal} {\bibinfo
  {journal} {Nature}\ }\textbf {\bibinfo {volume} {546}},\ \bibinfo {pages} {270} (\bibinfo {year} {2017})}\BibitemShut {NoStop}%
\bibitem [{\citenamefont {Chen}\ \emph {et~al.}(2019)\citenamefont {Chen}, \citenamefont {Sun}, \citenamefont {Wang}, \citenamefont {Gu}, \citenamefont {Xu}, \citenamefont {Wu},\ and\ \citenamefont {Gao}}]{Chen2019}%
  \BibitemOpen
  \bibfield  {author} {\bibinfo {author} {\bibfnamefont {W.}~\bibnamefont {Chen}}, \bibinfo {author} {\bibfnamefont {Z.}~\bibnamefont {Sun}}, \bibinfo {author} {\bibfnamefont {Z.}~\bibnamefont {Wang}}, \bibinfo {author} {\bibfnamefont {L.}~\bibnamefont {Gu}}, \bibinfo {author} {\bibfnamefont {X.}~\bibnamefont {Xu}}, \bibinfo {author} {\bibfnamefont {S.}~\bibnamefont {Wu}},\ and\ \bibinfo {author} {\bibfnamefont {C.}~\bibnamefont {Gao}},\ }\href {https://doi.org/10.1126/science.aav1937} {\bibfield  {journal} {\bibinfo  {journal} {Science}\ }\textbf {\bibinfo {volume} {366}},\ \bibinfo {pages} {983} (\bibinfo {year} {2019})}\BibitemShut {NoStop}%
\bibitem [{\citenamefont {Bedoya-Pinto}\ \emph {et~al.}(2021)\citenamefont {Bedoya-Pinto}, \citenamefont {Ji}, \citenamefont {Pandeya}, \citenamefont {Gargiani}, \citenamefont {Valvidares}, \citenamefont {Sessi}, \citenamefont {Taylor}, \citenamefont {Radu}, \citenamefont {Chang},\ and\ \citenamefont {Parkin}}]{Bedoya-Pinto2021}%
  \BibitemOpen
  \bibfield  {author} {\bibinfo {author} {\bibfnamefont {A.}~\bibnamefont {Bedoya-Pinto}}, \bibinfo {author} {\bibfnamefont {J.-R.}\ \bibnamefont {Ji}}, \bibinfo {author} {\bibfnamefont {A.~K.}\ \bibnamefont {Pandeya}}, \bibinfo {author} {\bibfnamefont {P.}~\bibnamefont {Gargiani}}, \bibinfo {author} {\bibfnamefont {M.}~\bibnamefont {Valvidares}}, \bibinfo {author} {\bibfnamefont {P.}~\bibnamefont {Sessi}}, \bibinfo {author} {\bibfnamefont {J.~M.}\ \bibnamefont {Taylor}}, \bibinfo {author} {\bibfnamefont {F.}~\bibnamefont {Radu}}, \bibinfo {author} {\bibfnamefont {K.}~\bibnamefont {Chang}},\ and\ \bibinfo {author} {\bibfnamefont {S.~S.~P.}\ \bibnamefont {Parkin}},\ }\href {https://doi.org/10.1126/science.abd5146} {\bibfield  {journal} {\bibinfo  {journal} {Science}\ }\textbf {\bibinfo {volume} {374}},\ \bibinfo {pages} {616} (\bibinfo {year} {2021})}\BibitemShut {NoStop}%
\bibitem [{\citenamefont {Chua}\ \emph {et~al.}(2021)\citenamefont {Chua}, \citenamefont {Zhou}, \citenamefont {Yu}, \citenamefont {Yu}, \citenamefont {Gou}, \citenamefont {Zhu}, \citenamefont {Zhang}, \citenamefont {Liu}, \citenamefont {Breese}, \citenamefont {Chen}, \citenamefont {Loh}, \citenamefont {Feng}, \citenamefont {Yang}, \citenamefont {Huang},\ and\ \citenamefont {Wee}}]{Chua2021}%
  \BibitemOpen
  \bibfield  {author} {\bibinfo {author} {\bibfnamefont {R.}~\bibnamefont {Chua}}, \bibinfo {author} {\bibfnamefont {J.}~\bibnamefont {Zhou}}, \bibinfo {author} {\bibfnamefont {X.}~\bibnamefont {Yu}}, \bibinfo {author} {\bibfnamefont {W.}~\bibnamefont {Yu}}, \bibinfo {author} {\bibfnamefont {J.}~\bibnamefont {Gou}}, \bibinfo {author} {\bibfnamefont {R.}~\bibnamefont {Zhu}}, \bibinfo {author} {\bibfnamefont {L.}~\bibnamefont {Zhang}}, \bibinfo {author} {\bibfnamefont {M.}~\bibnamefont {Liu}}, \bibinfo {author} {\bibfnamefont {M.~B.~H.}\ \bibnamefont {Breese}}, \bibinfo {author} {\bibfnamefont {W.}~\bibnamefont {Chen}}, \bibinfo {author} {\bibfnamefont {K.~P.}\ \bibnamefont {Loh}}, \bibinfo {author} {\bibfnamefont {Y.~P.}\ \bibnamefont {Feng}}, \bibinfo {author} {\bibfnamefont {M.}~\bibnamefont {Yang}}, \bibinfo {author} {\bibfnamefont {Y.~L.}\ \bibnamefont {Huang}},\ and\ \bibinfo {author} {\bibfnamefont {A.~T.~S.}\ \bibnamefont {Wee}},\ }\href {https://doi.org/https://doi.org/10.1002/adma.202103360}
  {\bibfield  {journal} {\bibinfo  {journal} {Adv. Mater.}\ }\textbf {\bibinfo {volume} {33}},\ \bibinfo {pages} {2103360} (\bibinfo {year} {2021})}\BibitemShut {NoStop}%
\bibitem [{\citenamefont {Lasek}\ \emph {et~al.}(2022)\citenamefont {Lasek}, \citenamefont {Coelho}, \citenamefont {Gargiani}, \citenamefont {Valvidares}, \citenamefont {Mohseni}, \citenamefont {Meyerheim}, \citenamefont {Kostanovskiy}, \citenamefont {Zberecki},\ and\ \citenamefont {Batzill}}]{Lasek2022}%
  \BibitemOpen
  \bibfield  {author} {\bibinfo {author} {\bibfnamefont {K.}~\bibnamefont {Lasek}}, \bibinfo {author} {\bibfnamefont {P.~M.}\ \bibnamefont {Coelho}}, \bibinfo {author} {\bibfnamefont {P.}~\bibnamefont {Gargiani}}, \bibinfo {author} {\bibfnamefont {M.}~\bibnamefont {Valvidares}}, \bibinfo {author} {\bibfnamefont {K.}~\bibnamefont {Mohseni}}, \bibinfo {author} {\bibfnamefont {H.~L.}\ \bibnamefont {Meyerheim}}, \bibinfo {author} {\bibfnamefont {I.}~\bibnamefont {Kostanovskiy}}, \bibinfo {author} {\bibfnamefont {K.}~\bibnamefont {Zberecki}},\ and\ \bibinfo {author} {\bibfnamefont {M.}~\bibnamefont {Batzill}},\ }\href {https://doi.org/10.1063/5.0070079} {\bibfield  {journal} {\bibinfo  {journal} {Appl. Phys. Rev.}\ }\textbf {\bibinfo {volume} {9}},\ \bibinfo {pages} {011409} (\bibinfo {year} {2022})}\BibitemShut {NoStop}%
\bibitem [{\citenamefont {Saha}\ \emph {et~al.}(2022)\citenamefont {Saha}, \citenamefont {Meyerheim}, \citenamefont {G{\"o}bel}, \citenamefont {Hazra}, \citenamefont {Deniz}, \citenamefont {Mohseni}, \citenamefont {Antonov}, \citenamefont {Ernst}, \citenamefont {Knyazev}, \citenamefont {Bedoya-Pinto}, \citenamefont {Mertig},\ and\ \citenamefont {Parkin}}]{Saha2022}%
  \BibitemOpen
  \bibfield  {author} {\bibinfo {author} {\bibfnamefont {R.}~\bibnamefont {Saha}}, \bibinfo {author} {\bibfnamefont {H.~L.}\ \bibnamefont {Meyerheim}}, \bibinfo {author} {\bibfnamefont {B.}~\bibnamefont {G{\"o}bel}}, \bibinfo {author} {\bibfnamefont {B.~K.}\ \bibnamefont {Hazra}}, \bibinfo {author} {\bibfnamefont {H.}~\bibnamefont {Deniz}}, \bibinfo {author} {\bibfnamefont {K.}~\bibnamefont {Mohseni}}, \bibinfo {author} {\bibfnamefont {V.}~\bibnamefont {Antonov}}, \bibinfo {author} {\bibfnamefont {A.}~\bibnamefont {Ernst}}, \bibinfo {author} {\bibfnamefont {D.}~\bibnamefont {Knyazev}}, \bibinfo {author} {\bibfnamefont {A.}~\bibnamefont {Bedoya-Pinto}}, \bibinfo {author} {\bibfnamefont {I.}~\bibnamefont {Mertig}},\ and\ \bibinfo {author} {\bibfnamefont {S.~S.~P.}\ \bibnamefont {Parkin}},\ }\href {https://doi.org/https://doi.org/10.1038/s41467-022-31319-y} {\bibfield  {journal} {\bibinfo  {journal} {Nat. Commun.}\ }\textbf {\bibinfo {volume} {13}},\ \bibinfo {pages} {3965} (\bibinfo {year} {2022})}\BibitemShut
  {NoStop}%
\bibitem [{\citenamefont {Xian}\ \emph {et~al.}(2022)\citenamefont {Xian}, \citenamefont {Wang}, \citenamefont {Nie}, \citenamefont {Li}, \citenamefont {Han}, \citenamefont {Lin}, \citenamefont {Zhang}, \citenamefont {Liu}, \citenamefont {Zhang}, \citenamefont {Miao}, \citenamefont {Yi}, \citenamefont {Wu}, \citenamefont {Chen}, \citenamefont {Han}, \citenamefont {Xia}, \citenamefont {Ji},\ and\ \citenamefont {Fu}}]{Xian2022}%
  \BibitemOpen
  \bibfield  {author} {\bibinfo {author} {\bibfnamefont {J.-J.}\ \bibnamefont {Xian}}, \bibinfo {author} {\bibfnamefont {C.}~\bibnamefont {Wang}}, \bibinfo {author} {\bibfnamefont {J.-H.}\ \bibnamefont {Nie}}, \bibinfo {author} {\bibfnamefont {R.}~\bibnamefont {Li}}, \bibinfo {author} {\bibfnamefont {M.}~\bibnamefont {Han}}, \bibinfo {author} {\bibfnamefont {J.}~\bibnamefont {Lin}}, \bibinfo {author} {\bibfnamefont {W.-H.}\ \bibnamefont {Zhang}}, \bibinfo {author} {\bibfnamefont {Z.-Y.}\ \bibnamefont {Liu}}, \bibinfo {author} {\bibfnamefont {Z.-M.}\ \bibnamefont {Zhang}}, \bibinfo {author} {\bibfnamefont {M.-P.}\ \bibnamefont {Miao}}, \bibinfo {author} {\bibfnamefont {Y.}~\bibnamefont {Yi}}, \bibinfo {author} {\bibfnamefont {S.}~\bibnamefont {Wu}}, \bibinfo {author} {\bibfnamefont {X.}~\bibnamefont {Chen}}, \bibinfo {author} {\bibfnamefont {J.}~\bibnamefont {Han}}, \bibinfo {author} {\bibfnamefont {Z.}~\bibnamefont {Xia}}, \bibinfo {author} {\bibfnamefont {W.}~\bibnamefont {Ji}},\ and\ \bibinfo {author}
  {\bibfnamefont {Y.-S.}\ \bibnamefont {Fu}},\ }\href {https://doi.org/https://doi.org/10.1038/s41467-021-27834-z} {\bibfield  {journal} {\bibinfo  {journal} {Nat. Commun.}\ }\textbf {\bibinfo {volume} {13}},\ \bibinfo {pages} {257} (\bibinfo {year} {2022})}\BibitemShut {NoStop}%
\bibitem [{\citenamefont {Zhang}\ \emph {et~al.}(2023)\citenamefont {Zhang}, \citenamefont {Liu}, \citenamefont {Zhang}, \citenamefont {Yuan}, \citenamefont {Wen}, \citenamefont {Li}, \citenamefont {Zheng}, \citenamefont {Zhang}, \citenamefont {Hou}, \citenamefont {Yin}, \citenamefont {Liu}, \citenamefont {Peng},\ and\ \citenamefont {Zhang}}]{Zhang2023}%
  \BibitemOpen
  \bibfield  {author} {\bibinfo {author} {\bibfnamefont {C.}~\bibnamefont {Zhang}}, \bibinfo {author} {\bibfnamefont {C.}~\bibnamefont {Liu}}, \bibinfo {author} {\bibfnamefont {J.}~\bibnamefont {Zhang}}, \bibinfo {author} {\bibfnamefont {Y.}~\bibnamefont {Yuan}}, \bibinfo {author} {\bibfnamefont {Y.}~\bibnamefont {Wen}}, \bibinfo {author} {\bibfnamefont {Y.}~\bibnamefont {Li}}, \bibinfo {author} {\bibfnamefont {D.}~\bibnamefont {Zheng}}, \bibinfo {author} {\bibfnamefont {Q.}~\bibnamefont {Zhang}}, \bibinfo {author} {\bibfnamefont {Z.}~\bibnamefont {Hou}}, \bibinfo {author} {\bibfnamefont {G.}~\bibnamefont {Yin}}, \bibinfo {author} {\bibfnamefont {K.}~\bibnamefont {Liu}}, \bibinfo {author} {\bibfnamefont {Y.}~\bibnamefont {Peng}},\ and\ \bibinfo {author} {\bibfnamefont {X.-X.}\ \bibnamefont {Zhang}},\ }\href {https://doi.org/https://doi.org/10.1002/adma.202205967} {\bibfield  {journal} {\bibinfo  {journal} {Adv. Mater.}\ }\textbf {\bibinfo {volume} {35}},\ \bibinfo {pages} {2205967} (\bibinfo {year}
  {2023})}\BibitemShut {NoStop}%
\bibitem [{\citenamefont {Khatun}\ \emph {et~al.}(2024)\citenamefont {Khatun}, \citenamefont {Alanwoko}, \citenamefont {Pathirage}, \citenamefont {de~Oliveira}, \citenamefont {Tromer}, \citenamefont {Autreto}, \citenamefont {Galvao},\ and\ \citenamefont {Batzill}}]{Khatun24}%
  \BibitemOpen
  \bibfield  {author} {\bibinfo {author} {\bibfnamefont {S.}~\bibnamefont {Khatun}}, \bibinfo {author} {\bibfnamefont {O.}~\bibnamefont {Alanwoko}}, \bibinfo {author} {\bibfnamefont {V.}~\bibnamefont {Pathirage}}, \bibinfo {author} {\bibfnamefont {C.~C.}\ \bibnamefont {de~Oliveira}}, \bibinfo {author} {\bibfnamefont {R.~M.}\ \bibnamefont {Tromer}}, \bibinfo {author} {\bibfnamefont {P.~A.~S.}\ \bibnamefont {Autreto}}, \bibinfo {author} {\bibfnamefont {D.~S.}\ \bibnamefont {Galvao}},\ and\ \bibinfo {author} {\bibfnamefont {M.}~\bibnamefont {Batzill}},\ }\href {https://doi.org/https://doi.org/10.1002/adfm.202315112} {\bibfield  {journal} {\bibinfo  {journal} {Adv. Funct. Mater.}\ ,\ \bibinfo {pages} {2315112}} (\bibinfo {year} {2024})}\BibitemShut {NoStop}%
\bibitem [{\citenamefont {Cui}\ \emph {et~al.}(2024)\citenamefont {Cui}, \citenamefont {He}, \citenamefont {Wu}, \citenamefont {Zhang}, \citenamefont {Lu}, \citenamefont {Li}, \citenamefont {Hu}, \citenamefont {Pan}, \citenamefont {Zhu}, \citenamefont {Huan}, \citenamefont {Li}, \citenamefont {Duan}, \citenamefont {Ji}, \citenamefont {Zhao},\ and\ \citenamefont {Zhang}}]{Cui2024}%
  \BibitemOpen
  \bibfield  {author} {\bibinfo {author} {\bibfnamefont {F.}~\bibnamefont {Cui}}, \bibinfo {author} {\bibfnamefont {K.}~\bibnamefont {He}}, \bibinfo {author} {\bibfnamefont {S.}~\bibnamefont {Wu}}, \bibinfo {author} {\bibfnamefont {H.}~\bibnamefont {Zhang}}, \bibinfo {author} {\bibfnamefont {Y.}~\bibnamefont {Lu}}, \bibinfo {author} {\bibfnamefont {Z.}~\bibnamefont {Li}}, \bibinfo {author} {\bibfnamefont {J.}~\bibnamefont {Hu}}, \bibinfo {author} {\bibfnamefont {S.}~\bibnamefont {Pan}}, \bibinfo {author} {\bibfnamefont {L.}~\bibnamefont {Zhu}}, \bibinfo {author} {\bibfnamefont {Y.}~\bibnamefont {Huan}}, \bibinfo {author} {\bibfnamefont {B.}~\bibnamefont {Li}}, \bibinfo {author} {\bibfnamefont {X.}~\bibnamefont {Duan}}, \bibinfo {author} {\bibfnamefont {Q.}~\bibnamefont {Ji}}, \bibinfo {author} {\bibfnamefont {X.}~\bibnamefont {Zhao}},\ and\ \bibinfo {author} {\bibfnamefont {Y.}~\bibnamefont {Zhang}},\ }\href {https://doi.org/10.1021/acsnano.3c10609} {\bibfield  {journal} {\bibinfo  {journal} {ACS Nano}\
  }\textbf {\bibinfo {volume} {18}},\ \bibinfo {pages} {6276} (\bibinfo {year} {2024})}\BibitemShut {NoStop}%
\bibitem [{\citenamefont {Lu}\ \emph {et~al.}(2024)\citenamefont {Lu}, \citenamefont {He}, \citenamefont {Pan}, \citenamefont {Guo}, \citenamefont {Yang}, \citenamefont {Wu}, \citenamefont {Song}, \citenamefont {Li}, \citenamefont {Liu}, \citenamefont {Liu}, \citenamefont {Shen}, \citenamefont {Chen}, \citenamefont {Ma}, \citenamefont {Tang},\ and\ \citenamefont {Xie}}]{Lu2024}%
  \BibitemOpen
  \bibfield  {author} {\bibinfo {author} {\bibfnamefont {S.}~\bibnamefont {Lu}}, \bibinfo {author} {\bibfnamefont {Z.}~\bibnamefont {He}}, \bibinfo {author} {\bibfnamefont {Y.}~\bibnamefont {Pan}}, \bibinfo {author} {\bibfnamefont {Z.}~\bibnamefont {Guo}}, \bibinfo {author} {\bibfnamefont {Y.}~\bibnamefont {Yang}}, \bibinfo {author} {\bibfnamefont {F.}~\bibnamefont {Wu}}, \bibinfo {author} {\bibfnamefont {Y.}~\bibnamefont {Song}}, \bibinfo {author} {\bibfnamefont {Z.}~\bibnamefont {Li}}, \bibinfo {author} {\bibfnamefont {Z.}~\bibnamefont {Liu}}, \bibinfo {author} {\bibfnamefont {Z.}~\bibnamefont {Liu}}, \bibinfo {author} {\bibfnamefont {D.}~\bibnamefont {Shen}}, \bibinfo {author} {\bibfnamefont {L.}~\bibnamefont {Chen}}, \bibinfo {author} {\bibfnamefont {Y.}~\bibnamefont {Ma}}, \bibinfo {author} {\bibfnamefont {S.}~\bibnamefont {Tang}},\ and\ \bibinfo {author} {\bibfnamefont {X.}~\bibnamefont {Xie}},\ }\href {https://doi.org/https://doi.org/10.1002/adfm.202406339} {\bibfield  {journal} {\bibinfo  {journal}
  {Adv. Funct. Mater.}\ }\textbf {\bibinfo {volume} {34}},\ \bibinfo {pages} {2406339} (\bibinfo {year} {2024})}\BibitemShut {NoStop}%
\bibitem [{\citenamefont {Kushwaha}\ \emph {et~al.}(2025)\citenamefont {Kushwaha}, \citenamefont {Armitage}, \citenamefont {Edwards}, \citenamefont {Trzaska}, \citenamefont {Bencok}, \citenamefont {Biswas}, \citenamefont {Lee}, \citenamefont {Sanders}, \citenamefont {van~der Laan}, \citenamefont {Wahl}, \citenamefont {King},\ and\ \citenamefont {Rajan}}]{Kushwaha2024}%
  \BibitemOpen
  \bibfield  {author} {\bibinfo {author} {\bibfnamefont {N.}~\bibnamefont {Kushwaha}}, \bibinfo {author} {\bibfnamefont {O.}~\bibnamefont {Armitage}}, \bibinfo {author} {\bibfnamefont {B.}~\bibnamefont {Edwards}}, \bibinfo {author} {\bibfnamefont {L.}~\bibnamefont {Trzaska}}, \bibinfo {author} {\bibfnamefont {P.}~\bibnamefont {Bencok}}, \bibinfo {author} {\bibfnamefont {D.}~\bibnamefont {Biswas}}, \bibinfo {author} {\bibfnamefont {T.-L.}\ \bibnamefont {Lee}}, \bibinfo {author} {\bibfnamefont {C.}~\bibnamefont {Sanders}}, \bibinfo {author} {\bibfnamefont {G.}~\bibnamefont {van~der Laan}}, \bibinfo {author} {\bibfnamefont {P.}~\bibnamefont {Wahl}}, \bibinfo {author} {\bibfnamefont {P.~D.}\ \bibnamefont {King}},\ and\ \bibinfo {author} {\bibfnamefont {A.}~\bibnamefont {Rajan}},\ }\href {https://doi.org/10.1038/s41535-025-00772-5} {\bibfield  {journal} {\bibinfo  {journal} {npj Quantum Mater.}\ }\textbf {\bibinfo {volume} {10}},\ \bibinfo {pages} {50} (\bibinfo {year} {2025})}\BibitemShut {NoStop}%
\bibitem [{\citenamefont {Chu}\ \emph {et~al.}(2019)\citenamefont {Chu}, \citenamefont {Zhang}, \citenamefont {Wen}, \citenamefont {Qiao}, \citenamefont {Wu}, \citenamefont {He}, \citenamefont {Yin}, \citenamefont {Cheng}, \citenamefont {Wang}, \citenamefont {Wang}, \citenamefont {Xiong}, \citenamefont {Li},\ and\ \citenamefont {He}}]{Chu2019}%
  \BibitemOpen
  \bibfield  {author} {\bibinfo {author} {\bibfnamefont {J.}~\bibnamefont {Chu}}, \bibinfo {author} {\bibfnamefont {Y.}~\bibnamefont {Zhang}}, \bibinfo {author} {\bibfnamefont {Y.}~\bibnamefont {Wen}}, \bibinfo {author} {\bibfnamefont {R.}~\bibnamefont {Qiao}}, \bibinfo {author} {\bibfnamefont {C.}~\bibnamefont {Wu}}, \bibinfo {author} {\bibfnamefont {P.}~\bibnamefont {He}}, \bibinfo {author} {\bibfnamefont {L.}~\bibnamefont {Yin}}, \bibinfo {author} {\bibfnamefont {R.}~\bibnamefont {Cheng}}, \bibinfo {author} {\bibfnamefont {F.}~\bibnamefont {Wang}}, \bibinfo {author} {\bibfnamefont {Z.}~\bibnamefont {Wang}}, \bibinfo {author} {\bibfnamefont {J.}~\bibnamefont {Xiong}}, \bibinfo {author} {\bibfnamefont {Y.}~\bibnamefont {Li}},\ and\ \bibinfo {author} {\bibfnamefont {J.}~\bibnamefont {He}},\ }\href {https://doi.org/10.1021/acs.nanolett.9b00386} {\bibfield  {journal} {\bibinfo  {journal} {Nano Lett.}\ }\textbf {\bibinfo {volume} {19}},\ \bibinfo {pages} {2154} (\bibinfo {year} {2019})}\BibitemShut {NoStop}%
\bibitem [{\citenamefont {Cui}\ \emph {et~al.}(2020)\citenamefont {Cui}, \citenamefont {Zhao}, \citenamefont {Xu}, \citenamefont {Tang}, \citenamefont {Shang}, \citenamefont {Shi}, \citenamefont {Huan}, \citenamefont {Liao}, \citenamefont {Chen}, \citenamefont {Hou}, \citenamefont {Zhang}, \citenamefont {Pennycook},\ and\ \citenamefont {Zhang}}]{Cui2019}%
  \BibitemOpen
  \bibfield  {author} {\bibinfo {author} {\bibfnamefont {F.}~\bibnamefont {Cui}}, \bibinfo {author} {\bibfnamefont {X.}~\bibnamefont {Zhao}}, \bibinfo {author} {\bibfnamefont {J.}~\bibnamefont {Xu}}, \bibinfo {author} {\bibfnamefont {B.}~\bibnamefont {Tang}}, \bibinfo {author} {\bibfnamefont {Q.}~\bibnamefont {Shang}}, \bibinfo {author} {\bibfnamefont {J.}~\bibnamefont {Shi}}, \bibinfo {author} {\bibfnamefont {Y.}~\bibnamefont {Huan}}, \bibinfo {author} {\bibfnamefont {J.}~\bibnamefont {Liao}}, \bibinfo {author} {\bibfnamefont {Q.}~\bibnamefont {Chen}}, \bibinfo {author} {\bibfnamefont {Y.}~\bibnamefont {Hou}}, \bibinfo {author} {\bibfnamefont {Q.}~\bibnamefont {Zhang}}, \bibinfo {author} {\bibfnamefont {S.~J.}\ \bibnamefont {Pennycook}},\ and\ \bibinfo {author} {\bibfnamefont {Y.}~\bibnamefont {Zhang}},\ }\href {https://doi.org/https://doi.org/10.1002/adma.201905896} {\bibfield  {journal} {\bibinfo  {journal} {Adv. Mater.}\ }\textbf {\bibinfo {volume} {32}},\ \bibinfo {pages} {1905896} (\bibinfo {year}
  {2020})}\BibitemShut {NoStop}%
\bibitem [{\citenamefont {Zhou}\ \emph {et~al.}(2019)\citenamefont {Zhou}, \citenamefont {Wang}, \citenamefont {Han}, \citenamefont {Wang}, \citenamefont {Li}, \citenamefont {Gan},\ and\ \citenamefont {Zhai}}]{Zhou2019}%
  \BibitemOpen
  \bibfield  {author} {\bibinfo {author} {\bibfnamefont {S.}~\bibnamefont {Zhou}}, \bibinfo {author} {\bibfnamefont {R.}~\bibnamefont {Wang}}, \bibinfo {author} {\bibfnamefont {J.}~\bibnamefont {Han}}, \bibinfo {author} {\bibfnamefont {D.}~\bibnamefont {Wang}}, \bibinfo {author} {\bibfnamefont {H.}~\bibnamefont {Li}}, \bibinfo {author} {\bibfnamefont {L.}~\bibnamefont {Gan}},\ and\ \bibinfo {author} {\bibfnamefont {T.}~\bibnamefont {Zhai}},\ }\href {https://doi.org/https://doi.org/10.1002/adfm.201805880} {\bibfield  {journal} {\bibinfo  {journal} {Adv. Funct. Mater.}\ }\textbf {\bibinfo {volume} {29}},\ \bibinfo {pages} {1805880} (\bibinfo {year} {2019})}\BibitemShut {NoStop}%
\bibitem [{\citenamefont {Xie}\ \emph {et~al.}(2021)\citenamefont {Xie}, \citenamefont {Wang}, \citenamefont {Li}, \citenamefont {Li}, \citenamefont {Zhang}, \citenamefont {Zhu}, \citenamefont {Guo}, \citenamefont {Wang},\ and\ \citenamefont {Zhang}}]{Xie2021}%
  \BibitemOpen
  \bibfield  {author} {\bibinfo {author} {\bibfnamefont {L.}~\bibnamefont {Xie}}, \bibinfo {author} {\bibfnamefont {J.}~\bibnamefont {Wang}}, \bibinfo {author} {\bibfnamefont {J.}~\bibnamefont {Li}}, \bibinfo {author} {\bibfnamefont {C.}~\bibnamefont {Li}}, \bibinfo {author} {\bibfnamefont {Y.}~\bibnamefont {Zhang}}, \bibinfo {author} {\bibfnamefont {B.}~\bibnamefont {Zhu}}, \bibinfo {author} {\bibfnamefont {Y.}~\bibnamefont {Guo}}, \bibinfo {author} {\bibfnamefont {Z.}~\bibnamefont {Wang}},\ and\ \bibinfo {author} {\bibfnamefont {K.}~\bibnamefont {Zhang}},\ }\href {https://doi.org/https://doi.org/10.1002/aelm.202000962} {\bibfield  {journal} {\bibinfo  {journal} {Adv. Electron. Mater.}\ }\textbf {\bibinfo {volume} {7}},\ \bibinfo {pages} {2000962} (\bibinfo {year} {2021})}\BibitemShut {NoStop}%
\bibitem [{\citenamefont {Moinuddin}\ \emph {et~al.}(2021)\citenamefont {Moinuddin}, \citenamefont {Srinivasan},\ and\ \citenamefont {Sharma}}]{Moinuddin2021}%
  \BibitemOpen
  \bibfield  {author} {\bibinfo {author} {\bibfnamefont {M.~G.}\ \bibnamefont {Moinuddin}}, \bibinfo {author} {\bibfnamefont {S.}~\bibnamefont {Srinivasan}},\ and\ \bibinfo {author} {\bibfnamefont {S.~K.}\ \bibnamefont {Sharma}},\ }\href {https://doi.org/https://doi.org/10.1002/aelm.202001116} {\bibfield  {journal} {\bibinfo  {journal} {Adv. Electron. Mater.}\ }\textbf {\bibinfo {volume} {7}},\ \bibinfo {pages} {2001116} (\bibinfo {year} {2021})}\BibitemShut {NoStop}%
\bibitem [{\citenamefont {Cui}\ \emph {et~al.}(2022)\citenamefont {Cui}, \citenamefont {Zhao}, \citenamefont {Tang}, \citenamefont {Zhu}, \citenamefont {Huan}, \citenamefont {Chen}, \citenamefont {Liu},\ and\ \citenamefont {Zhang}}]{Cui2022}%
  \BibitemOpen
  \bibfield  {author} {\bibinfo {author} {\bibfnamefont {F.}~\bibnamefont {Cui}}, \bibinfo {author} {\bibfnamefont {X.}~\bibnamefont {Zhao}}, \bibinfo {author} {\bibfnamefont {B.}~\bibnamefont {Tang}}, \bibinfo {author} {\bibfnamefont {L.}~\bibnamefont {Zhu}}, \bibinfo {author} {\bibfnamefont {Y.}~\bibnamefont {Huan}}, \bibinfo {author} {\bibfnamefont {Q.}~\bibnamefont {Chen}}, \bibinfo {author} {\bibfnamefont {Z.}~\bibnamefont {Liu}},\ and\ \bibinfo {author} {\bibfnamefont {Y.}~\bibnamefont {Zhang}},\ }\href {https://doi.org/https://doi.org/10.1002/smll.202105744} {\bibfield  {journal} {\bibinfo  {journal} {Small}\ }\textbf {\bibinfo {volume} {18}},\ \bibinfo {pages} {2105744} (\bibinfo {year} {2022})}\BibitemShut {NoStop}%
\bibitem [{\citenamefont {Liu}\ \emph {et~al.}(2022)\citenamefont {Liu}, \citenamefont {Tseng}, \citenamefont {Huang}, \citenamefont {Lo}, \citenamefont {Hou}, \citenamefont {Wang}, \citenamefont {Yasuhara},\ and\ \citenamefont {Wu}}]{Liu2022}%
  \BibitemOpen
  \bibfield  {author} {\bibinfo {author} {\bibfnamefont {C.-L.}\ \bibnamefont {Liu}}, \bibinfo {author} {\bibfnamefont {Y.-T.}\ \bibnamefont {Tseng}}, \bibinfo {author} {\bibfnamefont {C.-W.}\ \bibnamefont {Huang}}, \bibinfo {author} {\bibfnamefont {H.-Y.}\ \bibnamefont {Lo}}, \bibinfo {author} {\bibfnamefont {A.-Y.}\ \bibnamefont {Hou}}, \bibinfo {author} {\bibfnamefont {C.-H.}\ \bibnamefont {Wang}}, \bibinfo {author} {\bibfnamefont {A.}~\bibnamefont {Yasuhara}},\ and\ \bibinfo {author} {\bibfnamefont {W.-W.}\ \bibnamefont {Wu}},\ }\href {https://doi.org/10.1021/acs.nanolett.2c02974} {\bibfield  {journal} {\bibinfo  {journal} {Nano Lett.}\ }\textbf {\bibinfo {volume} {22}},\ \bibinfo {pages} {7944} (\bibinfo {year} {2022})}\BibitemShut {NoStop}%
\bibitem [{\citenamefont {Yao}\ \emph {et~al.}(2023)\citenamefont {Yao}, \citenamefont {Liu}, \citenamefont {Zhou}, \citenamefont {Yang}, \citenamefont {Huang}, \citenamefont {Fu}, \citenamefont {Yuan}, \citenamefont {Nie}, \citenamefont {Dai}, \citenamefont {Xu},\ and\ \citenamefont {Gao}}]{Yao2023}%
  \BibitemOpen
  \bibfield  {author} {\bibinfo {author} {\bibfnamefont {B.}~\bibnamefont {Yao}}, \bibinfo {author} {\bibfnamefont {W.}~\bibnamefont {Liu}}, \bibinfo {author} {\bibfnamefont {X.}~\bibnamefont {Zhou}}, \bibinfo {author} {\bibfnamefont {J.}~\bibnamefont {Yang}}, \bibinfo {author} {\bibfnamefont {X.}~\bibnamefont {Huang}}, \bibinfo {author} {\bibfnamefont {Z.}~\bibnamefont {Fu}}, \bibinfo {author} {\bibfnamefont {G.}~\bibnamefont {Yuan}}, \bibinfo {author} {\bibfnamefont {Y.}~\bibnamefont {Nie}}, \bibinfo {author} {\bibfnamefont {Y.}~\bibnamefont {Dai}}, \bibinfo {author} {\bibfnamefont {J.}~\bibnamefont {Xu}},\ and\ \bibinfo {author} {\bibfnamefont {L.}~\bibnamefont {Gao}},\ }\href {https://doi.org/10.1088/1361-648X/acd509} {\bibfield  {journal} {\bibinfo  {journal} {J. Phys. Condens. Matter}\ }\textbf {\bibinfo {volume} {35}},\ \bibinfo {pages} {335302} (\bibinfo {year} {2023})}\BibitemShut {NoStop}%
\bibitem [{\citenamefont {Song}\ \emph {et~al.}(2024)\citenamefont {Song}, \citenamefont {Zhao}, \citenamefont {Du}, \citenamefont {Li}, \citenamefont {Li}, \citenamefont {Feng}, \citenamefont {Yang}, \citenamefont {Wen}, \citenamefont {Huang}, \citenamefont {Peng}, \citenamefont {Sun}, \citenamefont {Jiang}, \citenamefont {He},\ and\ \citenamefont {Shi}}]{Song2024}%
  \BibitemOpen
  \bibfield  {author} {\bibinfo {author} {\bibfnamefont {L.}~\bibnamefont {Song}}, \bibinfo {author} {\bibfnamefont {Y.}~\bibnamefont {Zhao}}, \bibinfo {author} {\bibfnamefont {R.}~\bibnamefont {Du}}, \bibinfo {author} {\bibfnamefont {H.}~\bibnamefont {Li}}, \bibinfo {author} {\bibfnamefont {X.}~\bibnamefont {Li}}, \bibinfo {author} {\bibfnamefont {W.}~\bibnamefont {Feng}}, \bibinfo {author} {\bibfnamefont {J.}~\bibnamefont {Yang}}, \bibinfo {author} {\bibfnamefont {X.}~\bibnamefont {Wen}}, \bibinfo {author} {\bibfnamefont {L.}~\bibnamefont {Huang}}, \bibinfo {author} {\bibfnamefont {Y.}~\bibnamefont {Peng}}, \bibinfo {author} {\bibfnamefont {H.}~\bibnamefont {Sun}}, \bibinfo {author} {\bibfnamefont {Y.}~\bibnamefont {Jiang}}, \bibinfo {author} {\bibfnamefont {J.}~\bibnamefont {He}},\ and\ \bibinfo {author} {\bibfnamefont {J.}~\bibnamefont {Shi}},\ }\href {https://doi.org/10.1021/acs.nanolett.3c05105} {\bibfield  {journal} {\bibinfo  {journal} {Nano Lett.}\ }\textbf {\bibinfo {volume} {24}},\ \bibinfo {pages}
  {2408} (\bibinfo {year} {2024})}\BibitemShut {NoStop}%
\bibitem [{\citenamefont {Habib}\ \emph {et~al.}(2019{\natexlab{a}})\citenamefont {Habib}, \citenamefont {Wang}, \citenamefont {Obaidulla}, \citenamefont {Khan}, \citenamefont {Pi},\ and\ \citenamefont {Xu}}]{Habib2019}%
  \BibitemOpen
  \bibfield  {author} {\bibinfo {author} {\bibfnamefont {M.~R.}\ \bibnamefont {Habib}}, \bibinfo {author} {\bibfnamefont {S.}~\bibnamefont {Wang}}, \bibinfo {author} {\bibfnamefont {S.~M.}\ \bibnamefont {Obaidulla}}, \bibinfo {author} {\bibfnamefont {Y.}~\bibnamefont {Khan}}, \bibinfo {author} {\bibfnamefont {X.}~\bibnamefont {Pi}},\ and\ \bibinfo {author} {\bibfnamefont {M.}~\bibnamefont {Xu}},\ }\href {https://doi.org/https://doi.org/10.1002/pssb.201800597} {\bibfield  {journal} {\bibinfo  {journal} {Phys. Status Solidi B}\ }\textbf {\bibinfo {volume} {256}},\ \bibinfo {pages} {1800597} (\bibinfo {year} {2019}{\natexlab{a}})}\BibitemShut {NoStop}%
\bibitem [{\citenamefont {Rajendran~Nair}\ \emph {et~al.}(2022)\citenamefont {Rajendran~Nair}, \citenamefont {Abdelaziem}, \citenamefont {Zhao}, \citenamefont {Wang}, \citenamefont {Hu}, \citenamefont {Wu}, \citenamefont {Xun}, \citenamefont {Le~Goualher}, \citenamefont {Zhu}, \citenamefont {Yin}, \citenamefont {Valsaraj}, \citenamefont {Salim}, \citenamefont {Ke},\ and\ \citenamefont {Liu}}]{Nair2022}%
  \BibitemOpen
  \bibfield  {author} {\bibinfo {author} {\bibfnamefont {G.~K.}\ \bibnamefont {Rajendran~Nair}}, \bibinfo {author} {\bibfnamefont {A.}~\bibnamefont {Abdelaziem}}, \bibinfo {author} {\bibfnamefont {X.}~\bibnamefont {Zhao}}, \bibinfo {author} {\bibfnamefont {X.}~\bibnamefont {Wang}}, \bibinfo {author} {\bibfnamefont {D.}~\bibnamefont {Hu}}, \bibinfo {author} {\bibfnamefont {Y.}~\bibnamefont {Wu}}, \bibinfo {author} {\bibfnamefont {C.}~\bibnamefont {Xun}}, \bibinfo {author} {\bibfnamefont {F.}~\bibnamefont {Le~Goualher}}, \bibinfo {author} {\bibfnamefont {C.}~\bibnamefont {Zhu}}, \bibinfo {author} {\bibfnamefont {P.~L.~P.}\ \bibnamefont {Yin}}, \bibinfo {author} {\bibfnamefont {V.}~\bibnamefont {Valsaraj}}, \bibinfo {author} {\bibfnamefont {T.}~\bibnamefont {Salim}}, \bibinfo {author} {\bibfnamefont {L.}~\bibnamefont {Ke}},\ and\ \bibinfo {author} {\bibfnamefont {Z.}~\bibnamefont {Liu}},\ }\href {https://doi.org/https://doi.org/10.1002/pssr.202100495} {\bibfield  {journal} {\bibinfo  {journal} {Phys. Status
  Solidi RRL}\ }\textbf {\bibinfo {volume} {16}},\ \bibinfo {pages} {2100495} (\bibinfo {year} {2022})}\BibitemShut {NoStop}%
\bibitem [{\citenamefont {Xiao}\ \emph {et~al.}(2022{\natexlab{a}})\citenamefont {Xiao}, \citenamefont {Zhuang}, \citenamefont {Loh}, \citenamefont {Liang}, \citenamefont {Gayen}, \citenamefont {Ye}, \citenamefont {Bosman}, \citenamefont {Eda}, \citenamefont {Wang},\ and\ \citenamefont {Xu}}]{Xiao2022a}%
  \BibitemOpen
  \bibfield  {author} {\bibinfo {author} {\bibfnamefont {H.}~\bibnamefont {Xiao}}, \bibinfo {author} {\bibfnamefont {W.}~\bibnamefont {Zhuang}}, \bibinfo {author} {\bibfnamefont {L.}~\bibnamefont {Loh}}, \bibinfo {author} {\bibfnamefont {T.}~\bibnamefont {Liang}}, \bibinfo {author} {\bibfnamefont {A.}~\bibnamefont {Gayen}}, \bibinfo {author} {\bibfnamefont {P.}~\bibnamefont {Ye}}, \bibinfo {author} {\bibfnamefont {M.}~\bibnamefont {Bosman}}, \bibinfo {author} {\bibfnamefont {G.}~\bibnamefont {Eda}}, \bibinfo {author} {\bibfnamefont {X.}~\bibnamefont {Wang}},\ and\ \bibinfo {author} {\bibfnamefont {M.}~\bibnamefont {Xu}},\ }\href {https://doi.org/https://doi.org/10.1002/admi.202201353} {\bibfield  {journal} {\bibinfo  {journal} {Adv. Mater. Interfaces}\ }\textbf {\bibinfo {volume} {9}},\ \bibinfo {pages} {2201353} (\bibinfo {year} {2022}{\natexlab{a}})}\BibitemShut {NoStop}%
\bibitem [{\citenamefont {Yao}\ \emph {et~al.}(2024)\citenamefont {Yao}, \citenamefont {Wang}, \citenamefont {Li}, \citenamefont {Hong}, \citenamefont {He}, \citenamefont {Fu}, \citenamefont {Cai},\ and\ \citenamefont {Liu}}]{Yao2024}%
  \BibitemOpen
  \bibfield  {author} {\bibinfo {author} {\bibfnamefont {Y.}~\bibnamefont {Yao}}, \bibinfo {author} {\bibfnamefont {B.}~\bibnamefont {Wang}}, \bibinfo {author} {\bibfnamefont {Y.}~\bibnamefont {Li}}, \bibinfo {author} {\bibfnamefont {W.}~\bibnamefont {Hong}}, \bibinfo {author} {\bibfnamefont {X.}~\bibnamefont {He}}, \bibinfo {author} {\bibfnamefont {Z.}~\bibnamefont {Fu}}, \bibinfo {author} {\bibfnamefont {Q.}~\bibnamefont {Cai}},\ and\ \bibinfo {author} {\bibfnamefont {W.}~\bibnamefont {Liu}},\ }\href {https://doi.org/10.1039/D4TC00903G} {\bibfield  {journal} {\bibinfo  {journal} {J. Mater. Chem. C}\ }\textbf {\bibinfo {volume} {12}},\ \bibinfo {pages} {11513} (\bibinfo {year} {2024})}\BibitemShut {NoStop}%
\bibitem [{\citenamefont {Khan}\ \emph {et~al.}(2024)\citenamefont {Khan}, \citenamefont {Yadav}, \citenamefont {Wadhwa}, \citenamefont {Sunaina}, \citenamefont {Ankush},\ and\ \citenamefont {Jha}}]{Khan2024}%
  \BibitemOpen
  \bibfield  {author} {\bibinfo {author} {\bibfnamefont {N.}~\bibnamefont {Khan}}, \bibinfo {author} {\bibfnamefont {K.~K.}\ \bibnamefont {Yadav}}, \bibinfo {author} {\bibfnamefont {R.}~\bibnamefont {Wadhwa}}, \bibinfo {author} {\bibnamefont {Sunaina}}, \bibinfo {author} {\bibnamefont {Ankush}},\ and\ \bibinfo {author} {\bibfnamefont {M.}~\bibnamefont {Jha}},\ }\href {https://doi.org/https://doi.org/10.1016/j.ijhydene.2023.12.187} {\bibfield  {journal} {\bibinfo  {journal} {Int. J. Hydrogen Energy}\ }\textbf {\bibinfo {volume} {56}},\ \bibinfo {pages} {1294} (\bibinfo {year} {2024})}\BibitemShut {NoStop}%
\bibitem [{\citenamefont {Shivayogimath}\ \emph {et~al.}(2019)\citenamefont {Shivayogimath}, \citenamefont {Thomsen}, \citenamefont {Mackenzie}, \citenamefont {Geisler}, \citenamefont {Stan}, \citenamefont {Holt}, \citenamefont {Bianchi}, \citenamefont {Crovetto}, \citenamefont {Whelan}, \citenamefont {Carvalho}, \citenamefont {Neto}, \citenamefont {Hofmann}, \citenamefont {Stenger}, \citenamefont {B{\o}ggild},\ and\ \citenamefont {Booth}}]{Shivayogimath2019}%
  \BibitemOpen
  \bibfield  {author} {\bibinfo {author} {\bibfnamefont {A.}~\bibnamefont {Shivayogimath}}, \bibinfo {author} {\bibfnamefont {J.~D.}\ \bibnamefont {Thomsen}}, \bibinfo {author} {\bibfnamefont {D.~M.~A.}\ \bibnamefont {Mackenzie}}, \bibinfo {author} {\bibfnamefont {M.}~\bibnamefont {Geisler}}, \bibinfo {author} {\bibfnamefont {R.-M.}\ \bibnamefont {Stan}}, \bibinfo {author} {\bibfnamefont {A.~J.}\ \bibnamefont {Holt}}, \bibinfo {author} {\bibfnamefont {M.}~\bibnamefont {Bianchi}}, \bibinfo {author} {\bibfnamefont {A.}~\bibnamefont {Crovetto}}, \bibinfo {author} {\bibfnamefont {P.~R.}\ \bibnamefont {Whelan}}, \bibinfo {author} {\bibfnamefont {A.}~\bibnamefont {Carvalho}}, \bibinfo {author} {\bibfnamefont {A.~H.~C.}\ \bibnamefont {Neto}}, \bibinfo {author} {\bibfnamefont {P.}~\bibnamefont {Hofmann}}, \bibinfo {author} {\bibfnamefont {N.}~\bibnamefont {Stenger}}, \bibinfo {author} {\bibfnamefont {P.}~\bibnamefont {B{\o}ggild}},\ and\ \bibinfo {author} {\bibfnamefont {T.~J.}\ \bibnamefont {Booth}},\ }\href
  {https://doi.org/10.1038/s41467-019-11075-2} {\bibfield  {journal} {\bibinfo  {journal} {Nat. Commun.}\ }\textbf {\bibinfo {volume} {10}},\ \bibinfo {pages} {2957} (\bibinfo {year} {2019})}\BibitemShut {NoStop}%
\bibitem [{\citenamefont {Chen}\ \emph {et~al.}(2021)\citenamefont {Chen}, \citenamefont {Deng}, \citenamefont {Yan}, \citenamefont {Shi}, \citenamefont {Chang}, \citenamefont {Ding}, \citenamefont {Sun}, \citenamefont {Yang},\ and\ \citenamefont {Liu}}]{Chen2021}%
  \BibitemOpen
  \bibfield  {author} {\bibinfo {author} {\bibfnamefont {K.}~\bibnamefont {Chen}}, \bibinfo {author} {\bibfnamefont {J.}~\bibnamefont {Deng}}, \bibinfo {author} {\bibfnamefont {Y.}~\bibnamefont {Yan}}, \bibinfo {author} {\bibfnamefont {Q.}~\bibnamefont {Shi}}, \bibinfo {author} {\bibfnamefont {T.}~\bibnamefont {Chang}}, \bibinfo {author} {\bibfnamefont {X.}~\bibnamefont {Ding}}, \bibinfo {author} {\bibfnamefont {J.}~\bibnamefont {Sun}}, \bibinfo {author} {\bibfnamefont {S.}~\bibnamefont {Yang}},\ and\ \bibinfo {author} {\bibfnamefont {J.~Z.}\ \bibnamefont {Liu}},\ }\href {https://doi.org/10.1038/s41524-021-00547-z} {\bibfield  {journal} {\bibinfo  {journal} {Npj Comput. Mater.}\ }\textbf {\bibinfo {volume} {7}},\ \bibinfo {pages} {79} (\bibinfo {year} {2021})}\BibitemShut {NoStop}%
\bibitem [{\citenamefont {Habib}\ \emph {et~al.}(2019{\natexlab{b}})\citenamefont {Habib}, \citenamefont {Wang}, \citenamefont {Obaidulla}, \citenamefont {Khan}, \citenamefont {Pi},\ and\ \citenamefont {Xu}}]{Habibb2019}%
  \BibitemOpen
  \bibfield  {author} {\bibinfo {author} {\bibfnamefont {M.~R.}\ \bibnamefont {Habib}}, \bibinfo {author} {\bibfnamefont {S.}~\bibnamefont {Wang}}, \bibinfo {author} {\bibfnamefont {S.~M.}\ \bibnamefont {Obaidulla}}, \bibinfo {author} {\bibfnamefont {Y.}~\bibnamefont {Khan}}, \bibinfo {author} {\bibfnamefont {X.}~\bibnamefont {Pi}},\ and\ \bibinfo {author} {\bibfnamefont {M.}~\bibnamefont {Xu}},\ }\href {https://doi.org/https://doi.org/10.1002/pssb.201800597} {\bibfield  {journal} {\bibinfo  {journal} {Phys. Status Solidi B}\ }\textbf {\bibinfo {volume} {256}},\ \bibinfo {pages} {1800597} (\bibinfo {year} {2019}{\natexlab{b}})}\BibitemShut {NoStop}%
\bibitem [{\citenamefont {Lei}\ \emph {et~al.}(2024)\citenamefont {Lei}, \citenamefont {Li}, \citenamefont {Zhou}, \citenamefont {Wang}, \citenamefont {Xiong}, \citenamefont {Chen},\ and\ \citenamefont {Ouyang}}]{Lei2024}%
  \BibitemOpen
  \bibfield  {author} {\bibinfo {author} {\bibfnamefont {B.}~\bibnamefont {Lei}}, \bibinfo {author} {\bibfnamefont {A.}~\bibnamefont {Li}}, \bibinfo {author} {\bibfnamefont {W.}~\bibnamefont {Zhou}}, \bibinfo {author} {\bibfnamefont {Y.}~\bibnamefont {Wang}}, \bibinfo {author} {\bibfnamefont {W.}~\bibnamefont {Xiong}}, \bibinfo {author} {\bibfnamefont {Y.}~\bibnamefont {Chen}},\ and\ \bibinfo {author} {\bibfnamefont {F.}~\bibnamefont {Ouyang}},\ }\href {https://doi.org/10.1007/s11467-023-1387-y} {\bibfield  {journal} {\bibinfo  {journal} {Front. Phys.}\ }\textbf {\bibinfo {volume} {19}},\ \bibinfo {pages} {43200} (\bibinfo {year} {2024})}\BibitemShut {NoStop}%
\bibitem [{\citenamefont {Liu}\ \emph {et~al.}(2024{\natexlab{b}})\citenamefont {Liu}, \citenamefont {Yang}, \citenamefont {Wei}, \citenamefont {Sun},\ and\ \citenamefont {Zhao}}]{Liu2024b}%
  \BibitemOpen
  \bibfield  {author} {\bibinfo {author} {\bibfnamefont {H.}~\bibnamefont {Liu}}, \bibinfo {author} {\bibfnamefont {L.}~\bibnamefont {Yang}}, \bibinfo {author} {\bibfnamefont {X.}~\bibnamefont {Wei}}, \bibinfo {author} {\bibfnamefont {S.}~\bibnamefont {Sun}},\ and\ \bibinfo {author} {\bibfnamefont {Y.}~\bibnamefont {Zhao}},\ }\href {https://doi.org/10.1007/s11224-023-02241-w} {\bibfield  {journal} {\bibinfo  {journal} {Struct. Chem.}\ }\textbf {\bibinfo {volume} {35}},\ \bibinfo {pages} {923} (\bibinfo {year} {2024}{\natexlab{b}})}\BibitemShut {NoStop}%
\bibitem [{\citenamefont {Sun}\ \emph {et~al.}(2020)\citenamefont {Sun}, \citenamefont {Hong}, \citenamefont {Zhou}, \citenamefont {Yuan},\ and\ \citenamefont {Zhang}}]{Sun2020}%
  \BibitemOpen
  \bibfield  {author} {\bibinfo {author} {\bibfnamefont {F.}~\bibnamefont {Sun}}, \bibinfo {author} {\bibfnamefont {A.}~\bibnamefont {Hong}}, \bibinfo {author} {\bibfnamefont {W.}~\bibnamefont {Zhou}}, \bibinfo {author} {\bibfnamefont {C.}~\bibnamefont {Yuan}},\ and\ \bibinfo {author} {\bibfnamefont {W.}~\bibnamefont {Zhang}},\ }\href {https://doi.org/https://doi.org/10.1016/j.mtcomm.2020.101707} {\bibfield  {journal} {\bibinfo  {journal} {Mater. Today Commun.}\ }\textbf {\bibinfo {volume} {25}},\ \bibinfo {pages} {101707} (\bibinfo {year} {2020})}\BibitemShut {NoStop}%
\bibitem [{\citenamefont {Wang}\ \emph {et~al.}(2018)\citenamefont {Wang}, \citenamefont {Zhou}, \citenamefont {Pan}, \citenamefont {Qiao}, \citenamefont {Kong}, \citenamefont {Kaun},\ and\ \citenamefont {Ji}}]{Wang2018}%
  \BibitemOpen
  \bibfield  {author} {\bibinfo {author} {\bibfnamefont {C.}~\bibnamefont {Wang}}, \bibinfo {author} {\bibfnamefont {X.}~\bibnamefont {Zhou}}, \bibinfo {author} {\bibfnamefont {Y.}~\bibnamefont {Pan}}, \bibinfo {author} {\bibfnamefont {J.}~\bibnamefont {Qiao}}, \bibinfo {author} {\bibfnamefont {X.}~\bibnamefont {Kong}}, \bibinfo {author} {\bibfnamefont {C.-C.}\ \bibnamefont {Kaun}},\ and\ \bibinfo {author} {\bibfnamefont {W.}~\bibnamefont {Ji}},\ }\href {https://doi.org/10.1103/PhysRevB.97.245409} {\bibfield  {journal} {\bibinfo  {journal} {Phys. Rev. B}\ }\textbf {\bibinfo {volume} {97}},\ \bibinfo {pages} {245409} (\bibinfo {year} {2018})}\BibitemShut {NoStop}%
\bibitem [{\citenamefont {Zhuang}\ \emph {et~al.}(2014)\citenamefont {Zhuang}, \citenamefont {Johannes}, \citenamefont {Blonsky},\ and\ \citenamefont {Hennig}}]{Zhuang2014}%
  \BibitemOpen
  \bibfield  {author} {\bibinfo {author} {\bibfnamefont {H.~L.}\ \bibnamefont {Zhuang}}, \bibinfo {author} {\bibfnamefont {M.~D.}\ \bibnamefont {Johannes}}, \bibinfo {author} {\bibfnamefont {M.~N.}\ \bibnamefont {Blonsky}},\ and\ \bibinfo {author} {\bibfnamefont {R.~G.}\ \bibnamefont {Hennig}},\ }\href {https://doi.org/10.1063/1.4861659} {\bibfield  {journal} {\bibinfo  {journal} {Appl. Phys. Lett.}\ }\textbf {\bibinfo {volume} {104}},\ \bibinfo {pages} {022116} (\bibinfo {year} {2014})}\BibitemShut {NoStop}%
\bibitem [{\citenamefont {Gao}\ and\ \citenamefont {Yao}(2014)}]{Gao2014}%
  \BibitemOpen
  \bibfield  {author} {\bibinfo {author} {\bibfnamefont {G.~Y.}\ \bibnamefont {Gao}}\ and\ \bibinfo {author} {\bibfnamefont {K.~L.}\ \bibnamefont {Yao}},\ }\href {https://doi.org/10.1063/1.4901173} {\bibfield  {journal} {\bibinfo  {journal} {Appl. Phys. Lett.}\ }\textbf {\bibinfo {volume} {105}},\ \bibinfo {pages} {182405} (\bibinfo {year} {2014})}\BibitemShut {NoStop}%
\bibitem [{\citenamefont {Li}\ \emph {et~al.}(2023)\citenamefont {Li}, \citenamefont {Wu}, \citenamefont {Deng}, \citenamefont {Yin}, \citenamefont {Han}, \citenamefont {Tian},\ and\ \citenamefont {Zhang}}]{Li2023}%
  \BibitemOpen
  \bibfield  {author} {\bibinfo {author} {\bibfnamefont {Y.}~\bibnamefont {Li}}, \bibinfo {author} {\bibfnamefont {Y.}~\bibnamefont {Wu}}, \bibinfo {author} {\bibfnamefont {L.}~\bibnamefont {Deng}}, \bibinfo {author} {\bibfnamefont {X.}~\bibnamefont {Yin}}, \bibinfo {author} {\bibfnamefont {X.}~\bibnamefont {Han}}, \bibinfo {author} {\bibfnamefont {F.}~\bibnamefont {Tian}},\ and\ \bibinfo {author} {\bibfnamefont {X.}~\bibnamefont {Zhang}},\ }\href {https://doi.org/10.1063/5.0145789} {\bibfield  {journal} {\bibinfo  {journal} {J. Appl. Phys.}\ }\textbf {\bibinfo {volume} {133}},\ \bibinfo {pages} {134301} (\bibinfo {year} {2023})}\BibitemShut {NoStop}%
\bibitem [{\citenamefont {Zhang}\ \emph {et~al.}(2021)\citenamefont {Zhang}, \citenamefont {Ding}, \citenamefont {Chen}, \citenamefont {Guo}, \citenamefont {Pan}, \citenamefont {Li}, \citenamefont {Zhao}, \citenamefont {Liu},\ and\ \citenamefont {Xie}}]{Zhang2021}%
  \BibitemOpen
  \bibfield  {author} {\bibinfo {author} {\bibfnamefont {Y.}~\bibnamefont {Zhang}}, \bibinfo {author} {\bibfnamefont {W.}~\bibnamefont {Ding}}, \bibinfo {author} {\bibfnamefont {Z.}~\bibnamefont {Chen}}, \bibinfo {author} {\bibfnamefont {J.}~\bibnamefont {Guo}}, \bibinfo {author} {\bibfnamefont {H.}~\bibnamefont {Pan}}, \bibinfo {author} {\bibfnamefont {X.}~\bibnamefont {Li}}, \bibinfo {author} {\bibfnamefont {Z.}~\bibnamefont {Zhao}}, \bibinfo {author} {\bibfnamefont {Y.}~\bibnamefont {Liu}},\ and\ \bibinfo {author} {\bibfnamefont {W.}~\bibnamefont {Xie}},\ }\href {https://doi.org/10.1021/acs.jpcc.0c11449} {\bibfield  {journal} {\bibinfo  {journal} {J. Phys. Chem. C}\ }\textbf {\bibinfo {volume} {125}},\ \bibinfo {pages} {8398} (\bibinfo {year} {2021})}\BibitemShut {NoStop}%
\bibitem [{\citenamefont {Zhang}\ \emph {et~al.}(2019)\citenamefont {Zhang}, \citenamefont {Wang}, \citenamefont {Guo}, \citenamefont {Zhang}, \citenamefont {Chen},\ and\ \citenamefont {Wang}}]{Zhang2019}%
  \BibitemOpen
  \bibfield  {author} {\bibinfo {author} {\bibfnamefont {X.}~\bibnamefont {Zhang}}, \bibinfo {author} {\bibfnamefont {B.}~\bibnamefont {Wang}}, \bibinfo {author} {\bibfnamefont {Y.}~\bibnamefont {Guo}}, \bibinfo {author} {\bibfnamefont {Y.}~\bibnamefont {Zhang}}, \bibinfo {author} {\bibfnamefont {Y.}~\bibnamefont {Chen}},\ and\ \bibinfo {author} {\bibfnamefont {J.}~\bibnamefont {Wang}},\ }\href {https://doi.org/10.1039/C9NH00038K} {\bibfield  {journal} {\bibinfo  {journal} {Nanoscale Horiz.}\ }\textbf {\bibinfo {volume} {4}},\ \bibinfo {pages} {859} (\bibinfo {year} {2019})}\BibitemShut {NoStop}%
\bibitem [{\citenamefont {Massalski}\ \emph {et~al.}(2007)\citenamefont {Massalski}, \citenamefont {{Editor-in-Chief}.~Okamoto}, \citenamefont {Subramanian}, \citenamefont {Kacprzak},\ and\ \citenamefont {Editors.}}]{Massalski2007}%
  \BibitemOpen
  \bibfield  {author} {\bibinfo {author} {\bibfnamefont {T.~B.}\ \bibnamefont {Massalski}}, \bibinfo {author} {\bibfnamefont {H.}~\bibnamefont {{Editor-in-Chief}.~Okamoto}}, \bibinfo {author} {\bibfnamefont {P.~R.}\ \bibnamefont {Subramanian}}, \bibinfo {author} {\bibfnamefont {L.}~\bibnamefont {Kacprzak}},\ and\ \bibinfo {author} {\bibnamefont {Editors.}},\ }\href@noop {} {\emph {\bibinfo {title} {Binary Alloy Phase Diagrams–Second Edition}}}\ (\bibinfo  {publisher} {ASM International},\ \bibinfo {address} {Materials Park, Ohio, USA},\ \bibinfo {year} {2007})\BibitemShut {NoStop}%
\bibitem [{\citenamefont {Jellinek}(1957)}]{Jellinek1957}%
  \BibitemOpen
  \bibfield  {author} {\bibinfo {author} {\bibfnamefont {F.}~\bibnamefont {Jellinek}},\ }\href@noop {} {\bibfield  {journal} {\bibinfo  {journal} {Acta Crystallogr.}\ }\textbf {\bibinfo {volume} {10}},\ \bibinfo {pages} {620} (\bibinfo {year} {1957})}\BibitemShut {NoStop}%
\bibitem [{\citenamefont {Popma}\ and\ \citenamefont {{Van Bruggen}}(1969)}]{Popma1969}%
  \BibitemOpen
  \bibfield  {author} {\bibinfo {author} {\bibfnamefont {T.}~\bibnamefont {Popma}}\ and\ \bibinfo {author} {\bibfnamefont {C.}~\bibnamefont {{Van Bruggen}}},\ }\href {https://doi.org/https://doi.org/10.1016/0022-1902(69)80055-2} {\bibfield  {journal} {\bibinfo  {journal} {J. Inorg. Nucl. Chem.}\ }\textbf {\bibinfo {volume} {31}},\ \bibinfo {pages} {73} (\bibinfo {year} {1969})}\BibitemShut {NoStop}%
\bibitem [{\citenamefont {Rau}(1977)}]{Rau1977}%
  \BibitemOpen
  \bibfield  {author} {\bibinfo {author} {\bibfnamefont {H.}~\bibnamefont {Rau}},\ }\href {https://doi.org/https://doi.org/10.1016/0022-5088(77)90194-1} {\bibfield  {journal} {\bibinfo  {journal} {J. Less-Common Met.}\ }\textbf {\bibinfo {volume} {55}},\ \bibinfo {pages} {205} (\bibinfo {year} {1977})}\BibitemShut {NoStop}%
\bibitem [{\citenamefont {Yuzuri}\ and\ \citenamefont {Nakamura}(1964)}]{Yuzuri1964}%
  \BibitemOpen
  \bibfield  {author} {\bibinfo {author} {\bibfnamefont {M.}~\bibnamefont {Yuzuri}}\ and\ \bibinfo {author} {\bibfnamefont {Y.}~\bibnamefont {Nakamura}},\ }\href {https://doi.org/10.1143/JPSJ.19.1350} {\bibfield  {journal} {\bibinfo  {journal} {J. Phys. Soc. Jpn.}\ }\textbf {\bibinfo {volume} {19}},\ \bibinfo {pages} {1350} (\bibinfo {year} {1964})}\BibitemShut {NoStop}%
\bibitem [{\citenamefont {Mikami}\ \emph {et~al.}(1972)\citenamefont {Mikami}, \citenamefont {Igaki},\ and\ \citenamefont {Ōhashi}}]{Mikami1972}%
  \BibitemOpen
  \bibfield  {author} {\bibinfo {author} {\bibfnamefont {M.}~\bibnamefont {Mikami}}, \bibinfo {author} {\bibfnamefont {K.}~\bibnamefont {Igaki}},\ and\ \bibinfo {author} {\bibfnamefont {N.}~\bibnamefont {Ōhashi}},\ }\href {https://doi.org/10.1143/JPSJ.32.1217} {\bibfield  {journal} {\bibinfo  {journal} {J. Phys. Soc. Jpn.}\ }\textbf {\bibinfo {volume} {32}},\ \bibinfo {pages} {1217} (\bibinfo {year} {1972})}\BibitemShut {NoStop}%
\bibitem [{\citenamefont {{Van Laar}}(1967)}]{vanLaar1967}%
  \BibitemOpen
  \bibfield  {author} {\bibinfo {author} {\bibfnamefont {B.}~\bibnamefont {{Van Laar}}},\ }\href {https://doi.org/https://doi.org/10.1016/0375-9601(67)90320-9} {\bibfield  {journal} {\bibinfo  {journal} {Phys. Lett. A}\ }\textbf {\bibinfo {volume} {25}},\ \bibinfo {pages} {27} (\bibinfo {year} {1967})}\BibitemShut {NoStop}%
\bibitem [{\citenamefont {Popma}\ \emph {et~al.}(1971)\citenamefont {Popma}, \citenamefont {Haas},\ and\ \citenamefont {{Van Laar}}}]{Popma1971}%
  \BibitemOpen
  \bibfield  {author} {\bibinfo {author} {\bibfnamefont {T.}~\bibnamefont {Popma}}, \bibinfo {author} {\bibfnamefont {C.}~\bibnamefont {Haas}},\ and\ \bibinfo {author} {\bibfnamefont {B.}~\bibnamefont {{Van Laar}}},\ }\href {https://doi.org/https://doi.org/10.1016/0022-3697(71)90007-2} {\bibfield  {journal} {\bibinfo  {journal} {J. Phys. Chem. Sol.}\ }\textbf {\bibinfo {volume} {32}},\ \bibinfo {pages} {581} (\bibinfo {year} {1971})}\BibitemShut {NoStop}%
\bibitem [{\citenamefont {Maignan}\ \emph {et~al.}(2012)\citenamefont {Maignan}, \citenamefont {Bréard}, \citenamefont {Guilmeau},\ and\ \citenamefont {Gascoin}}]{Maignan1912}%
  \BibitemOpen
  \bibfield  {author} {\bibinfo {author} {\bibfnamefont {A.}~\bibnamefont {Maignan}}, \bibinfo {author} {\bibfnamefont {Y.}~\bibnamefont {Bréard}}, \bibinfo {author} {\bibfnamefont {E.}~\bibnamefont {Guilmeau}},\ and\ \bibinfo {author} {\bibfnamefont {F.}~\bibnamefont {Gascoin}},\ }\href {https://doi.org/10.1063/1.4736417} {\bibfield  {journal} {\bibinfo  {journal} {J. Appl. Phys.}\ }\textbf {\bibinfo {volume} {112}},\ \bibinfo {pages} {013716} (\bibinfo {year} {2012})}\BibitemShut {NoStop}%
\bibitem [{\citenamefont {Anedda}\ \emph {et~al.}(1982)\citenamefont {Anedda}, \citenamefont {Fortin}, \citenamefont {Ledda},\ and\ \citenamefont {Serpi}}]{Anedda1982}%
  \BibitemOpen
  \bibfield  {author} {\bibinfo {author} {\bibfnamefont {A.}~\bibnamefont {Anedda}}, \bibinfo {author} {\bibfnamefont {E.}~\bibnamefont {Fortin}}, \bibinfo {author} {\bibfnamefont {F.}~\bibnamefont {Ledda}},\ and\ \bibinfo {author} {\bibfnamefont {A.}~\bibnamefont {Serpi}},\ }\href {https://doi.org/https://doi.org/10.1002/pssb.2221140267} {\bibfield  {journal} {\bibinfo  {journal} {Phys. Status Solidi B}\ }\textbf {\bibinfo {volume} {114}},\ \bibinfo {pages} {K143} (\bibinfo {year} {1982})}\BibitemShut {NoStop}%
\bibitem [{\citenamefont {Fan}\ \emph {et~al.}(2024)\citenamefont {Fan}, \citenamefont {Chen}, \citenamefont {Xu}, \citenamefont {Zou}, \citenamefont {Ouyang}, \citenamefont {Deng}, \citenamefont {Wang}, \citenamefont {Zhao},\ and\ \citenamefont {Zhou}}]{Fan2024}%
  \BibitemOpen
  \bibfield  {author} {\bibinfo {author} {\bibfnamefont {X.}~\bibnamefont {Fan}}, \bibinfo {author} {\bibfnamefont {Z.}~\bibnamefont {Chen}}, \bibinfo {author} {\bibfnamefont {D.}~\bibnamefont {Xu}}, \bibinfo {author} {\bibfnamefont {L.}~\bibnamefont {Zou}}, \bibinfo {author} {\bibfnamefont {F.}~\bibnamefont {Ouyang}}, \bibinfo {author} {\bibfnamefont {S.}~\bibnamefont {Deng}}, \bibinfo {author} {\bibfnamefont {X.}~\bibnamefont {Wang}}, \bibinfo {author} {\bibfnamefont {J.}~\bibnamefont {Zhao}},\ and\ \bibinfo {author} {\bibfnamefont {Y.}~\bibnamefont {Zhou}},\ }\href {https://doi.org/https://doi.org/10.1002/adfm.202404750} {\bibfield  {journal} {\bibinfo  {journal} {Adv. Funct. Mater.}\ }\textbf {\bibinfo {volume} {34}},\ \bibinfo {pages} {2404750} (\bibinfo {year} {2024})}\BibitemShut {NoStop}%
\bibitem [{\citenamefont {Kato}(1990)}]{Kato1990}%
  \BibitemOpen
  \bibfield  {author} {\bibinfo {author} {\bibfnamefont {K.}~\bibnamefont {Kato}},\ }\href {https://doi.org/10.1107/S0108768189008931} {\bibfield  {journal} {\bibinfo  {journal} {Acta Crystallogr. B}\ }\textbf {\bibinfo {volume} {46}},\ \bibinfo {pages} {39} (\bibinfo {year} {1990})}\BibitemShut {NoStop}%
\bibitem [{\citenamefont {Lafond}\ \emph {et~al.}(1992)\citenamefont {Lafond}, \citenamefont {Fragnaud}, \citenamefont {Evain},\ and\ \citenamefont {Meerschaut}}]{Lafond1992}%
  \BibitemOpen
  \bibfield  {author} {\bibinfo {author} {\bibfnamefont {A.}~\bibnamefont {Lafond}}, \bibinfo {author} {\bibfnamefont {P.}~\bibnamefont {Fragnaud}}, \bibinfo {author} {\bibfnamefont {M.}~\bibnamefont {Evain}},\ and\ \bibinfo {author} {\bibfnamefont {A.}~\bibnamefont {Meerschaut}},\ }\href {https://doi.org/https://doi.org/10.1016/0025-5408(92)90078-E} {\bibfield  {journal} {\bibinfo  {journal} {Mater. Res. Bull.}\ }\textbf {\bibinfo {volume} {27}},\ \bibinfo {pages} {705} (\bibinfo {year} {1992})}\BibitemShut {NoStop}%
\bibitem [{\citenamefont {Lafond}\ \emph {et~al.}(1994)\citenamefont {Lafond}, \citenamefont {Deudon}, \citenamefont {Meerschaut},\ and\ \citenamefont {Sulpice}}]{Lafond1994}%
  \BibitemOpen
  \bibfield  {author} {\bibinfo {author} {\bibfnamefont {A.}~\bibnamefont {Lafond}}, \bibinfo {author} {\bibfnamefont {C.}~\bibnamefont {Deudon}}, \bibinfo {author} {\bibfnamefont {A.}~\bibnamefont {Meerschaut}},\ and\ \bibinfo {author} {\bibfnamefont {A.}~\bibnamefont {Sulpice}},\ }\href {https://pascal-francis.inist.fr/vibad/index.php?action=getRecordDetail&idt=3369201} {\bibfield  {journal} {\bibinfo  {journal} {Eur. J. Solid State Inorg. Chem.}\ }\textbf {\bibinfo {volume} {31}},\ \bibinfo {pages} {967} (\bibinfo {year} {1994})}\BibitemShut {NoStop}%
\bibitem [{\citenamefont {Xiao}\ \emph {et~al.}(2022{\natexlab{b}})\citenamefont {Xiao}, \citenamefont {Xiao}, \citenamefont {Chen},\ and\ \citenamefont {Wang}}]{Xiao2022b}%
  \BibitemOpen
  \bibfield  {author} {\bibinfo {author} {\bibfnamefont {G.}~\bibnamefont {Xiao}}, \bibinfo {author} {\bibfnamefont {W.-Z.}\ \bibnamefont {Xiao}}, \bibinfo {author} {\bibfnamefont {Q.}~\bibnamefont {Chen}},\ and\ \bibinfo {author} {\bibfnamefont {L.-l.}\ \bibnamefont {Wang}},\ }\href {https://doi.org/10.1039/D2TC03711D} {\bibfield  {journal} {\bibinfo  {journal} {J. Mater. Chem. C}\ }\textbf {\bibinfo {volume} {10}},\ \bibinfo {pages} {17665} (\bibinfo {year} {2022}{\natexlab{b}})}\BibitemShut {NoStop}%
\bibitem [{\citenamefont {Lu}\ \emph {et~al.}(2023)\citenamefont {Lu}, \citenamefont {Yang}, \citenamefont {Sun}, \citenamefont {Robertson},\ and\ \citenamefont {Zhao}}]{Lu2023}%
  \BibitemOpen
  \bibfield  {author} {\bibinfo {author} {\bibfnamefont {H.}~\bibnamefont {Lu}}, \bibinfo {author} {\bibfnamefont {T.}~\bibnamefont {Yang}}, \bibinfo {author} {\bibfnamefont {Z.}~\bibnamefont {Sun}}, \bibinfo {author} {\bibfnamefont {J.}~\bibnamefont {Robertson}},\ and\ \bibinfo {author} {\bibfnamefont {W.}~\bibnamefont {Zhao}},\ }\href {https://arxiv.org/abs/2312.04497} {} (\bibinfo {year} {2023}),\ \Eprint {https://arxiv.org/abs/2312.04497} {arXiv:2312.04497 [cond-mat.mtrl-sci]} \BibitemShut {NoStop}%
\bibitem [{\citenamefont {Lee}\ \emph {et~al.}(2021)\citenamefont {Lee}, \citenamefont {Dismukes}, \citenamefont {Telford}, \citenamefont {Wiscons}, \citenamefont {Wang}, \citenamefont {Xu}, \citenamefont {Nuckolls}, \citenamefont {Dean}, \citenamefont {Roy},\ and\ \citenamefont {Zhu}}]{Lee2021}%
  \BibitemOpen
  \bibfield  {author} {\bibinfo {author} {\bibfnamefont {K.}~\bibnamefont {Lee}}, \bibinfo {author} {\bibfnamefont {A.~H.}\ \bibnamefont {Dismukes}}, \bibinfo {author} {\bibfnamefont {E.~J.}\ \bibnamefont {Telford}}, \bibinfo {author} {\bibfnamefont {R.~A.}\ \bibnamefont {Wiscons}}, \bibinfo {author} {\bibfnamefont {J.}~\bibnamefont {Wang}}, \bibinfo {author} {\bibfnamefont {X.}~\bibnamefont {Xu}}, \bibinfo {author} {\bibfnamefont {C.}~\bibnamefont {Nuckolls}}, \bibinfo {author} {\bibfnamefont {C.~R.}\ \bibnamefont {Dean}}, \bibinfo {author} {\bibfnamefont {X.}~\bibnamefont {Roy}},\ and\ \bibinfo {author} {\bibfnamefont {X.}~\bibnamefont {Zhu}},\ }\href {https://doi.org/10.1021/acs.nanolett.1c00219} {\bibfield  {journal} {\bibinfo  {journal} {Nano Lett.}\ }\textbf {\bibinfo {volume} {21}},\ \bibinfo {pages} {3511} (\bibinfo {year} {2021})}\BibitemShut {NoStop}%
\bibitem [{\citenamefont {Wang}\ \emph {et~al.}(2022)\citenamefont {Wang}, \citenamefont {Bedoya-Pinto}, \citenamefont {Blei}, \citenamefont {Dismukes}, \citenamefont {Hamo}, \citenamefont {Jenkins}, \citenamefont {Koperski}, \citenamefont {Liu}, \citenamefont {Sun}, \citenamefont {Telford}, \citenamefont {Kim}, \citenamefont {Augustin}, \citenamefont {Vool}, \citenamefont {Yin}, \citenamefont {Li}, \citenamefont {Falin}, \citenamefont {Dean}, \citenamefont {Casanova}, \citenamefont {Evans}, \citenamefont {Chshiev}, \citenamefont {Mishchenko}, \citenamefont {Petrovic}, \citenamefont {He}, \citenamefont {Zhao}, \citenamefont {Tsen}, \citenamefont {Gerardot}, \citenamefont {Brotons-Gisbert}, \citenamefont {Guguchia}, \citenamefont {Roy}, \citenamefont {Tongay}, \citenamefont {Wang}, \citenamefont {Hasan}, \citenamefont {Wrachtrup}, \citenamefont {Yacoby}, \citenamefont {Fert}, \citenamefont {Parkin}, \citenamefont {Novoselov}, \citenamefont {Dai}, \citenamefont {Balicas},\ and\ \citenamefont
  {Santos}}]{Wang2022}%
  \BibitemOpen
  \bibfield  {author} {\bibinfo {author} {\bibfnamefont {Q.~H.}\ \bibnamefont {Wang}}, \bibinfo {author} {\bibfnamefont {A.}~\bibnamefont {Bedoya-Pinto}}, \bibinfo {author} {\bibfnamefont {M.}~\bibnamefont {Blei}}, \bibinfo {author} {\bibfnamefont {A.~H.}\ \bibnamefont {Dismukes}}, \bibinfo {author} {\bibfnamefont {A.}~\bibnamefont {Hamo}}, \bibinfo {author} {\bibfnamefont {S.}~\bibnamefont {Jenkins}}, \bibinfo {author} {\bibfnamefont {M.}~\bibnamefont {Koperski}}, \bibinfo {author} {\bibfnamefont {Y.}~\bibnamefont {Liu}}, \bibinfo {author} {\bibfnamefont {Q.-C.}\ \bibnamefont {Sun}}, \bibinfo {author} {\bibfnamefont {E.~J.}\ \bibnamefont {Telford}}, \bibinfo {author} {\bibfnamefont {H.~H.}\ \bibnamefont {Kim}}, \bibinfo {author} {\bibfnamefont {M.}~\bibnamefont {Augustin}}, \bibinfo {author} {\bibfnamefont {U.}~\bibnamefont {Vool}}, \bibinfo {author} {\bibfnamefont {J.-X.}\ \bibnamefont {Yin}}, \bibinfo {author} {\bibfnamefont {L.~H.}\ \bibnamefont {Li}}, \bibinfo {author} {\bibfnamefont {A.}~\bibnamefont
  {Falin}}, \bibinfo {author} {\bibfnamefont {C.~R.}\ \bibnamefont {Dean}}, \bibinfo {author} {\bibfnamefont {F.}~\bibnamefont {Casanova}}, \bibinfo {author} {\bibfnamefont {R.~F.~L.}\ \bibnamefont {Evans}}, \bibinfo {author} {\bibfnamefont {M.}~\bibnamefont {Chshiev}}, \bibinfo {author} {\bibfnamefont {A.}~\bibnamefont {Mishchenko}}, \bibinfo {author} {\bibfnamefont {C.}~\bibnamefont {Petrovic}}, \bibinfo {author} {\bibfnamefont {R.}~\bibnamefont {He}}, \bibinfo {author} {\bibfnamefont {L.}~\bibnamefont {Zhao}}, \bibinfo {author} {\bibfnamefont {A.~W.}\ \bibnamefont {Tsen}}, \bibinfo {author} {\bibfnamefont {B.~D.}\ \bibnamefont {Gerardot}}, \bibinfo {author} {\bibfnamefont {M.}~\bibnamefont {Brotons-Gisbert}}, \bibinfo {author} {\bibfnamefont {Z.}~\bibnamefont {Guguchia}}, \bibinfo {author} {\bibfnamefont {X.}~\bibnamefont {Roy}}, \bibinfo {author} {\bibfnamefont {S.}~\bibnamefont {Tongay}}, \bibinfo {author} {\bibfnamefont {Z.}~\bibnamefont {Wang}}, \bibinfo {author} {\bibfnamefont {M.~Z.}\ \bibnamefont
  {Hasan}}, \bibinfo {author} {\bibfnamefont {J.}~\bibnamefont {Wrachtrup}}, \bibinfo {author} {\bibfnamefont {A.}~\bibnamefont {Yacoby}}, \bibinfo {author} {\bibfnamefont {A.}~\bibnamefont {Fert}}, \bibinfo {author} {\bibfnamefont {S.}~\bibnamefont {Parkin}}, \bibinfo {author} {\bibfnamefont {K.~S.}\ \bibnamefont {Novoselov}}, \bibinfo {author} {\bibfnamefont {P.}~\bibnamefont {Dai}}, \bibinfo {author} {\bibfnamefont {L.}~\bibnamefont {Balicas}},\ and\ \bibinfo {author} {\bibfnamefont {E.~J.~G.}\ \bibnamefont {Santos}},\ }\href {https://doi.org/10.1021/acsnano.1c09150} {\bibfield  {journal} {\bibinfo  {journal} {ACS Nano}\ }\textbf {\bibinfo {volume} {16}},\ \bibinfo {pages} {6960} (\bibinfo {year} {2022})}\BibitemShut {NoStop}%
\bibitem [{\citenamefont {Hall}\ \emph {et~al.}(2018)\citenamefont {Hall}, \citenamefont {Pieli{\'c}}, \citenamefont {Murray}, \citenamefont {Jolie}, \citenamefont {Wekking}, \citenamefont {Busse}, \citenamefont {Kralj},\ and\ \citenamefont {Michely}}]{Hall2018}%
  \BibitemOpen
  \bibfield  {author} {\bibinfo {author} {\bibfnamefont {J.}~\bibnamefont {Hall}}, \bibinfo {author} {\bibfnamefont {B.}~\bibnamefont {Pieli{\'c}}}, \bibinfo {author} {\bibfnamefont {C.}~\bibnamefont {Murray}}, \bibinfo {author} {\bibfnamefont {W.}~\bibnamefont {Jolie}}, \bibinfo {author} {\bibfnamefont {T.}~\bibnamefont {Wekking}}, \bibinfo {author} {\bibfnamefont {C.}~\bibnamefont {Busse}}, \bibinfo {author} {\bibfnamefont {M.}~\bibnamefont {Kralj}},\ and\ \bibinfo {author} {\bibfnamefont {T.}~\bibnamefont {Michely}},\ }\href@noop {} {\bibfield  {journal} {\bibinfo  {journal} {2D Mater.}\ }\textbf {\bibinfo {volume} {5}},\ \bibinfo {pages} {025005} (\bibinfo {year} {2018})}\BibitemShut {NoStop}%
\bibitem [{\citenamefont {Busse}\ \emph {et~al.}(2011)\citenamefont {Busse}, \citenamefont {Lazi\ifmmode~\acute{c}\else \'{c}\fi{}}, \citenamefont {Djemour}, \citenamefont {Coraux}, \citenamefont {Gerber}, \citenamefont {Atodiresei}, \citenamefont {Caciuc}, \citenamefont {Brako}, \citenamefont {N'Diaye}, \citenamefont {Bl\"ugel}, \citenamefont {Zegenhagen},\ and\ \citenamefont {Michely}}]{Busse2011}%
  \BibitemOpen
  \bibfield  {author} {\bibinfo {author} {\bibfnamefont {C.}~\bibnamefont {Busse}}, \bibinfo {author} {\bibfnamefont {P.}~\bibnamefont {Lazi\ifmmode~\acute{c}\else \'{c}\fi{}}}, \bibinfo {author} {\bibfnamefont {R.}~\bibnamefont {Djemour}}, \bibinfo {author} {\bibfnamefont {J.}~\bibnamefont {Coraux}}, \bibinfo {author} {\bibfnamefont {T.}~\bibnamefont {Gerber}}, \bibinfo {author} {\bibfnamefont {N.}~\bibnamefont {Atodiresei}}, \bibinfo {author} {\bibfnamefont {V.}~\bibnamefont {Caciuc}}, \bibinfo {author} {\bibfnamefont {R.}~\bibnamefont {Brako}}, \bibinfo {author} {\bibfnamefont {A.~T.}\ \bibnamefont {N'Diaye}}, \bibinfo {author} {\bibfnamefont {S.}~\bibnamefont {Bl\"ugel}}, \bibinfo {author} {\bibfnamefont {J.}~\bibnamefont {Zegenhagen}},\ and\ \bibinfo {author} {\bibfnamefont {T.}~\bibnamefont {Michely}},\ }\href {https://doi.org/10.1103/PhysRevLett.107.036101} {\bibfield  {journal} {\bibinfo  {journal} {Phys. Rev. Lett.}\ }\textbf {\bibinfo {volume} {107}},\ \bibinfo {pages} {036101} (\bibinfo {year}
  {2011})}\BibitemShut {NoStop}%
\bibitem [{\citenamefont {Kraus}\ \emph {et~al.}(2022)\citenamefont {Kraus}, \citenamefont {Huttmann}, \citenamefont {Fischer}, \citenamefont {Knispel}, \citenamefont {Bischof}, \citenamefont {Herman}, \citenamefont {Bianchi}, \citenamefont {Stan}, \citenamefont {Holt}, \citenamefont {Caciuc}, \citenamefont {Tsukamoto}, \citenamefont {Wende}, \citenamefont {Hofmann}, \citenamefont {Atodiresei},\ and\ \citenamefont {Michely}}]{Kraus2022}%
  \BibitemOpen
  \bibfield  {author} {\bibinfo {author} {\bibfnamefont {S.}~\bibnamefont {Kraus}}, \bibinfo {author} {\bibfnamefont {F.}~\bibnamefont {Huttmann}}, \bibinfo {author} {\bibfnamefont {J.}~\bibnamefont {Fischer}}, \bibinfo {author} {\bibfnamefont {T.}~\bibnamefont {Knispel}}, \bibinfo {author} {\bibfnamefont {K.}~\bibnamefont {Bischof}}, \bibinfo {author} {\bibfnamefont {A.}~\bibnamefont {Herman}}, \bibinfo {author} {\bibfnamefont {M.}~\bibnamefont {Bianchi}}, \bibinfo {author} {\bibfnamefont {R.-M.}\ \bibnamefont {Stan}}, \bibinfo {author} {\bibfnamefont {A.~J.}\ \bibnamefont {Holt}}, \bibinfo {author} {\bibfnamefont {V.}~\bibnamefont {Caciuc}}, \bibinfo {author} {\bibfnamefont {S.}~\bibnamefont {Tsukamoto}}, \bibinfo {author} {\bibfnamefont {H.}~\bibnamefont {Wende}}, \bibinfo {author} {\bibfnamefont {P.}~\bibnamefont {Hofmann}}, \bibinfo {author} {\bibfnamefont {N.}~\bibnamefont {Atodiresei}},\ and\ \bibinfo {author} {\bibfnamefont {T.}~\bibnamefont {Michely}},\ }\href
  {https://doi.org/10.1103/PhysRevB.105.165405} {\bibfield  {journal} {\bibinfo  {journal} {Phys. Rev. B}\ }\textbf {\bibinfo {volume} {105}},\ \bibinfo {pages} {165405} (\bibinfo {year} {2022})}\BibitemShut {NoStop}%
\bibitem [{\citenamefont {Knispel}\ \emph {et~al.}(2025)\citenamefont {Knispel}, \citenamefont {Mohrenstecher}, \citenamefont {Speckmann}, \citenamefont {Safeer}, \citenamefont {van Efferen}, \citenamefont {Boix}, \citenamefont {Grüneis}, \citenamefont {Jolie}, \citenamefont {Preobrajenski}, \citenamefont {Knudsen}, \citenamefont {Atodiresei}, \citenamefont {Michely},\ and\ \citenamefont {Fischer}}]{Knispel2025}%
  \BibitemOpen
  \bibfield  {author} {\bibinfo {author} {\bibfnamefont {T.}~\bibnamefont {Knispel}}, \bibinfo {author} {\bibfnamefont {D.}~\bibnamefont {Mohrenstecher}}, \bibinfo {author} {\bibfnamefont {C.}~\bibnamefont {Speckmann}}, \bibinfo {author} {\bibfnamefont {A.}~\bibnamefont {Safeer}}, \bibinfo {author} {\bibfnamefont {C.}~\bibnamefont {van Efferen}}, \bibinfo {author} {\bibfnamefont {V.}~\bibnamefont {Boix}}, \bibinfo {author} {\bibfnamefont {A.}~\bibnamefont {Grüneis}}, \bibinfo {author} {\bibfnamefont {W.}~\bibnamefont {Jolie}}, \bibinfo {author} {\bibfnamefont {A.}~\bibnamefont {Preobrajenski}}, \bibinfo {author} {\bibfnamefont {J.}~\bibnamefont {Knudsen}}, \bibinfo {author} {\bibfnamefont {N.}~\bibnamefont {Atodiresei}}, \bibinfo {author} {\bibfnamefont {T.}~\bibnamefont {Michely}},\ and\ \bibinfo {author} {\bibfnamefont {J.}~\bibnamefont {Fischer}},\ }\href {https://doi.org/https://doi.org/10.1002/smll.202408044} {\bibfield  {journal} {\bibinfo  {journal} {Small}\ ,\ \bibinfo {pages} {2408044}} (\bibinfo
  {year} {2025})}\BibitemShut {NoStop}%
\bibitem [{\citenamefont {Dreher}\ \emph {et~al.}(2021)\citenamefont {Dreher}, \citenamefont {Wan}, \citenamefont {Chikina}, \citenamefont {Bianchi}, \citenamefont {Guo}, \citenamefont {Harsh}, \citenamefont {Ma{\~n}as-Valero}, \citenamefont {Coronado}, \citenamefont {Martínez-Galera}, \citenamefont {Hofmann}, \citenamefont {Miwa},\ and\ \citenamefont {Ugeda}}]{Dreher2021}%
  \BibitemOpen
  \bibfield  {author} {\bibinfo {author} {\bibfnamefont {P.}~\bibnamefont {Dreher}}, \bibinfo {author} {\bibfnamefont {W.}~\bibnamefont {Wan}}, \bibinfo {author} {\bibfnamefont {A.}~\bibnamefont {Chikina}}, \bibinfo {author} {\bibfnamefont {M.}~\bibnamefont {Bianchi}}, \bibinfo {author} {\bibfnamefont {H.}~\bibnamefont {Guo}}, \bibinfo {author} {\bibfnamefont {R.}~\bibnamefont {Harsh}}, \bibinfo {author} {\bibfnamefont {S.}~\bibnamefont {Ma{\~n}as-Valero}}, \bibinfo {author} {\bibfnamefont {E.}~\bibnamefont {Coronado}}, \bibinfo {author} {\bibfnamefont {A.~J.}\ \bibnamefont {Martínez-Galera}}, \bibinfo {author} {\bibfnamefont {P.}~\bibnamefont {Hofmann}}, \bibinfo {author} {\bibfnamefont {J.~A.}\ \bibnamefont {Miwa}},\ and\ \bibinfo {author} {\bibfnamefont {M.~M.}\ \bibnamefont {Ugeda}},\ }\href {https://doi.org/10.1021/acsnano.1c06012} {\bibfield  {journal} {\bibinfo  {journal} {ACS Nano}\ }\textbf {\bibinfo {volume} {15}},\ \bibinfo {pages} {19430} (\bibinfo {year} {2021})}\BibitemShut {NoStop}%
\bibitem [{\citenamefont {van Efferen}\ \emph {et~al.}(2024)\citenamefont {van Efferen}, \citenamefont {Hall}, \citenamefont {Atodiresei}, \citenamefont {Boix}, \citenamefont {Safeer}, \citenamefont {Wekking}, \citenamefont {Vinogradov}, \citenamefont {Preobrajenski}, \citenamefont {Knudsen}, \citenamefont {Fischer}, \citenamefont {Jolie},\ and\ \citenamefont {Michely}}]{VanEfferen2024}%
  \BibitemOpen
  \bibfield  {author} {\bibinfo {author} {\bibfnamefont {C.}~\bibnamefont {van Efferen}}, \bibinfo {author} {\bibfnamefont {J.}~\bibnamefont {Hall}}, \bibinfo {author} {\bibfnamefont {N.}~\bibnamefont {Atodiresei}}, \bibinfo {author} {\bibfnamefont {V.}~\bibnamefont {Boix}}, \bibinfo {author} {\bibfnamefont {A.}~\bibnamefont {Safeer}}, \bibinfo {author} {\bibfnamefont {T.}~\bibnamefont {Wekking}}, \bibinfo {author} {\bibfnamefont {N.~A.}\ \bibnamefont {Vinogradov}}, \bibinfo {author} {\bibfnamefont {A.~B.}\ \bibnamefont {Preobrajenski}}, \bibinfo {author} {\bibfnamefont {J.}~\bibnamefont {Knudsen}}, \bibinfo {author} {\bibfnamefont {J.}~\bibnamefont {Fischer}}, \bibinfo {author} {\bibfnamefont {W.}~\bibnamefont {Jolie}},\ and\ \bibinfo {author} {\bibfnamefont {T.}~\bibnamefont {Michely}},\ }\href {https://doi.org/10.1021/acsnano.3c05907} {\bibfield  {journal} {\bibinfo  {journal} {ACS Nano}\ }\textbf {\bibinfo {volume} {18}},\ \bibinfo {pages} {14161} (\bibinfo {year} {2024})}\BibitemShut {NoStop}%
\bibitem [{\citenamefont {Haastrup}\ \emph {et~al.}(2018)\citenamefont {Haastrup}, \citenamefont {Strange}, \citenamefont {Pandey}, \citenamefont {Deilmann}, \citenamefont {Schmidt}, \citenamefont {Hinsche}, \citenamefont {Gjerding}, \citenamefont {Torelli}, \citenamefont {Larsen}, \citenamefont {Riis-Jensen}, \citenamefont {Gath}, \citenamefont {Jacobsen}, \citenamefont {Jørgen~Mortensen}, \citenamefont {Olsen},\ and\ \citenamefont {Thygesen}}]{Haastrup2018}%
  \BibitemOpen
  \bibfield  {author} {\bibinfo {author} {\bibfnamefont {S.}~\bibnamefont {Haastrup}}, \bibinfo {author} {\bibfnamefont {M.}~\bibnamefont {Strange}}, \bibinfo {author} {\bibfnamefont {M.}~\bibnamefont {Pandey}}, \bibinfo {author} {\bibfnamefont {T.}~\bibnamefont {Deilmann}}, \bibinfo {author} {\bibfnamefont {P.~S.}\ \bibnamefont {Schmidt}}, \bibinfo {author} {\bibfnamefont {N.~F.}\ \bibnamefont {Hinsche}}, \bibinfo {author} {\bibfnamefont {M.~N.}\ \bibnamefont {Gjerding}}, \bibinfo {author} {\bibfnamefont {D.}~\bibnamefont {Torelli}}, \bibinfo {author} {\bibfnamefont {P.~M.}\ \bibnamefont {Larsen}}, \bibinfo {author} {\bibfnamefont {A.~C.}\ \bibnamefont {Riis-Jensen}}, \bibinfo {author} {\bibfnamefont {J.}~\bibnamefont {Gath}}, \bibinfo {author} {\bibfnamefont {K.~W.}\ \bibnamefont {Jacobsen}}, \bibinfo {author} {\bibfnamefont {J.}~\bibnamefont {Jørgen~Mortensen}}, \bibinfo {author} {\bibfnamefont {T.}~\bibnamefont {Olsen}},\ and\ \bibinfo {author} {\bibfnamefont {K.~S.}\ \bibnamefont {Thygesen}},\ }\href
  {https://doi.org/10.1088/2053-1583/aacfc1} {\bibfield  {journal} {\bibinfo  {journal} {2D Mater.}\ }\textbf {\bibinfo {volume} {5}},\ \bibinfo {pages} {042002} (\bibinfo {year} {2018})}\BibitemShut {NoStop}%
\bibitem [{\citenamefont {Gjerding}\ \emph {et~al.}(2021)\citenamefont {Gjerding}, \citenamefont {Taghizadeh}, \citenamefont {Rasmussen}, \citenamefont {Ali}, \citenamefont {Bertoldo}, \citenamefont {Deilmann}, \citenamefont {Knøsgaard}, \citenamefont {Kruse}, \citenamefont {Larsen}, \citenamefont {Manti}, \citenamefont {Pedersen}, \citenamefont {Petralanda}, \citenamefont {Skovhus}, \citenamefont {Svendsen}, \citenamefont {Mortensen}, \citenamefont {Olsen},\ and\ \citenamefont {Thygesen}}]{Gjerding2021}%
  \BibitemOpen
  \bibfield  {author} {\bibinfo {author} {\bibfnamefont {M.~N.}\ \bibnamefont {Gjerding}}, \bibinfo {author} {\bibfnamefont {A.}~\bibnamefont {Taghizadeh}}, \bibinfo {author} {\bibfnamefont {A.}~\bibnamefont {Rasmussen}}, \bibinfo {author} {\bibfnamefont {S.}~\bibnamefont {Ali}}, \bibinfo {author} {\bibfnamefont {F.}~\bibnamefont {Bertoldo}}, \bibinfo {author} {\bibfnamefont {T.}~\bibnamefont {Deilmann}}, \bibinfo {author} {\bibfnamefont {N.~R.}\ \bibnamefont {Knøsgaard}}, \bibinfo {author} {\bibfnamefont {M.}~\bibnamefont {Kruse}}, \bibinfo {author} {\bibfnamefont {A.~H.}\ \bibnamefont {Larsen}}, \bibinfo {author} {\bibfnamefont {S.}~\bibnamefont {Manti}}, \bibinfo {author} {\bibfnamefont {T.~G.}\ \bibnamefont {Pedersen}}, \bibinfo {author} {\bibfnamefont {U.}~\bibnamefont {Petralanda}}, \bibinfo {author} {\bibfnamefont {T.}~\bibnamefont {Skovhus}}, \bibinfo {author} {\bibfnamefont {M.~K.}\ \bibnamefont {Svendsen}}, \bibinfo {author} {\bibfnamefont {J.~J.}\ \bibnamefont {Mortensen}}, \bibinfo {author}
  {\bibfnamefont {T.}~\bibnamefont {Olsen}},\ and\ \bibinfo {author} {\bibfnamefont {K.~S.}\ \bibnamefont {Thygesen}},\ }\href {https://doi.org/10.1088/2053-1583/ac1059} {\bibfield  {journal} {\bibinfo  {journal} {2D Mater.}\ }\textbf {\bibinfo {volume} {8}},\ \bibinfo {pages} {044002} (\bibinfo {year} {2021})}\BibitemShut {NoStop}%
\bibitem [{\citenamefont {Jain}\ \emph {et~al.}(2013)\citenamefont {Jain}, \citenamefont {Ong}, \citenamefont {Hautier}, \citenamefont {Chen}, \citenamefont {Richards}, \citenamefont {Dacek}, \citenamefont {Cholia}, \citenamefont {Gunter}, \citenamefont {Skinner}, \citenamefont {Ceder},\ and\ \citenamefont {Persson}}]{Jain13}%
  \BibitemOpen
  \bibfield  {author} {\bibinfo {author} {\bibfnamefont {A.}~\bibnamefont {Jain}}, \bibinfo {author} {\bibfnamefont {S.~P.}\ \bibnamefont {Ong}}, \bibinfo {author} {\bibfnamefont {G.}~\bibnamefont {Hautier}}, \bibinfo {author} {\bibfnamefont {W.}~\bibnamefont {Chen}}, \bibinfo {author} {\bibfnamefont {W.~D.}\ \bibnamefont {Richards}}, \bibinfo {author} {\bibfnamefont {S.}~\bibnamefont {Dacek}}, \bibinfo {author} {\bibfnamefont {S.}~\bibnamefont {Cholia}}, \bibinfo {author} {\bibfnamefont {D.}~\bibnamefont {Gunter}}, \bibinfo {author} {\bibfnamefont {D.}~\bibnamefont {Skinner}}, \bibinfo {author} {\bibfnamefont {G.}~\bibnamefont {Ceder}},\ and\ \bibinfo {author} {\bibfnamefont {K.~A.}\ \bibnamefont {Persson}},\ }\href {https://doi.org/10.1063/1.4812323} {\bibfield  {journal} {\bibinfo  {journal} {APL Mater.}\ }\textbf {\bibinfo {volume} {1}},\ \bibinfo {pages} {011002} (\bibinfo {year} {2013})}\BibitemShut {NoStop}%
\bibitem [{\citenamefont {van Efferen}\ \emph {et~al.}(2022)\citenamefont {van Efferen}, \citenamefont {Murray}, \citenamefont {Fischer}, \citenamefont {Busse}, \citenamefont {Komsa}, \citenamefont {Michely},\ and\ \citenamefont {Jolie}}]{VanEfferen2022}%
  \BibitemOpen
  \bibfield  {author} {\bibinfo {author} {\bibfnamefont {C.}~\bibnamefont {van Efferen}}, \bibinfo {author} {\bibfnamefont {C.}~\bibnamefont {Murray}}, \bibinfo {author} {\bibfnamefont {J.}~\bibnamefont {Fischer}}, \bibinfo {author} {\bibfnamefont {C.}~\bibnamefont {Busse}}, \bibinfo {author} {\bibfnamefont {H.-P.}\ \bibnamefont {Komsa}}, \bibinfo {author} {\bibfnamefont {T.}~\bibnamefont {Michely}},\ and\ \bibinfo {author} {\bibfnamefont {W.}~\bibnamefont {Jolie}},\ }\href {https://doi.org/10.1088/2053-1583/ac5d0f} {\bibfield  {journal} {\bibinfo  {journal} {2D Mater.}\ }\textbf {\bibinfo {volume} {9}},\ \bibinfo {pages} {025026} (\bibinfo {year} {2022})}\BibitemShut {NoStop}%
\bibitem [{\citenamefont {Knispel}\ \emph {et~al.}(2024)\citenamefont {Knispel}, \citenamefont {Berges}, \citenamefont {Schobert}, \citenamefont {van Loon}, \citenamefont {Jolie}, \citenamefont {Wehling}, \citenamefont {Michely},\ and\ \citenamefont {Fischer}}]{Knispel2024}%
  \BibitemOpen
  \bibfield  {author} {\bibinfo {author} {\bibfnamefont {T.}~\bibnamefont {Knispel}}, \bibinfo {author} {\bibfnamefont {J.}~\bibnamefont {Berges}}, \bibinfo {author} {\bibfnamefont {A.}~\bibnamefont {Schobert}}, \bibinfo {author} {\bibfnamefont {E.~G. C.~P.}\ \bibnamefont {van Loon}}, \bibinfo {author} {\bibfnamefont {W.}~\bibnamefont {Jolie}}, \bibinfo {author} {\bibfnamefont {T.}~\bibnamefont {Wehling}}, \bibinfo {author} {\bibfnamefont {T.}~\bibnamefont {Michely}},\ and\ \bibinfo {author} {\bibfnamefont {J.}~\bibnamefont {Fischer}},\ }\href {https://doi.org/10.1021/acs.nanolett.3c02787} {\bibfield  {journal} {\bibinfo  {journal} {Nano Lett.}\ }\textbf {\bibinfo {volume} {24}},\ \bibinfo {pages} {1045} (\bibinfo {year} {2024})}\BibitemShut {NoStop}%
\bibitem [{\citenamefont {Binnig}\ \emph {et~al.}(1985)\citenamefont {Binnig}, \citenamefont {Frank}, \citenamefont {Fuchs}, \citenamefont {Garcia}, \citenamefont {Reihl}, \citenamefont {Rohrer}, \citenamefont {Salvan},\ and\ \citenamefont {Williams}}]{Binnig1985}%
  \BibitemOpen
  \bibfield  {author} {\bibinfo {author} {\bibfnamefont {G.}~\bibnamefont {Binnig}}, \bibinfo {author} {\bibfnamefont {K.~H.}\ \bibnamefont {Frank}}, \bibinfo {author} {\bibfnamefont {H.}~\bibnamefont {Fuchs}}, \bibinfo {author} {\bibfnamefont {N.}~\bibnamefont {Garcia}}, \bibinfo {author} {\bibfnamefont {B.}~\bibnamefont {Reihl}}, \bibinfo {author} {\bibfnamefont {H.}~\bibnamefont {Rohrer}}, \bibinfo {author} {\bibfnamefont {F.}~\bibnamefont {Salvan}},\ and\ \bibinfo {author} {\bibfnamefont {A.~R.}\ \bibnamefont {Williams}},\ }\href {https://doi.org/10.1103/PhysRevLett.55.991} {\bibfield  {journal} {\bibinfo  {journal} {Phys. Rev. Lett.}\ }\textbf {\bibinfo {volume} {55}},\ \bibinfo {pages} {991} (\bibinfo {year} {1985})}\BibitemShut {NoStop}%
\bibitem [{\citenamefont {Becker}\ \emph {et~al.}(1985)\citenamefont {Becker}, \citenamefont {Golovchenko},\ and\ \citenamefont {Swartzentruber}}]{Becker1985}%
  \BibitemOpen
  \bibfield  {author} {\bibinfo {author} {\bibfnamefont {R.~S.}\ \bibnamefont {Becker}}, \bibinfo {author} {\bibfnamefont {J.~A.}\ \bibnamefont {Golovchenko}},\ and\ \bibinfo {author} {\bibfnamefont {B.~S.}\ \bibnamefont {Swartzentruber}},\ }\href {https://doi.org/10.1103/PhysRevLett.55.987} {\bibfield  {journal} {\bibinfo  {journal} {Phys. Rev. Lett.}\ }\textbf {\bibinfo {volume} {55}},\ \bibinfo {pages} {987} (\bibinfo {year} {1985})}\BibitemShut {NoStop}%
\bibitem [{\citenamefont {Li}\ \emph {et~al.}(1998)\citenamefont {Li}, \citenamefont {Schneider}, \citenamefont {Berndt}, \citenamefont {Bryant},\ and\ \citenamefont {Crampin}}]{Li1998}%
  \BibitemOpen
  \bibfield  {author} {\bibinfo {author} {\bibfnamefont {J.}~\bibnamefont {Li}}, \bibinfo {author} {\bibfnamefont {W.-D.}\ \bibnamefont {Schneider}}, \bibinfo {author} {\bibfnamefont {R.}~\bibnamefont {Berndt}}, \bibinfo {author} {\bibfnamefont {O.~R.}\ \bibnamefont {Bryant}},\ and\ \bibinfo {author} {\bibfnamefont {S.}~\bibnamefont {Crampin}},\ }\href {https://doi.org/10.1103/PhysRevLett.81.4464} {\bibfield  {journal} {\bibinfo  {journal} {Phys. Rev. Lett.}\ }\textbf {\bibinfo {volume} {81}},\ \bibinfo {pages} {4464} (\bibinfo {year} {1998})}\BibitemShut {NoStop}%
\bibitem [{\citenamefont {Telford}\ \emph {et~al.}(2020)\citenamefont {Telford}, \citenamefont {Dismukes}, \citenamefont {Lee}, \citenamefont {Cheng}, \citenamefont {Wieteska}, \citenamefont {Bartholomew}, \citenamefont {Chen}, \citenamefont {Xu}, \citenamefont {Pasupathy}, \citenamefont {Zhu}, \citenamefont {Dean},\ and\ \citenamefont {Roy}}]{Telford2020}%
  \BibitemOpen
  \bibfield  {author} {\bibinfo {author} {\bibfnamefont {E.~J.}\ \bibnamefont {Telford}}, \bibinfo {author} {\bibfnamefont {A.~H.}\ \bibnamefont {Dismukes}}, \bibinfo {author} {\bibfnamefont {K.}~\bibnamefont {Lee}}, \bibinfo {author} {\bibfnamefont {M.}~\bibnamefont {Cheng}}, \bibinfo {author} {\bibfnamefont {A.}~\bibnamefont {Wieteska}}, \bibinfo {author} {\bibfnamefont {A.~K.}\ \bibnamefont {Bartholomew}}, \bibinfo {author} {\bibfnamefont {Y.-S.}\ \bibnamefont {Chen}}, \bibinfo {author} {\bibfnamefont {X.}~\bibnamefont {Xu}}, \bibinfo {author} {\bibfnamefont {A.~N.}\ \bibnamefont {Pasupathy}}, \bibinfo {author} {\bibfnamefont {X.}~\bibnamefont {Zhu}}, \bibinfo {author} {\bibfnamefont {C.~R.}\ \bibnamefont {Dean}},\ and\ \bibinfo {author} {\bibfnamefont {X.}~\bibnamefont {Roy}},\ }\href {https://doi.org/https://doi.org/10.1002/adma.202003240} {\bibfield  {journal} {\bibinfo  {journal} {Adv. Mater.}\ }\textbf {\bibinfo {volume} {32}},\ \bibinfo {pages} {2003240} (\bibinfo {year} {2020})}\BibitemShut
  {NoStop}%
\bibitem [{\citenamefont {Wilson}\ \emph {et~al.}(2021)\citenamefont {Wilson}, \citenamefont {Lee}, \citenamefont {Cenker}, \citenamefont {Xie}, \citenamefont {Dismukes}, \citenamefont {Telford}, \citenamefont {Fonseca}, \citenamefont {Sivakumar}, \citenamefont {Dean}, \citenamefont {Cao} \emph {et~al.}}]{Wilson2021}%
  \BibitemOpen
  \bibfield  {author} {\bibinfo {author} {\bibfnamefont {N.~P.}\ \bibnamefont {Wilson}}, \bibinfo {author} {\bibfnamefont {K.}~\bibnamefont {Lee}}, \bibinfo {author} {\bibfnamefont {J.}~\bibnamefont {Cenker}}, \bibinfo {author} {\bibfnamefont {K.}~\bibnamefont {Xie}}, \bibinfo {author} {\bibfnamefont {A.~H.}\ \bibnamefont {Dismukes}}, \bibinfo {author} {\bibfnamefont {E.~J.}\ \bibnamefont {Telford}}, \bibinfo {author} {\bibfnamefont {J.}~\bibnamefont {Fonseca}}, \bibinfo {author} {\bibfnamefont {S.}~\bibnamefont {Sivakumar}}, \bibinfo {author} {\bibfnamefont {C.}~\bibnamefont {Dean}}, \bibinfo {author} {\bibfnamefont {T.}~\bibnamefont {Cao}}, \emph {et~al.},\ }\href {https://doi.org/10.1038/s41563-021-01070-8} {\bibfield  {journal} {\bibinfo  {journal} {Nat. Mater.}\ }\textbf {\bibinfo {volume} {20}},\ \bibinfo {pages} {1657} (\bibinfo {year} {2021})}\BibitemShut {NoStop}%
\bibitem [{\citenamefont {Coraux}\ \emph {et~al.}(2009)\citenamefont {Coraux}, \citenamefont {N'Diaye}, \citenamefont {Engler}, \citenamefont {Busse}, \citenamefont {Wall}, \citenamefont {Buckanie}, \citenamefont {zu~Heringdorf}, \citenamefont {van Gastel}, \citenamefont {Poelsema},\ and\ \citenamefont {Michely}}]{Coraux2009}%
  \BibitemOpen
  \bibfield  {author} {\bibinfo {author} {\bibfnamefont {J.}~\bibnamefont {Coraux}}, \bibinfo {author} {\bibfnamefont {A.~T.}\ \bibnamefont {N'Diaye}}, \bibinfo {author} {\bibfnamefont {M.}~\bibnamefont {Engler}}, \bibinfo {author} {\bibfnamefont {C.}~\bibnamefont {Busse}}, \bibinfo {author} {\bibfnamefont {D.}~\bibnamefont {Wall}}, \bibinfo {author} {\bibfnamefont {N.}~\bibnamefont {Buckanie}}, \bibinfo {author} {\bibfnamefont {F.-J.~M.}\ \bibnamefont {zu~Heringdorf}}, \bibinfo {author} {\bibfnamefont {R.}~\bibnamefont {van Gastel}}, \bibinfo {author} {\bibfnamefont {B.}~\bibnamefont {Poelsema}},\ and\ \bibinfo {author} {\bibfnamefont {T.}~\bibnamefont {Michely}},\ }\href {https://doi.org/10.1088/1367-2630/11/2/023006} {\bibfield  {journal} {\bibinfo  {journal} {New J. Phys.}\ }\textbf {\bibinfo {volume} {11}},\ \bibinfo {pages} {023006} (\bibinfo {year} {2009})}\BibitemShut {NoStop}%
\bibitem [{\citenamefont {Horcas}\ \emph {et~al.}(2007)\citenamefont {Horcas}, \citenamefont {Fernández}, \citenamefont {Gómez-Rodríguez}, \citenamefont {Colchero}, \citenamefont {Gómez-Herrero},\ and\ \citenamefont {Baro}}]{Horcas_WSxM_2007}%
  \BibitemOpen
  \bibfield  {author} {\bibinfo {author} {\bibfnamefont {I.}~\bibnamefont {Horcas}}, \bibinfo {author} {\bibfnamefont {R.}~\bibnamefont {Fernández}}, \bibinfo {author} {\bibfnamefont {J.~M.}\ \bibnamefont {Gómez-Rodríguez}}, \bibinfo {author} {\bibfnamefont {J.}~\bibnamefont {Colchero}}, \bibinfo {author} {\bibfnamefont {J.}~\bibnamefont {Gómez-Herrero}},\ and\ \bibinfo {author} {\bibfnamefont {A.~M.}\ \bibnamefont {Baro}},\ }\href {https://doi.org/10.1063/1.2432410} {\bibfield  {journal} {\bibinfo  {journal} {Rev. Sci. Instrum.}\ }\textbf {\bibinfo {volume} {78}},\ \bibinfo {pages} {013705} (\bibinfo {year} {2007})}\BibitemShut {NoStop}%
\bibitem [{\citenamefont {Kresse}\ and\ \citenamefont {Joubert}(1999)}]{vasp1}%
  \BibitemOpen
  \bibfield  {author} {\bibinfo {author} {\bibfnamefont {G.}~\bibnamefont {Kresse}}\ and\ \bibinfo {author} {\bibfnamefont {D.}~\bibnamefont {Joubert}},\ }\href {https://doi.org/10.1103/physrevb.59.1758} {\bibfield  {journal} {\bibinfo  {journal} {Phys. Rev. B}\ }\textbf {\bibinfo {volume} {59}},\ \bibinfo {pages} {1758} (\bibinfo {year} {1999})}\BibitemShut {NoStop}%
\bibitem [{\citenamefont {Kresse}\ and\ \citenamefont {Furthmüller}(1996)}]{vasp2}%
  \BibitemOpen
  \bibfield  {author} {\bibinfo {author} {\bibfnamefont {G.}~\bibnamefont {Kresse}}\ and\ \bibinfo {author} {\bibfnamefont {J.}~\bibnamefont {Furthmüller}},\ }\href {https://doi.org/10.1103/physrevb.54.11169} {\bibfield  {journal} {\bibinfo  {journal} {Phys. Rev. B}\ }\textbf {\bibinfo {volume} {54}},\ \bibinfo {pages} {11169} (\bibinfo {year} {1996})}\BibitemShut {NoStop}%
\bibitem [{\citenamefont {Perdew}\ \emph {et~al.}(1996)\citenamefont {Perdew}, \citenamefont {Burke},\ and\ \citenamefont {Ernzerhof}}]{perdew_1996}%
  \BibitemOpen
  \bibfield  {author} {\bibinfo {author} {\bibfnamefont {J.~P.}\ \bibnamefont {Perdew}}, \bibinfo {author} {\bibfnamefont {K.}~\bibnamefont {Burke}},\ and\ \bibinfo {author} {\bibfnamefont {M.}~\bibnamefont {Ernzerhof}},\ }\href {https://doi.org/10.1103/physrevlett.77.3865} {\bibfield  {journal} {\bibinfo  {journal} {Phys. Rev. Lett.}\ }\textbf {\bibinfo {volume} {77}},\ \bibinfo {pages} {3865} (\bibinfo {year} {1996})}\BibitemShut {NoStop}%
\bibitem [{\citenamefont {Dudarev}\ \emph {et~al.}(1998)\citenamefont {Dudarev}, \citenamefont {Botton}, \citenamefont {Savrasov}, \citenamefont {Humphreys},\ and\ \citenamefont {Sutton}}]{Dudarev1998}%
  \BibitemOpen
  \bibfield  {author} {\bibinfo {author} {\bibfnamefont {S.~L.}\ \bibnamefont {Dudarev}}, \bibinfo {author} {\bibfnamefont {G.~A.}\ \bibnamefont {Botton}}, \bibinfo {author} {\bibfnamefont {S.~Y.}\ \bibnamefont {Savrasov}}, \bibinfo {author} {\bibfnamefont {C.~J.}\ \bibnamefont {Humphreys}},\ and\ \bibinfo {author} {\bibfnamefont {A.~P.}\ \bibnamefont {Sutton}},\ }\href {https://doi.org/10.1103/PhysRevB.57.1505} {\bibfield  {journal} {\bibinfo  {journal} {Phys. Rev. B}\ }\textbf {\bibinfo {volume} {57}},\ \bibinfo {pages} {1505} (\bibinfo {year} {1998})}\BibitemShut {NoStop}%
\bibitem [{\citenamefont {Grimme}(2006)}]{DFT-D2}%
  \BibitemOpen
  \bibfield  {author} {\bibinfo {author} {\bibfnamefont {S.}~\bibnamefont {Grimme}},\ }\href {https://doi.org/https://doi.org/10.1002/jcc.20495} {\bibfield  {journal} {\bibinfo  {journal} {J. Comput. Chem.}\ }\textbf {\bibinfo {volume} {27}},\ \bibinfo {pages} {1787} (\bibinfo {year} {2006})}\BibitemShut {NoStop}%
\end{thebibliography}%


\begin{thebibliography}{5}%
\makeatletter
\providecommand \@ifxundefined [1]{%
 \@ifx{#1\undefined}
}%
\providecommand \@ifnum [1]{%
 \ifnum #1\expandafter \@firstoftwo
 \else \expandafter \@secondoftwo
 \fi
}%
\providecommand \@ifx [1]{%
 \ifx #1\expandafter \@firstoftwo
 \else \expandafter \@secondoftwo
 \fi
}%
\providecommand \natexlab [1]{#1}%
\providecommand \enquote  [1]{``#1''}%
\providecommand \bibnamefont  [1]{#1}%
\providecommand \bibfnamefont [1]{#1}%
\providecommand \citenamefont [1]{#1}%
\providecommand \href@noop [0]{\@secondoftwo}%
\providecommand \href [0]{\begingroup \@sanitize@url \@href}%
\providecommand \@href[1]{\@@startlink{#1}\@@href}%
\providecommand \@@href[1]{\endgroup#1\@@endlink}%
\providecommand \@sanitize@url [0]{\catcode `\\12\catcode `\$12\catcode `\&12\catcode `\#12\catcode `\^12\catcode `\_12\catcode `\%12\relax}%
\providecommand \@@startlink[1]{}%
\providecommand \@@endlink[0]{}%
\providecommand \url  [0]{\begingroup\@sanitize@url \@url }%
\providecommand \@url [1]{\endgroup\@href {#1}{\urlprefix }}%
\providecommand \urlprefix  [0]{URL }%
\providecommand \Eprint [0]{\href }%
\providecommand \doibase [0]{https://doi.org/}%
\providecommand \selectlanguage [0]{\@gobble}%
\providecommand \bibinfo  [0]{\@secondoftwo}%
\providecommand \bibfield  [0]{\@secondoftwo}%
\providecommand \translation [1]{[#1]}%
\providecommand \BibitemOpen [0]{}%
\providecommand \bibitemStop [0]{}%
\providecommand \bibitemNoStop [0]{.\EOS\space}%
\providecommand \EOS [0]{\spacefactor3000\relax}%
\providecommand \BibitemShut  [1]{\csname bibitem#1\endcsname}%
\let\auto@bib@innerbib\@empty
\bibitem [{\citenamefont {Binnig}\ \emph {et~al.}(1985)\citenamefont {Binnig}, \citenamefont {Frank}, \citenamefont {Fuchs}, \citenamefont {Garcia}, \citenamefont {Reihl}, \citenamefont {Rohrer}, \citenamefont {Salvan},\ and\ \citenamefont {Williams}}]{Binnig1985}%
  \BibitemOpen
  \bibfield  {author} {\bibinfo {author} {\bibfnamefont {G.}~\bibnamefont {Binnig}}, \bibinfo {author} {\bibfnamefont {K.~H.}\ \bibnamefont {Frank}}, \bibinfo {author} {\bibfnamefont {H.}~\bibnamefont {Fuchs}}, \bibinfo {author} {\bibfnamefont {N.}~\bibnamefont {Garcia}}, \bibinfo {author} {\bibfnamefont {B.}~\bibnamefont {Reihl}}, \bibinfo {author} {\bibfnamefont {H.}~\bibnamefont {Rohrer}}, \bibinfo {author} {\bibfnamefont {F.}~\bibnamefont {Salvan}},\ and\ \bibinfo {author} {\bibfnamefont {A.~R.}\ \bibnamefont {Williams}},\ }\href {https://doi.org/10.1103/PhysRevLett.55.991} {\bibfield  {journal} {\bibinfo  {journal} {Phys. Rev. Lett.}\ }\textbf {\bibinfo {volume} {55}},\ \bibinfo {pages} {991} (\bibinfo {year} {1985})}\BibitemShut {NoStop}%
\bibitem [{\citenamefont {Becker}\ \emph {et~al.}(1985)\citenamefont {Becker}, \citenamefont {Golovchenko},\ and\ \citenamefont {Swartzentruber}}]{Becker1985}%
  \BibitemOpen
  \bibfield  {author} {\bibinfo {author} {\bibfnamefont {R.~S.}\ \bibnamefont {Becker}}, \bibinfo {author} {\bibfnamefont {J.~A.}\ \bibnamefont {Golovchenko}},\ and\ \bibinfo {author} {\bibfnamefont {B.~S.}\ \bibnamefont {Swartzentruber}},\ }\href {https://doi.org/10.1103/PhysRevLett.55.987} {\bibfield  {journal} {\bibinfo  {journal} {Phys. Rev. Lett.}\ }\textbf {\bibinfo {volume} {55}},\ \bibinfo {pages} {987} (\bibinfo {year} {1985})}\BibitemShut {NoStop}%
\bibitem [{\citenamefont {Gundlach}(1966)}]{Gundlach1966}%
  \BibitemOpen
  \bibfield  {author} {\bibinfo {author} {\bibfnamefont {K.}~\bibnamefont {Gundlach}},\ }\href {https://doi.org/https://doi.org/10.1016/0038-1101(66)90071-2} {\bibfield  {journal} {\bibinfo  {journal} {Solid-State Electron.}\ }\textbf {\bibinfo {volume} {9}},\ \bibinfo {pages} {949} (\bibinfo {year} {1966})}\BibitemShut {NoStop}%
\bibitem [{\citenamefont {Fowler}\ and\ \citenamefont {Nordheim}(1928)}]{Fowler1928}%
  \BibitemOpen
  \bibfield  {author} {\bibinfo {author} {\bibfnamefont {R.~H.}\ \bibnamefont {Fowler}}\ and\ \bibinfo {author} {\bibfnamefont {L.}~\bibnamefont {Nordheim}},\ }\href@noop {} {\bibfield  {journal} {\bibinfo  {journal} {Proc. R. Soc. Lond. A}\ }\textbf {\bibinfo {volume} {119}},\ \bibinfo {pages} {173} (\bibinfo {year} {1928})}\BibitemShut {NoStop}%
\bibitem [{\citenamefont {Lin}\ \emph {et~al.}(2007)\citenamefont {Lin}, \citenamefont {Lu}, \citenamefont {Su}, \citenamefont {Shih}, \citenamefont {Wu}, \citenamefont {Yao}, \citenamefont {Chang},\ and\ \citenamefont {Tsong}}]{Lin2007}%
  \BibitemOpen
  \bibfield  {author} {\bibinfo {author} {\bibfnamefont {C.~L.}\ \bibnamefont {Lin}}, \bibinfo {author} {\bibfnamefont {S.~M.}\ \bibnamefont {Lu}}, \bibinfo {author} {\bibfnamefont {W.~B.}\ \bibnamefont {Su}}, \bibinfo {author} {\bibfnamefont {H.~T.}\ \bibnamefont {Shih}}, \bibinfo {author} {\bibfnamefont {B.~F.}\ \bibnamefont {Wu}}, \bibinfo {author} {\bibfnamefont {Y.~D.}\ \bibnamefont {Yao}}, \bibinfo {author} {\bibfnamefont {C.~S.}\ \bibnamefont {Chang}},\ and\ \bibinfo {author} {\bibfnamefont {T.~T.}\ \bibnamefont {Tsong}},\ }\href {https://doi.org/10.1103/PhysRevLett.99.216103} {\bibfield  {journal} {\bibinfo  {journal} {Phys. Rev. Lett.}\ }\textbf {\bibinfo {volume} {99}},\ \bibinfo {pages} {216103} (\bibinfo {year} {2007})}\BibitemShut {NoStop}%
\end{thebibliography}%

\end{document}


\DeclareGraphicsExtensions{.pdf}

\title{Supporting Information:\\
Which chromium-sulfur compounds exist as 2D material?}

\author{Affan Safeer}
\affiliation{II. Physikalisches Institut, Universit\"{a}t zu K\"{o}ln, Z\"{u}lpicher Stra\ss e 77, D-50937 K\"{o}ln, Germany}

\author{Mahdi Ghorbani-Asl}
\affiliation{Helmholtz-Zentrum Dresden-Rossendorf, Institute of Ion Beam Physics and Materials Research, D-01328 Dresden, Germany}

\author{Wouter Jolie}
\affiliation{II. Physikalisches Institut, Universit\"{a}t zu K\"{o}ln, Z\"{u}lpicher Stra\ss e 77, D-50937 K\"{o}ln, Germany}

\author{Arkady V. Krasheninnikov}
\affiliation{Helmholtz-Zentrum Dresden-Rossendorf, Institute of Ion Beam Physics and Materials Research, D-01328 Dresden, Germany}

\author{Thomas Michely}
\affiliation{II. Physikalisches Institut, Universit\"{a}t zu K\"{o}ln, Z\"{u}lpicher Stra\ss e 77, D-50937 K\"{o}ln, Germany}

\author{Jeison Fischer} 
\email{Corresponding author: jfischer@ph2.uni-koeln.de}
\affiliation{II. Physikalisches Institut, Universit\"{a}t zu K\"{o}ln, Z\"{u}lpicher Stra\ss e 77, D-50937 K\"{o}ln, Germany}

\maketitle
\newpage
\tableofcontents
\clearpage

\subsection*{Supplementary Note 1: Extended version of Figure 1.}
\begin{figure}[hbt!]
\includegraphics[width=\textwidth]{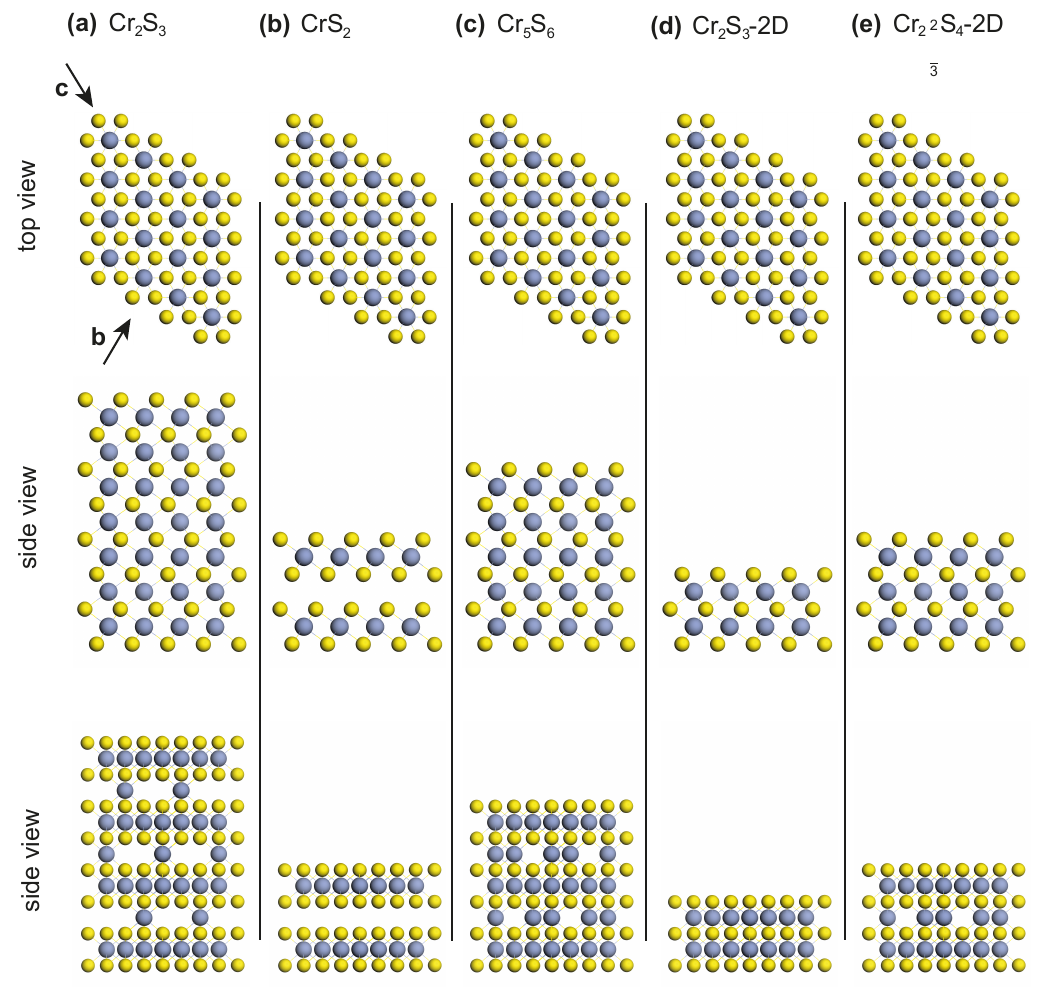}
\caption{\normalsize Crystal structure of Cr$_x$S$_y$ phases. (a) Rhombohedral Cr$_2$S$_3$. (b) T-phase CrS$_2$. (c) Cr$_5$S$_6$. (d) Cr$_2$S$_3$-2D. (e) Cr$_{2\frac{2}{3}}$S$_4$-2D. Side views of the middle row and bottom row are taken from the perspective along arrows \textit{b} and \textit{c} labeled in (a), respectively.}
  \label{fgr:SI_Fig_Sulfur_anneal} 
\end{figure}

\clearpage
\subsection*{Supplementary Note 2: Extended annealing sequence of Cr$_x$S$_y$-2D.}
\begin{figure}[hbt!]
\includegraphics[width=0.82\textwidth]{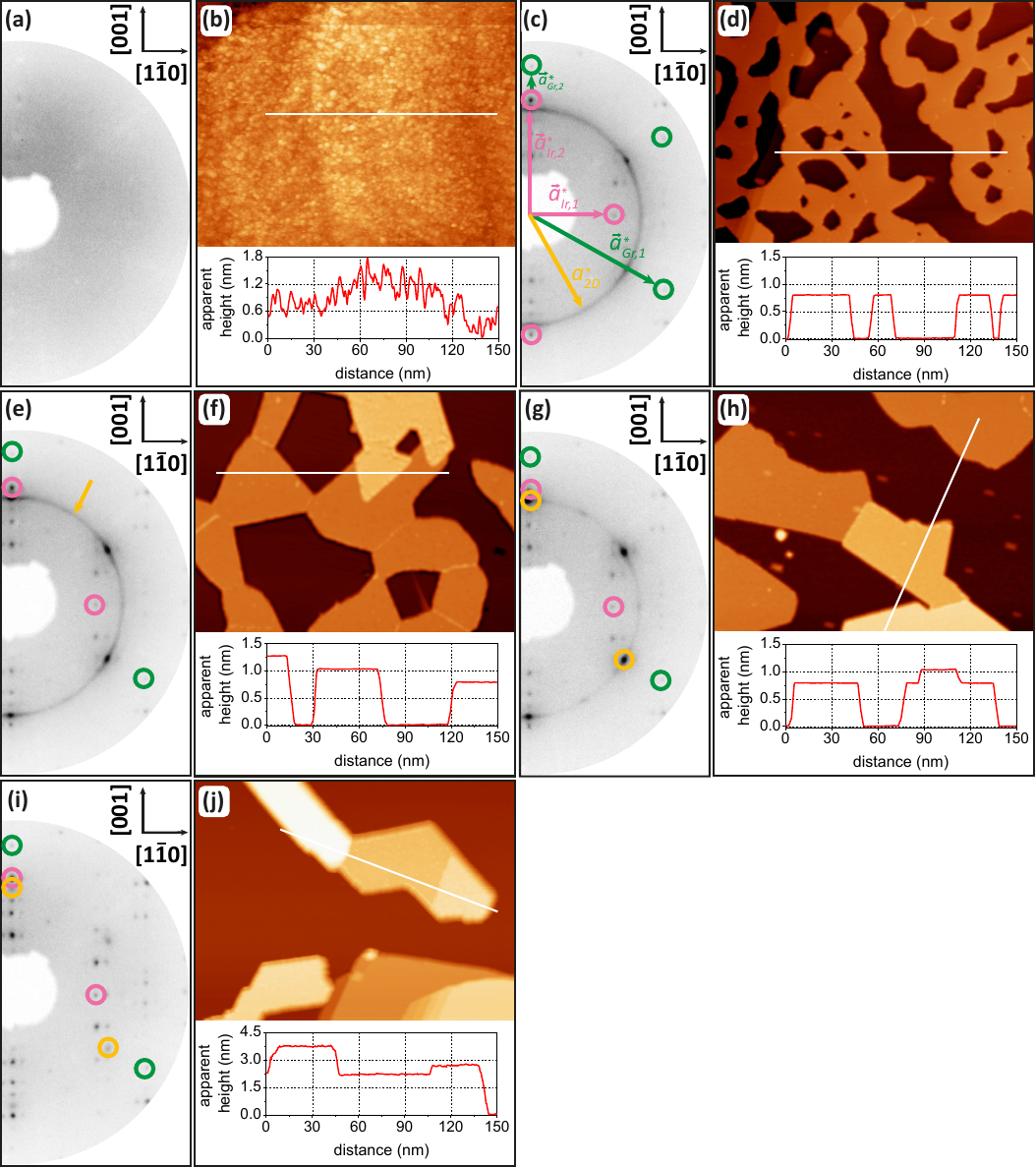}
\caption{\normalsize (a,b) Contrast-inverted 120\,eV LEED image and STM overview topography after room temperature deposition of Cr in S vapor on Gr/Ir(110). Contrast-inverted 120\,eV LEED patterns and STM overview topographies after annealing in sulfur vapor at (c,d) 750\,K, (e,f) 850\,K, (g,h) 950\,K, and (i,j) 1050\,K. Lower panels of STM topographies show height profiles along a white line in the corresponding STM topography. Some first-order Ir and Gr reflections are encircled magenta and green, respectively. First-order reflections of Cr$_x$S$_y$ are pointed by the yellow arrow or encircled yellow. STM images are obtained at room temperature with $V_b$ = 1 V, $I_t$ = 100 pA and have dimensions of 200\,nm $\times$ 150\,nm.} 
  \label{fgr:SI_Fig_Sulfur_anneal} 
\end{figure}

\clearpage
\subsection*{Supplementary Note 3: Annealing sequence of Cr$_x$S$_y$-2D without sulfur vapor during annealing}
\begin{figure}[hbt!]
  \includegraphics[width=0.82\textwidth]{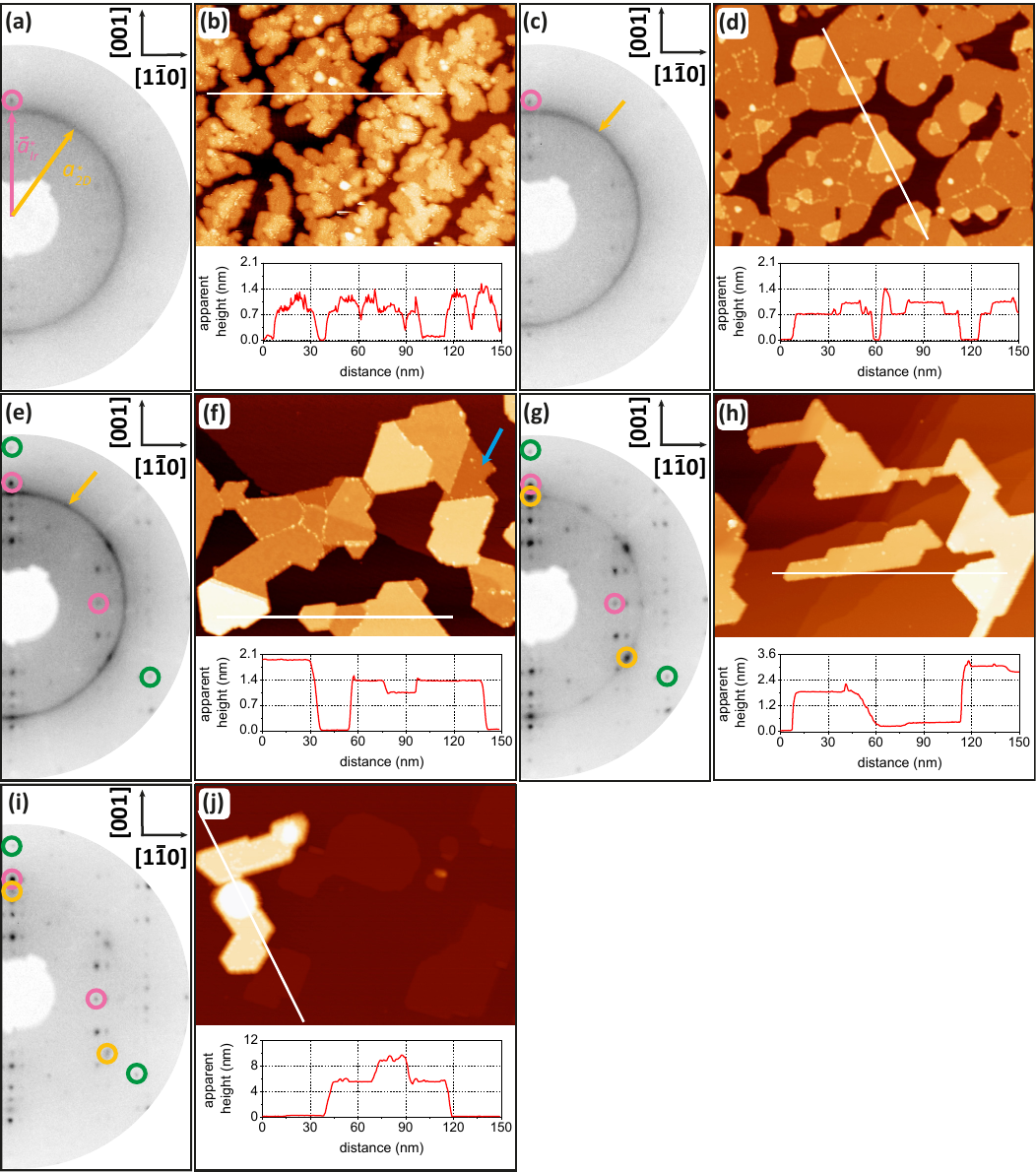}
  \caption{Contrast-inverted 120\,eV LEED patterns and STM overview topographies after annealing without sulfur to (a,b) 650\,K, (c,d) 750\,K, (e,f) 850\,K, (g,h) 950\,K, and (i,j) 1050\,K. Lower panels of STM topographies show height profiles along a white line in the corresponding STM topography, indicating the apparent height of the islands. Some first-order Ir and Gr reflections are encircled magenta and green, respectively. First-order reflections of Cr$_x$S$_y$ are pointed by the yellow arrow or encircled yellow. STM images are obtained at room temperature with $V_b = -1$\,V, $I_t = 100$\,pA and have dimensions of 200\,nm\,$\times 150$\,nm.}
  \label{fgr:SI_Fig_No_Sulfur_anneal}
\end{figure}

\clearpage
\subsection*{Supplementary Note 4: Annealing a high coverage sample to  850\,K}
\begin{figure}[hbt!]
  \includegraphics[width=\linewidth]{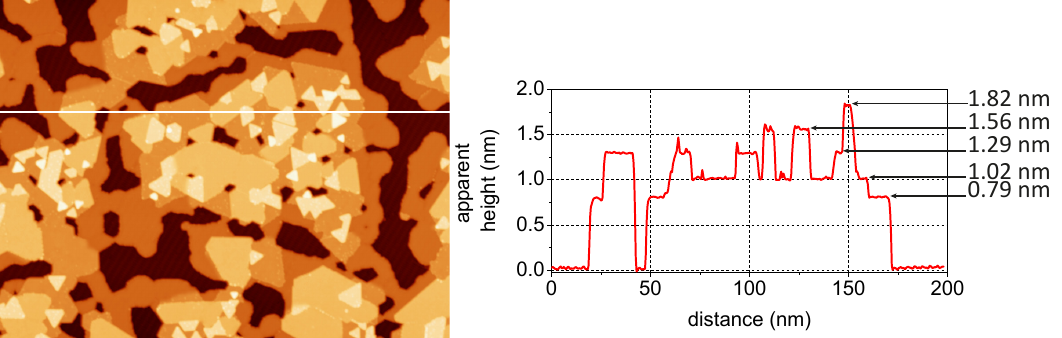}
  \caption{STM topography of larger coverage sample after annealing in sulfur vapor to 850\,K. Height profile in the right panel is taken along the white in STM topography. Image is obtained at 1.7\,K with $V_b$ = 1\,V, $I_t$ = 100\,pA and have dimensions of 200\,nm $\times$ 150\,nm.} 
  \label{fgr:SI_Fig_annealing_high_cov}
\end{figure}

\clearpage
\subsection*{Supplementary Note 5: Cr$_2$S$_3$-2D sample after exposing to ambient conditions}
\begin{figure}[hbt!]
  \includegraphics[width=\linewidth]{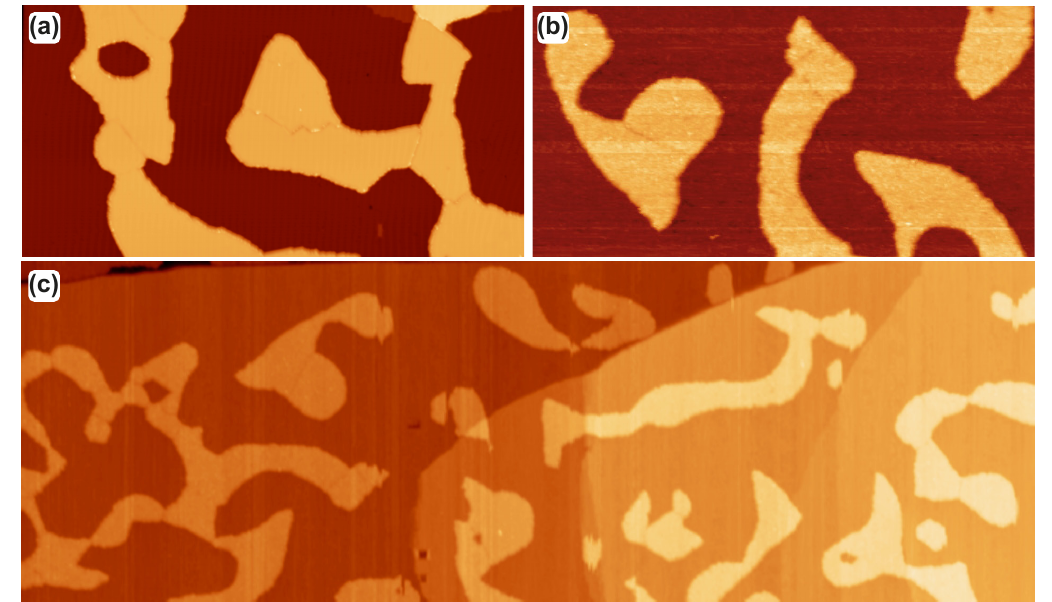}
  \caption{STM images of Cr$_2$S$_3$-2D on Gr/Ir(110) (a) before and (b) after exposure to ambient for 3 hours and subsequent pump-down without annealing. (c) Large size STM overview image of the sample in (b). STM images are acquired with $V_b$ = 1 V, $I_t$ = 100 pA at 1.7\,K. and have dimensions of (a) and (b) 200\,nm $\times$ 100\,nm and (c) 600\,nm $\times$ 180\,nm.}
  \label{fgr:SI_Fig_air_expose}
\end{figure}

\clearpage
\subsection*{Supplementary Note 6: Samples prepared by varying the growth conditions.}
\begin{figure}[hbt!]
  \includegraphics[width=\linewidth]{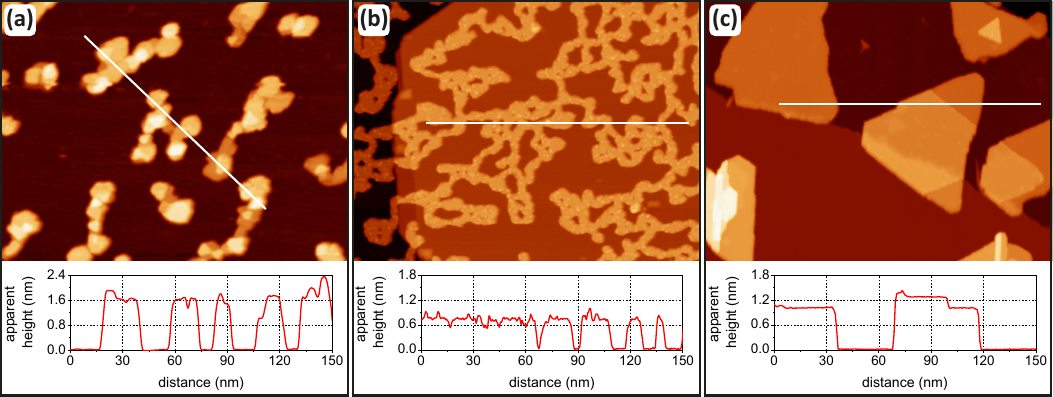}
  \caption{(a) STM image of a sample prepared by using $5\times10^{-9}$\,mbar sulfur during deposition at room temperature and followed by sulfur annealing at 750\,K. (b) STM topograph of the sample prepared by deposition of Cr in $5\times10^{-8}$\,mbar sulfur pressure on Gr/Ir(110) at 160\,K, followed by annealing at 700\,K. (c) STM topograph of sample grown at 600\,K on Gr/Ir(111) in $7\times10^{-8}$\,mbar sulfur pressure without additional annealing step. The height profiles in the lower panel correspond to the measurement along the white line in the STM image. Images (a) and (b) are acquired at room temperature with $V_b$ = 1 V, $I_t$ = 100 pA and (c) taken at 1.7\,K with $V_b$ = 1 V, $I_t$ = 50 pA. Image dimensions are 200\,nm $\times$ 150\,nm.}  
  \label{fgr:SI_Fig_diff_growth}
\end{figure}

\clearpage
\subsection*{Supplementary Note 7: Growth of Cr$_x$S$_y$-2D on Gr/Ir(111).}
\begin{figure}[hbt!]
  \includegraphics[width=0.9\textwidth]{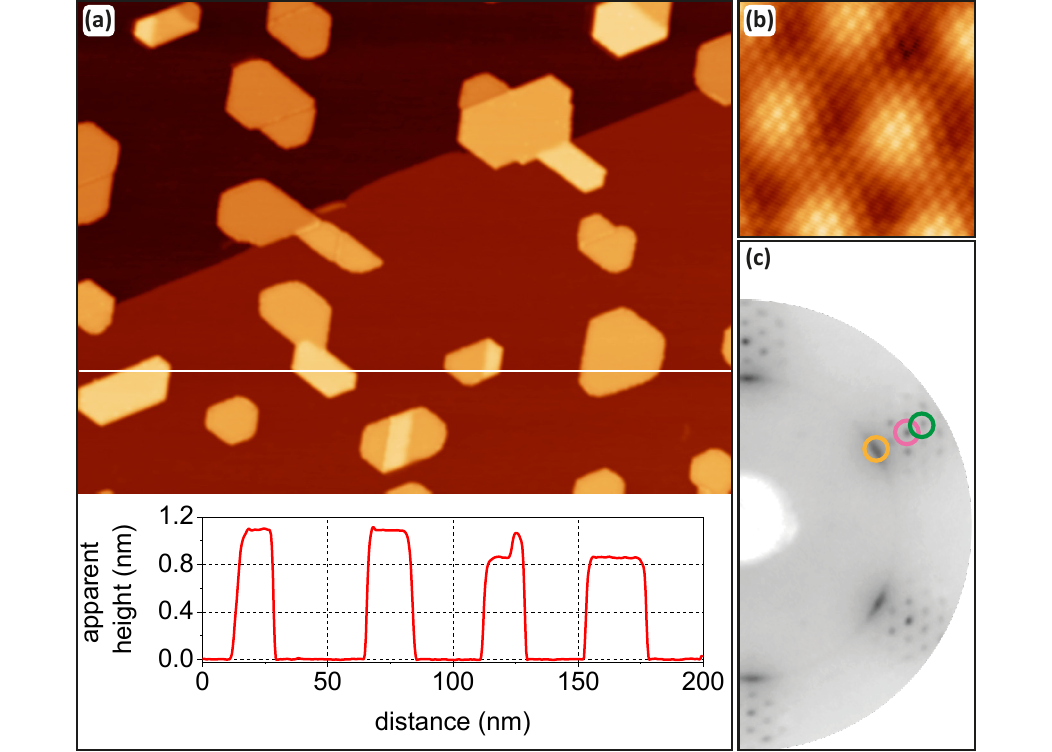}
  \caption{(a) STM topography of of Cr$_x$S$_y$-2D grown on Gr/Ir(111). Sample is prepared by Cr deposition in a sulfur pressure of $5\times10^{-8}$\,mbar, followed by annealing at 850\,K. The lower panel shows the height profile along the white line in the STM image. (b) Atomically resolved STM image of Cr$_2$S$_3$-2D. (c) Contrast-inverted 140\,eV LEED image corresponding to (a). First-order reflections of Ir, Gr, and Cr$_x$S$_y$-2D are encircled by magenta, green, and yellow, respectively.  Images are acquired at 1.7\,K with (a) $V_b$ = 1\,V, $I_t$ = 50\,pA and (b) $V_b$ = 100\,mV, $I_t$ = 100\,pA and have dimensions of (a) 200\,nm $\times$ 150\,nm and (b) 5\,nm $\times$ 5\,nm.}
  \label{fgr:SI_Fig_growth_GrIr111}
\end{figure}

\clearpage
\subsection*{Supplementary Note 8: Work function of Cr$_2$S$_3$-2D and Cr$_{2\frac{2}{3}}$S$_4$-2D}
\begin{figure}[hbt!]
  \includegraphics[width=\linewidth]{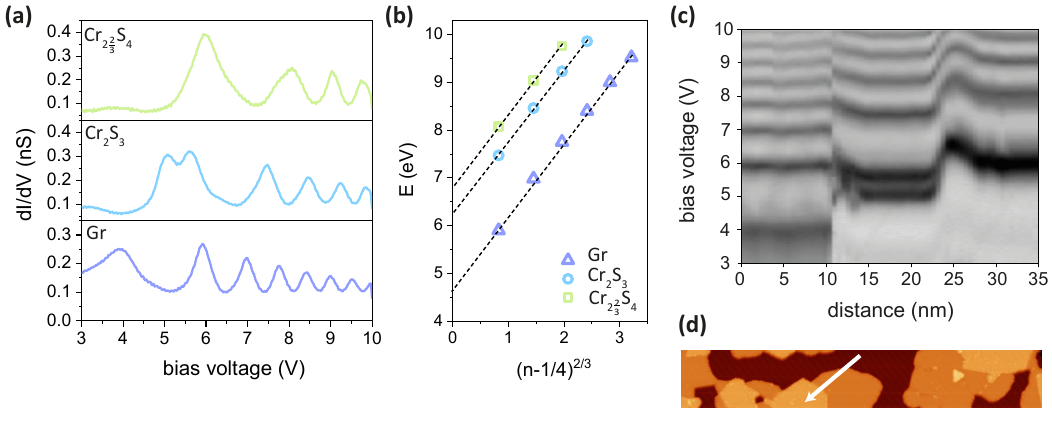}
  \caption{constant-current STS spectra of Gr, Cr$_2$S$_3$-2D, and Cr$_{2\frac{2}{3}}$S$_4$-2D taken in FER region. (b) FERs energies plotted as a function of $(n - 1/4)^{2/3}$. (c) constant-current STS line scan in the field emission resonance region along the white arrow in STM topograph shown in (d). The STS is performed with $V_{st}$ = 2\,V, $I_{st}$ = 50\,pA, $f_{mod}$ = 667\,Hz, and $V_{mod}$ = 40\,mV. 
 } 
  \label{fgr:SI_Fig_work_function}
  \end{figure}

To estimate the work function differences ($\Delta \Phi$) between Cr$_x$S$_y$-2D and Gr-Ir(110) we used field emission resonances (FERs)\cite{Binnig1985, Becker1985}, commonly referred to as Gundlach oscillations\cite{Gundlach1966}, which manifest in the Fowler-Nordheim tunneling regime \cite{Fowler1928}. To explore this, we recorded constant-current $\mathrm{d}I/\mathrm{d}V$ spectra on Cr$_{2\frac{2}{3}}$S$_4$-2D, Cr$_2$S$_3$-2D, and graphene (Gr), distant from the edges of the islands, by sweeping $V_b$ from 3\,V to 10\,V. The resulting spectra, shown in Figure~\ref{fgr:SI_Fig_work_function}(a), reveal distinct peaks associated with FER states arising from the formation of standing waves in the vacuum gap between the tip and the sample. The position of the FER states depends on the $\Phi$ of the sample. The work function can, in principle, be extracted from higher-order FER states using the method described by Lin \textit{et al}.\cite{Lin2007}. Using their method, we plotted the energies of FERs against $(n - 1/4)^{2/3}$, where n is the FER state order [see Figure~\ref{fgr:SI_Fig_work_function}(b)]. The y-axis intercepts of the linear fits of the resonances as a function of energy are estimates for the work function. We obtain for Cr$_2$S$_3$-2D and Cr$_{2\frac{2}{3}}$S$_4$-2D work functions of $\approx$ 1.5\,eV and $\approx$ 2.1\,eV higher than the work function of Gr, respectively. 

Moreover, a $\mathrm{d}I/\mathrm{d}V$ line scan, presented in Figure~\ref{fgr:SI_Fig_work_function}(c) and taken along the white arrow in Figure~\ref{fgr:SI_Fig_work_function}(d), confirmed that the spectral features remain consistent across the Cr$_{2\frac{2}{3}}$S$_4$-2D, Cr$_2$S$_3$-2D, and Gr regions except at the boundaries.

\clearpage
\subsection*{Supplementary Note 9: Bias dependence apparent height of Cr$_2$S$_3$-2D and Cr$_{2\frac{2}{3}}$S$_4$-2D.}
\begin{figure}[hbt!]
\includegraphics[width=\linewidth]{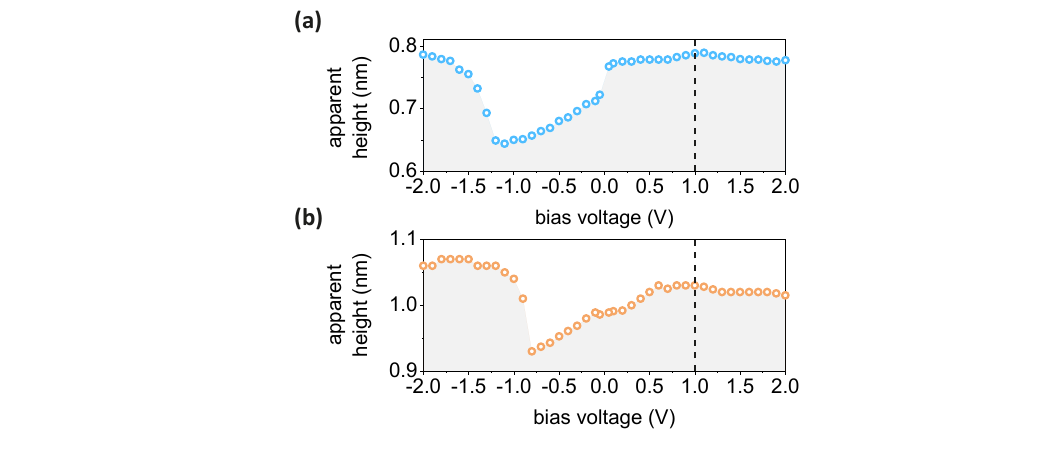}
\caption{Apparent height of (a) Cr$_2$S$_3$-2D and (b) Cr$_{2\frac{2}{3}}$S$_4$-2D on Gr/Ir(110) as a function of $V_\mathrm{b}$ with $I_\mathrm{t}$ = 50\,pA, measured simultaneously with the same tip on neighboring islands of the two compounds. Vertical dashed lines marked the $ V_\mathrm{b} = +1.0$\,V of the reference apparent height. The apparent heights averaged from  $ V_\mathrm{b} = -2.0$\,V to  $ V_\mathrm{b} = +2.0$\,V are  $(0.74 \pm 0.05)$\,nm for Cr$_2$S$_3$-2D and  $(1.02 \pm 0.04)$\,nm for  Cr$_{2\frac{2}{3}}$S$_4$-2D.}
\label{fgr:SI_Fig_app_height}
\end{figure}

\clearpage
\subsection*{Supplementary Note 10: Comparison of the $\mathrm{d}I/\mathrm{d}V$ spectra of Cr$_2$S$_3$-2D on Gr/Ir(110) with Cr$_2$S$_3$-2D on Gr/Ir(111)}
\begin{figure}[hbt!]
  \includegraphics[width=\linewidth]{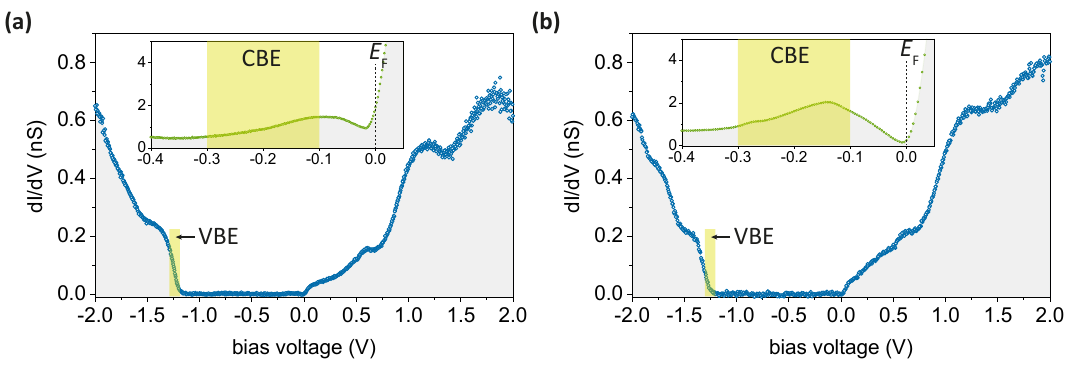}
  \caption{Electronic structure of single layer Cr$_2$S$_3$-2D on Gr/Ir(110) and on Gr/Ir(111). (a) and (b) are $\mathrm{d}I/\mathrm{d}V$ spectra of Cr$_2$S$_3$-2D Gr/Ir(110) and (b) $\mathrm{d}I/\mathrm{d}V$ spectra of Cr$_2$S$_3$-2D Gr/Ir(111). The insets in (a) and (b) are smaller range high-resolution $\mathrm{d}I/\mathrm{d}V$ spectra. Yellow boxes indicate estimated positions and uncertainties of valence band edges (VBEs) and conduction band edges (CBEs). The spectra are obtained at 1.7\,K with (a) $V_{st} = 2$\,V, $I_{st} = 1$\,nA, $f_{mod} = 667$\,Hz, and $V_{mod} = 20$\,mV, and for the insets $V_{st} = 100$\,mV, $I_{st} = 1$\,nA, $f_{mod} = 667$\,Hz, $V_{mod} = 20$\,mV.} 
  \label{fig:SI_Fig_STS_compare}
\end{figure}

\clearpage
\subsection*{Supplementary Note 11: DFT calculated Cr$_2$S$_3$-2D structures}
\begin{table}[h!]
     \begin{center}
     \begin{tabular}{|>{\centering\arraybackslash}p{24mm}|>{\centering\arraybackslash}m{48mm}|>{\centering\arraybackslash}m{16mm}|>{\centering\arraybackslash}m{20mm}|>{\centering\arraybackslash}m{15mm}|
     >{\centering\arraybackslash}m{15mm}|}
    \hline
    coordination & crystal structure side view & Cr-Cr position & $\Delta E$ (eV/Cr$_2$S$_3$) & d (nm) & \textit{a} (nm) \\ 
    \hline
    1H/1H & \includegraphics[width=38mm]{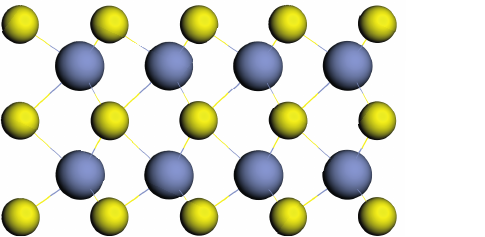} & aligned & 1.06 & 0.610 & 0.327  \\ 
    \hline
    1T/1T & \includegraphics[width=38mm]{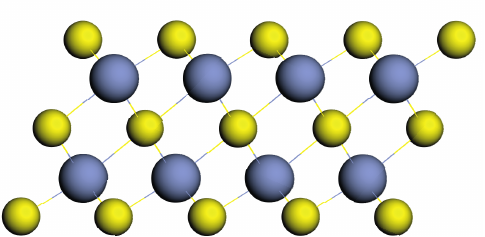} & shifted & 0.17 & 0.564 & 0.343 \\
    \hline
    1H/1H & \includegraphics[width=38mm]{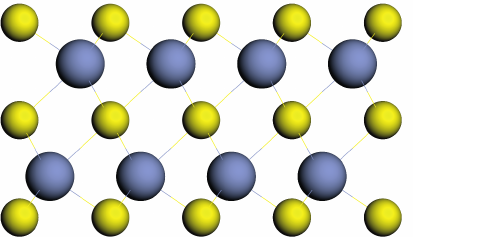} & shifted & 1.22 & 0.616 & 0.333  \\
    \hline
    1T/1T & \includegraphics[width=38mm]{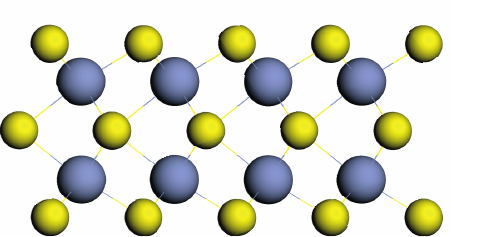} & aligned & 0 & 0.545 & 0.344 \\
    \hline
    1T/1H & \includegraphics[width=38mm]{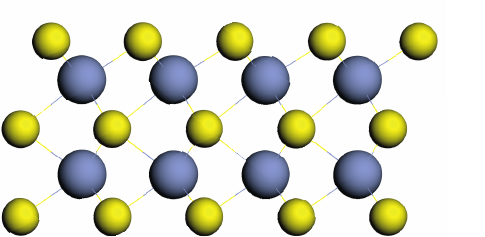} & aligned & 0.61 & 0.585 & 0.338  \\
    \hline
    1T/1H & \includegraphics[width=38mm]{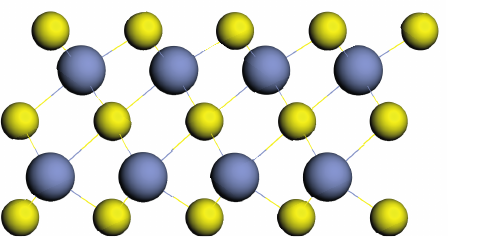} & shifted & 0.76 & 0.593 & 0.336  \\
    \hline
    1H/1T & \includegraphics[width=38mm]{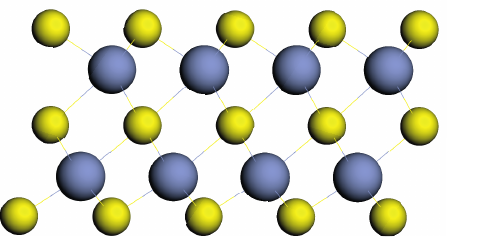} & shifted & 0.76 & 0.592 & 0.338 \\
    \hline
    1H/1T & \includegraphics[width=38mm]{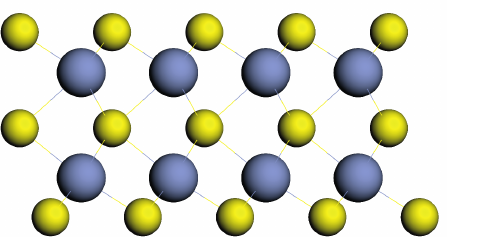} & aligned & 0.80 & 0.559 & 0.339 \\
    \hline
     \end{tabular}
      \caption{DFT calculated Cr$_2$S$_3$-2D structures for all possible stacking sequences. Cr coordination is either trigonal prismatic (H) or octahedral (T).}
      \label{Cr2S3_structures}
      \end{center}
      \end{table}

    \begin{table}[]
    \begin{tabular}{|c|cc|cc|cc|}
    \hline
    \multirow{2}{*}{$U$ (eV)} & \multicolumn{2}{c|}{bandgap (eV)}    & \multicolumn{2}{c|}{E$_\mathrm{AFM}$-E$_\mathrm{FM}$ (meV)} & \multicolumn{2}{c|}{magnetic state}  \\ \cline{2-7} 
                            & \multicolumn{1}{c|}{w/o SOC} & w SOC & \multicolumn{1}{c|}{w/o SOC}            & w SOC             & \multicolumn{1}{c|}{w/o SOC} & w SOC \\ \hline
    0                       & \multicolumn{1}{c|}{0.90}    &   0.88    & \multicolumn{1}{c|}{-51.78}             &         -40.93          & \multicolumn{1}{c|}{AFM}     & AFM   \\ \hline
    0.5                     & \multicolumn{1}{c|}{1.04}    & 1.02  & \multicolumn{1}{c|}{-30.76}             & -31.55            & \multicolumn{1}{c|}{AFM}     & AFM   \\ \hline
    1.0                     & \multicolumn{1}{c|}{1.19}    & 1.16  & \multicolumn{1}{c|}{-16.97}             & -16.44            & \multicolumn{1}{c|}{AFM}     & AFM   \\ \hline
    1.5                     & \multicolumn{1}{c|}{1.30}    & 1.28  & \multicolumn{1}{c|}{-8.03}              & -5.61             & \multicolumn{1}{c|}{AFM}     & AFM   \\ \hline
    2.0                     & \multicolumn{1}{c|}{1.32}    & 1.29  & \multicolumn{1}{c|}{-1.44}              & 2.53              & \multicolumn{1}{c|}{AFM}     & FM    \\ \hline
    2.5                     & \multicolumn{1}{c|}{0.93}    & 0.91  & \multicolumn{1}{c|}{3.53}               & 2.80              & \multicolumn{1}{c|}{FM}      & FM    \\ \hline
    3.0                     & \multicolumn{1}{c|}{0.89}    & 0.86  & \multicolumn{1}{c|}{6.39}               & 3.82              & \multicolumn{1}{c|}{FM}      & FM    \\ \hline
    3.5                     & \multicolumn{1}{c|}{0.84}    & 0.80   & \multicolumn{1}{c|}{8.05}               & 7.86              & \multicolumn{1}{c|}{FM}      & FM    \\ \hline
    \end{tabular}
    \caption{Bandgap and energy difference between AFM and FM state ($E_\mathrm{AFM}-E_\mathrm{FM}$) of Cr$_2$S$_3$-2D as function of Hubbard \textit{U}. The calculations are performed with and without the spin-orbit coupling (SOC) effect.}
    \label{Cr2.66S4_Bandgap_U}
    \end{table}
    
\clearpage
\subsection*{Supplementary Note 12: Optical properties.}
\begin{figure}[hbt!]
  \includegraphics[width=0.7\linewidth]{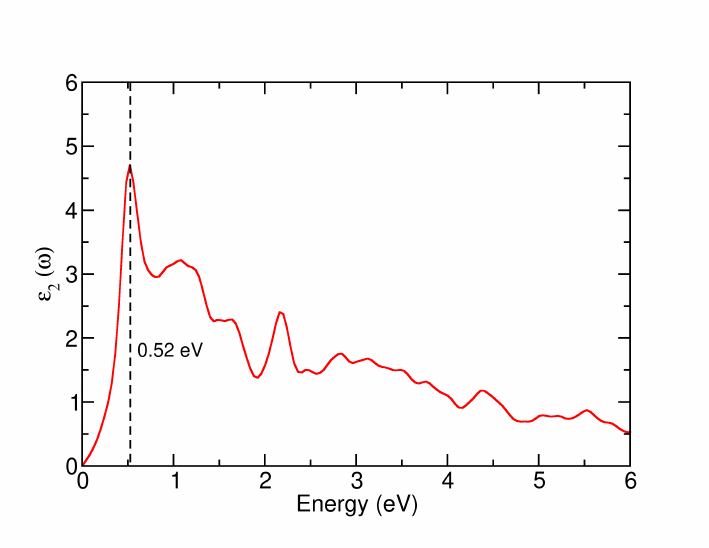}
  \caption{Absorption spectra calculated using the Green’s function–Bethe–Salpeter equation (GW-BSE) within the Tamm-Dancoff approximation. The exciton energy corresponding to the first peak is 0.52~eV. The fundamental gap from the electronic structure of the Cr$_2$S$_3$-2D in the AFM configuration is 0.90~eV. As a result, the exciton binding energy can be extracted as 0.38~eV.} 
  \label{fig:SI_Fig_Mag_config_Cr2.66S4}
\end{figure}

\clearpage
\subsection*{Supplementary Note 13: Comparison of the PDOS of free standing and graphene supported}
\begin{figure}[hbt!]
  \includegraphics[width=\linewidth]{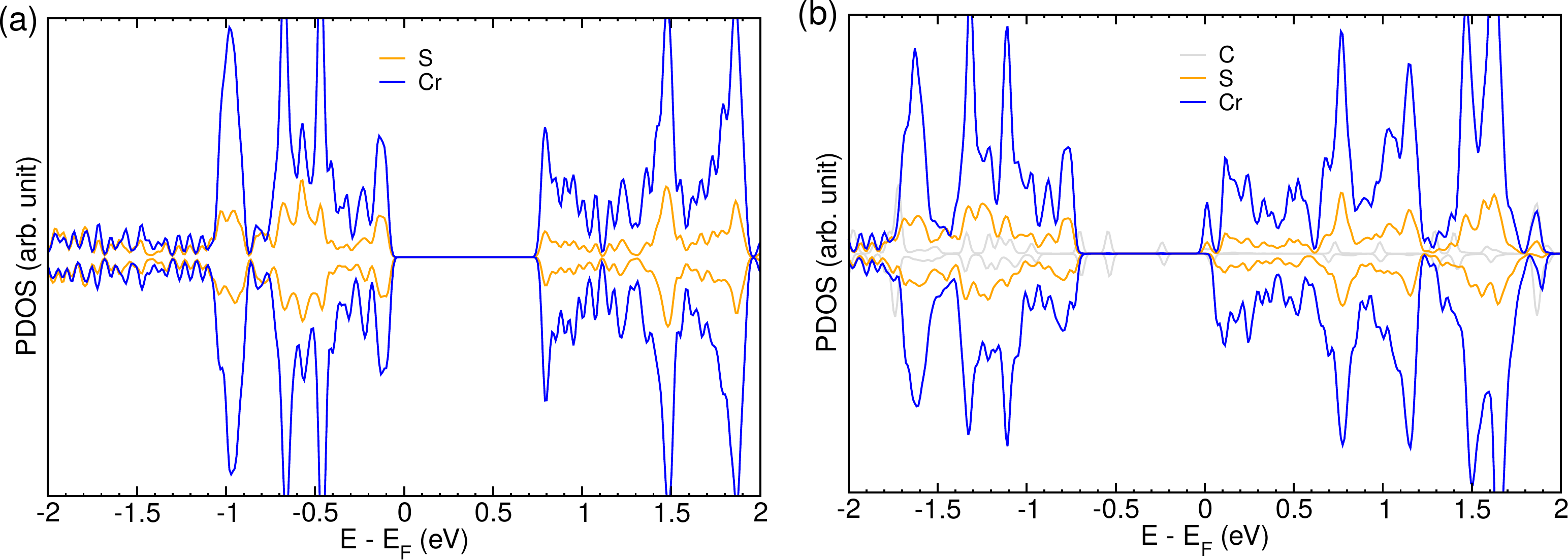}
  \caption{Projected density of states of (a) free-standing and (b) graphene-supported Cr$_2$S$_3$-2D.} 
  \label{fig:SI_Fig_DFT_PDOS}
\end{figure}

\clearpage
\subsection*{Supplementary Note 14: Magnetic configurations considered for  Cr$_2$S$_3$-2D.}
\begin{figure}[hbt!]
  \includegraphics[width=\linewidth]{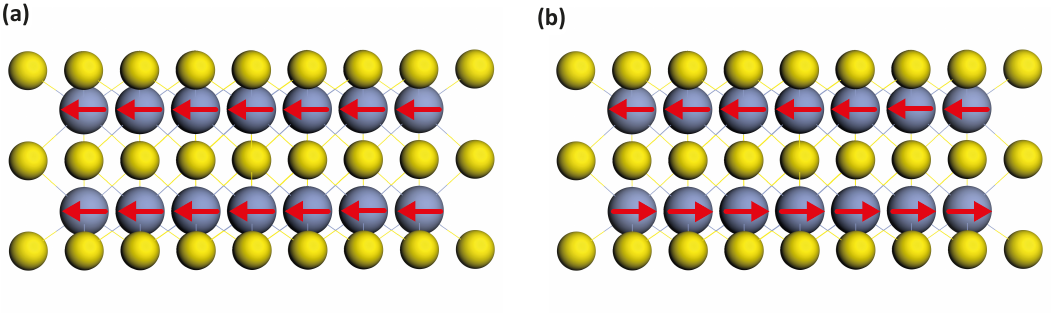}
  \caption{(a) Ferromagnetic (FM) and (b) A-type antiferromagnetic (AFM) configurations for Cr$_2$S$_3$-2D. For simplicity, we display magnetic moments pointing out-of-plane, our calculations did not include spin-orbit coupling and a statement on the magnetic anisotropy cannot be made.} 
  \label{fig:SI_Fig_Mag_config_Cr2S3}
\end{figure}

\clearpage
\subsection*{Supplementary Note 15: DFT calculated Cr$_{2\frac{2}{3}}$S$_4$-2D structures}
\begin{table}[h!]
     \begin{center}
     \begin{tabular}{|>{\centering\arraybackslash}m{22mm}|>{\centering\arraybackslash}m{30mm}|>{\centering\arraybackslash}m{30mm}|>{\centering\arraybackslash}m{30mm}|>{\centering\arraybackslash}m{30mm}|}
    \hline
    Stacking &  1T/1T/1T & 1T/1T/1T & 1T/1T/1T & 1T/1T/1T \\
    \hline
    Cr-Cr position &  aligned & aligned & shifted & shifted \\
    \hline
    1/3-Cr atoms missing plane &  middle & top & middle & top \\
    \hline
    Crystal structure & \includegraphics[width=35mm]{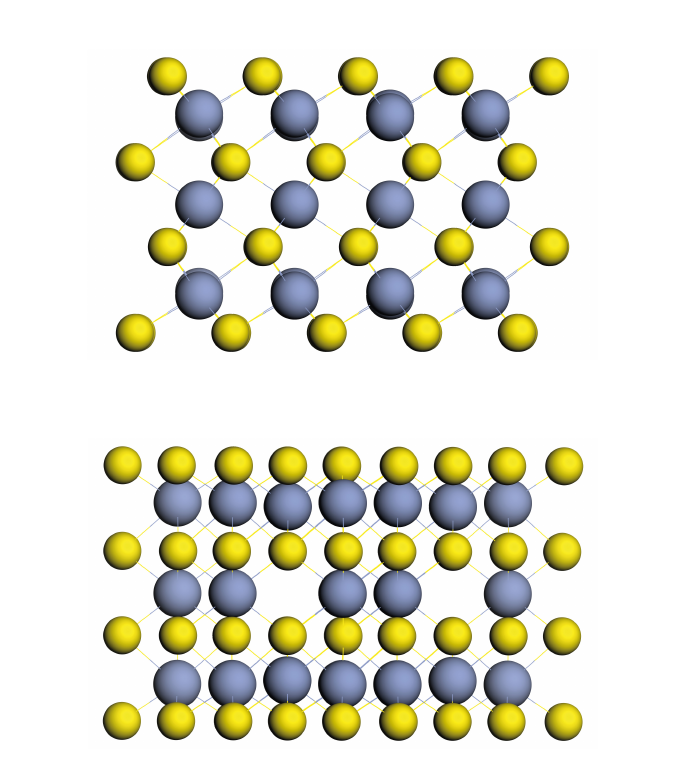} & \includegraphics[width=35mm]{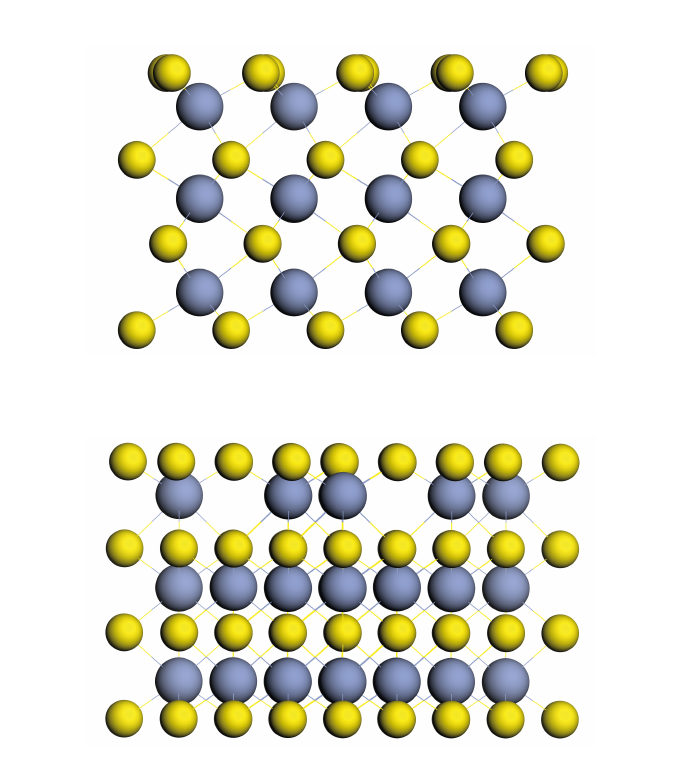} & \includegraphics[width=35mm]{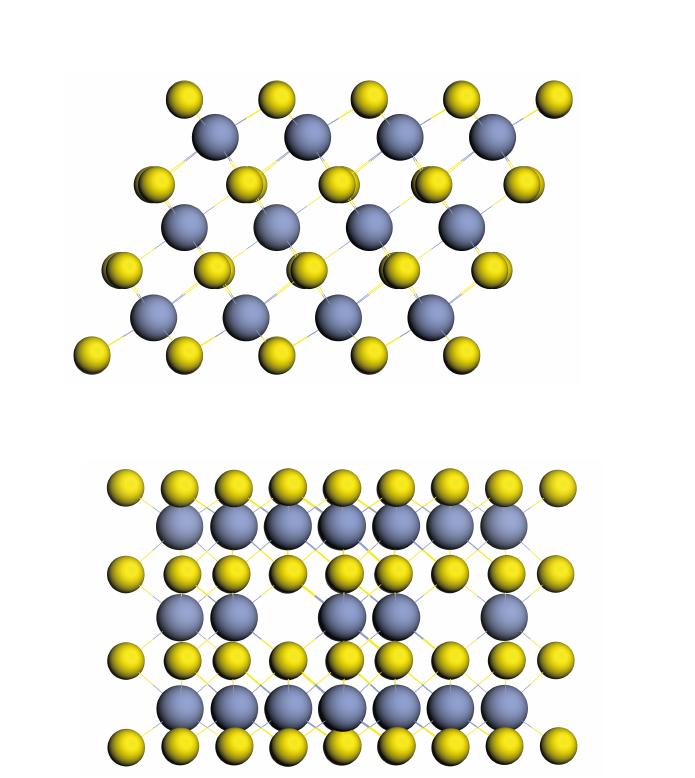} & \includegraphics[width=35mm]{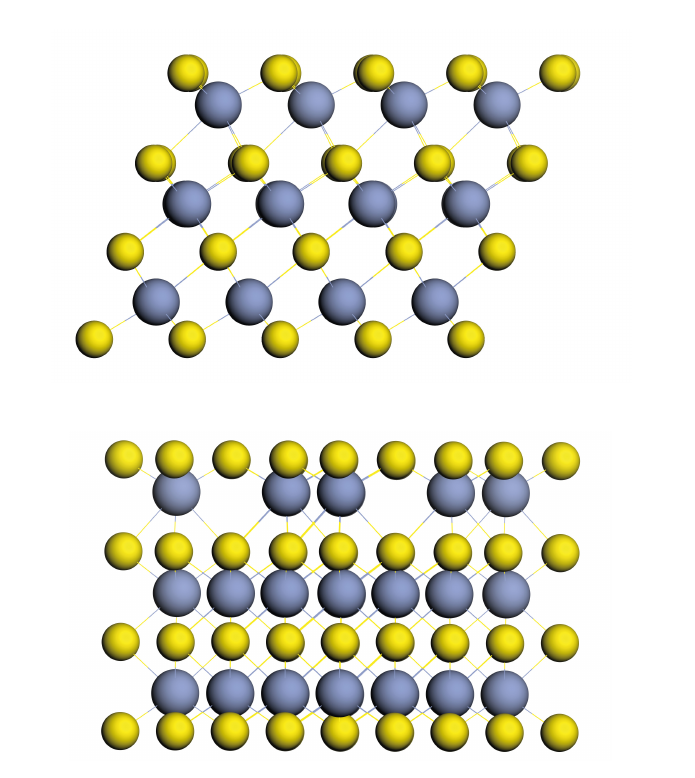} \\ 
    \hline
    \textit{a} (nm) & 0.347  & 0.347 & 0.346 & 0.346 \\ 
    \hline
    \textit{d} (nm) & 0.806 & 0.818 & 0.833 & 0.856 \\
    \hline
    \centering $\Delta E$ (eV/Cr$_{2\frac{2}{3}}$S$_4$) & 0 & 0.66 & 0.32 & 0.96 \\
    \hline
    \end{tabular}
    \caption{DFT calculated Cr$_{2\frac{2}{3}}$S$_4$-2D structures for all possible stacking sequences.}
      \label{Cr2S3_structures}
      \end{center}
      \end{table}

\begin{table}[h!]
     \begin{center}
     \begin{tabular}{|>{\centering\arraybackslash}m{20mm}|>{\centering\arraybackslash}m{30mm}|>{\centering\arraybackslash}m{30mm}|>{\centering\arraybackslash}m{30mm}|>{\centering\arraybackslash}m{30mm}|}
    \hline
    \textit{U} (eV) &  bandgap (eV) & E$_\mathrm{FiM-I}$-E$_\mathrm{FM}$ & E$_\mathrm{FiM-II}$-E$_\mathrm{FM}$ & magnetic state\\
    \hline
    0.0 &  0.33 & -69.22 & -55.96 & FiM-I\\
    \hline
    0.5 &  0.44 & -25.07 & -25.15 & FiM-II\\
    \hline    
    1.0 &  0.54 & 4.95 & -4.59 & FM\\
    \hline
    1.5 &  0.58 & 26.31 & 9.27 & FM\\
    \hline
    2.0 & 0.62 & 41.42 & 18.79 & FM\\
    \hline
    2.5 & 0.64  & 53.09 & 25.91 & FM\\
    \hline
    3.0 &  0.65 & 61.37 & 30.69 & FM\\
    \hline
    \end{tabular}
    \caption{Bandgap and energies difference between FiM-I/FiM-II and FM state (E$_\mathrm{FiM-I}$-E$_\mathrm{FM}$) for Cr$_{2\frac{2}{3}}$S$_4$-2D as a function on \textit{U}.}
      \label{Cr2.66S4_Bandgap_U}
      \end{center}
      \end{table}
      
\clearpage
\subsection*{Supplementary Note 16: Magnetic configurations considered for  Cr$_{2\frac{2}{3}}$S$_4$-2D.}
\begin{figure}[hbt!]
  \includegraphics[width=\linewidth]{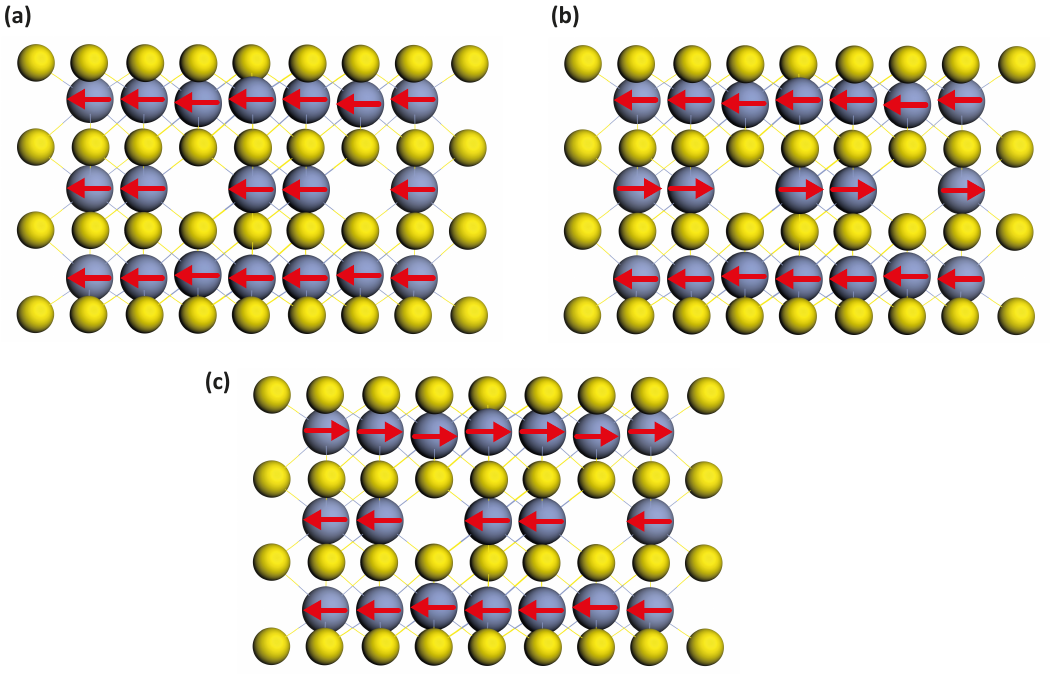}
  \caption{(a) Ferromagnetic (FM), (b) Ferrimagnetic-I (FiM-I) and Ferrimagnetic-II (FiM-II) for Cr$_{2\frac{2}{3}}$S$_4$-2D.} 
  \label{fig:SI_Fig_Mag_config_Cr2.66S4}
\end{figure}

\newpage
\bibliography{Ref}